# Star Formation in Self-gravitating Disks in Active Galactic Nuclei. III. Efficient Production of Iron and Infrared Spectral Energy Distributions

Jian-Min Wang[1,2,3], Shuo Zhai[1,4], Yan-Rong Li[1], Yu-Yang Songsheng[1], Luis C. Ho[5,6], Yong-Jie Chen[1,4], Jun-Rong Liu[1,4], Pu Du[1], and Ye-Fei Yuan[7]
[1] Key Laboratory for Particle Astrophysics, Institute of High Energy Physics, Chinese Academy of Sciences, 19B Yuquan Road, Beijing 100049, People's Republic of China
[2] School of Astronomy and Space Sciences, University of Chinese Academy of Sciences, 19A Yuquan Road, Beijing 100049, People's Republic of China
[3] National Astronomical Observatory of China, 20A Datun Road, Beijing 100020, People's Republic of China
[4] School of Physical Sciences, University of Chinese Academy of Sciences, 19A Yuquan Road, Beijing 100049, People's Republic of China
[5] Kavli Institute for Astronomy and Astrophysics, Peking University, Beijing 100871, People's Republic of China
[6] Department of Astronomy, School of Physics, Peking University, Beijing 100871, People's Republic of China
[7] Department of Astronomy, University of Science and Technology of China, Hefei 230026, People's Republic of China

## Abstract

Strong iron lines are a common feature of the optical spectra of active galactic nuclei (AGNs) and quasars from $z \sim 6-7$ to the local universe, and [Fe/Mg] ratios do not show cosmic evolution. During active episodes, accretion disks surrounding supermassive black holes (SMBHs) inevitably form stars in the self-gravitating part, and these stars accrete with high accretion rates. In this paper, we investigate the population evolution of accretion-modified stars (AMSs) to produce iron and magnesium in AGNs. The AMSs, as a new type of star, are allowed to have any metallicity but without significant loss from stellar winds, since the winds are choked by the dense medium of the disks and return to the core stars. Mass functions of the AMS population show a pile-up or cutoff pile-up shape in top-heavy or top-dominant forms if the stellar winds are strong, consistent with the narrow range of supernovae (SNe) explosions driven by the known pair-instability. This provides an efficient way to produce metals. Meanwhile, SN explosions support an inflated disk as a dusty torus. Furthermore, the evolving top-heavy initial mass functions lead to bright luminosity in infrared bands in dusty regions. This contributes a new component in infrared bands, which is independent of the emissions from the central part of accretion disks, appearing as a long-term trending of the NIR continuum compared to optical variations. Moreover, the model can be further tested through reverberation mapping of emission lines, including LIGO/LISA detections of gravitational waves and signatures from spatially resolved observations of GRAVITY+/VLTI.

*Unified Astronomy Thesaurus concepts:* Active galactic nuclei (16); Galaxy accretion disks (562); Supermassive black holes (1663)

## 1. Introduction

Accretion onto supermassive black holes (SMBHs) radiating powerful emissions has been widely accepted as the standard model of the energy sources of active galactic nuclei (AGNs; see an extensive review of Rees 1984). Reverberation mapping (RM) observations of AGNs provide solid evidence that emissions from accretion disks around the central SMBHs are the ionizing sources of gas emitting broad emission lines (Peterson 1993; Osterbrock & Ferland 2006), and also offer the opportunity of measuring SMBH masses (Peterson 2014). This standard scenario emphasizes the pivot roles of SMBHs in AGNs, but many related aspects, such as how to trigger accretion onto SMBHs and its consequences have not been sufficiently understood so far. One of the long-term issues is whether there is an unavoidable connection between the accretion and the chemical evolution of the central regions revealed by the broad emission line spectra. We digress for a moment to point out the intriguing situations tightly related to these issues. In the Galactic center (GC), a disk with a top-heavy stellar initial mass function (MF; with an index of −0.45) has been found with radii ranging from 0.03–0.5 pc ($10^5 \sim 10^6 R_g$ for GC SMBH $M_\bullet = 4 \times 10^6 M_\odot$, and $R_g$ is the gravitational radius of the SMBH; Bartko et al. 2010), which is notably flatter than the typical case of a top-heavy IMF in a very high-density region (gas density of $10^6 M_\odot \mathrm{pc}^{-3}$; see Equation (1).9 in Kroupa & Jerabkova 2021, or Equation (31) in the current paper). Moreover, about 200 young stars have been identified by SINFONI IFU at the Very Large Telescope (VLT) within 30″ (about 1 pc) distributing over warped counter-rotating disks (von Fellenberg et al. 2022). This finding implies that the disk could be a remnant of exceptionally top-heavy star formation in the nuclear regions (e.g., Levin & Beloborodov 2003), and provides thought-provoking clues to understanding the historical physics around SMBHs in the Milky Way. Except for the motivated evidence of the stellar disks in the GC, star formation around the central SMBH cannot be avoided in AGN disks in light of theoretical arguments (Collin & Zahn 1999; Bonnell & Rice 2008; Collin & Zahn 2008), forming the top-heavy stellar disk (Hobbs & Nayakshin 2009). Furthermore, spatially resolved observations of very nearby Seyfert galaxies through the SINFONI reveal that the circumnuclear regions (CNRs; within ∼10 pc) had undergone starburst once about a few tens of megayears ago (Davies et al. 2006, 2007). Quantitatively testing whether this also works in more compact regions of AGNs definitely advances the global scenario of SMBHs with accretion disks and their stellar environments.

Over three decades ago, it was realized that the vicinities of SMBHs are metal-rich, as high as several times the solar abundances (Hamann & Ferland 1992, 1993), and metallicity exhibits no significant cosmic evolution from high-$z$ to local (see a review of Hamann & Ferland 1999, hereafter). Several indicators of metallicity have been analyzed in light of UV line ratios, N V/C IV, He II/C IV, N V/He II (Hamann & Ferland 1992, 1993; Nagao et al. 2006; Shin et al. 2013; Du et al. 2014), Al III/He II, Si IV+OIV]/He II, and Si IV+OIV]/C IV (Marziani & Sulentic 2014; Negrete et al. 2018; Maiolino & Mannucci 2019; Sniegowska et al. 2021; Garnica et al. 2022). Metallicity inferred from these line ratios correlates with UV luminosity (Hamann & Ferland 1999), Eddington ratio (e.g., Shemmer & Netzer 2002), and SMBH masses (Warner et al. 2004); however, these correlations are still controversial (see Figure 17 in Sameshima et al. 2017). The noncosmic evolution of metallicity can be naturally explained by the hypothesis that SMBH activities are episodic over cosmic time (Small & Blandford 1992; Marconi et al. 2004; Wang et al. 2006) and metals episodically produced are swallowed by SMBHs in every episode (an episodic lifetime of $\sim 10^8$ yr in Marconi et al. 2004; Kelly & Shen 2013, in light of accretion growth). Independent measurements from proximity effects of quasar absorption lines (e.g., Bajtlik et al. 1988) have shown a relatively narrow range of the lifetimes of $z=2-3$ quasars, ranging from $\sim 1-30$ Myr (e.g., Khrykin et al. 2021). This implies that metals are produced very efficiently within a timescale shorter than the main-sequence lifetime of $\sim 10\,M_\odot$ stars; namely, there was quite a large number of massive stars (i.e., a top-heavy MF) that produced metals in galactic centers episodically (see an explanation of stellar disk in the galactic center in Panamarev & Kocsis 2022; Owen & Lin 2023). In this scenario, the chemical evolution of AGNs is independent of their host galaxies, leading to the noncosmic evolution of metallicity.

More specifically, evidence is abundant for the overabundance of iron discovered in high-$z$ quasars, providing tight constraints on the metal factories in galactic centers. The flux ratios of broad lines Fe II/Mg II, as a metallicity indicator, are still a matter of debate because there are other factors influencing these ratios, in particular, turbulence motion (e.g., Baldwin et al. 2004; Verner et al. 2004, 2013; Sarkar et al. 2021, hereafter). On the other hand, there is solid evidence that optical Fe II/H$\beta$ ($\mathcal{R}_{Fe}$) is a robust indicator of Eddington ratios (e.g., Boroson & Green 1992; Du & Wang 2019), and this supports that iron emissions cannot be fully in control by random nonabundance factors (turbulence, temperature, and density).[8] The Fe II/Mg II ratios could be a first-order proxy of the abundance of iron to $\alpha$ elements. In any case, iron abundance is an important parameter determining iron fluxes. Recently, JWST observations of $z \gtrsim 6$ quasars show that five out of eight objects have $\mathcal{R}_{Fe} \gtrsim 1$, which is significantly stronger than local PG quasars on average (Yang et al. 2023). However, the quasar J0100+2802 has weak Fe II lines (Eilers et al. 2023). This probably implies that we might have seen episodic production of iron depending on quasar ages.

Interestingly, there is growing potential evidence for supernovae (SNe) in AGNs (Villarroel et al. 2017, 2020).

Traditionally, iron is mainly produced by the explosion of type Ia supernovae (SN Ia) with a delay of $\sim 0.5$ Gyr to the star formation if the progenitors of SN Ia are accreting white dwarfs (e.g., Matteucci & Greggio 1986). The Fe II/Mg II ratios are expected to decline with redshifts. Many reliable measurements of Fe II/Mg II ratios have been done by a large number of astronomers with different samples (at different cosmic redshifts) in the last three decades (see Hamann & Ferland 1999; Dietrich et al. 2002; Iwamuro et al. 2002; Barth et al. 2003; Dietrich et al. 2003; Freudling et al. 2003; Jiang et al. 2007; Kurk et al. 2007; De Rosa et al. 2011, 2014; Mazzucchelli et al. 2017; Sameshima et al. 2017; Shin et al. 2019; Onoue et al. 2020; Sameshima et al. 2020; Schindler et al. 2020; Yang et al. 2021; Wang et al. 2022). However, no cosmic evolution of the Fe II/Mg II ratios has been found so far. The most distant quasar ($z=7.54$) shows a very iron-rich broad-line region (BLR; 20 times solar abundance of iron) implying that very massive Population III stars ($>100\,M_\odot$) and their pair-instability supernovae (PISNe) work in this quasar (Yoshii et al. 2022). Dietrich et al. (2002) suggested that major star formation happened before $z \sim 10$ to explain the rich iron. In light of numerous measurements, intensive star formation and massive stars play a key role in chemical evolution. On the other hand, massive stars (which cannot be Population III stars in quasars because of high metallicity) have strong stellar winds, and CO core masses are not massive enough to produce a significant amount of iron, because they hardly produce enough metals through SN explosions (e.g., Yusof et al. 2022). How to form massive stars and how to produce metals remain open questions so far; moreover, are there signatures of massive stars in spectral energy distributions (SEDs) of AGNs and quasars?

Accretion disks are known to power AGNs, but the process of how gas is supplied to these disks (the part releasing gravitational energy of the central SMBH) remains an open question due to the angular momentum of the fueling gas preventing accretion inward. Self-gravity of the outer part of AGN accretion disks was proposed by Paczyński (1978), where clumps formed by disk self-gravity collide with each other to transport angular momentum outward. Shortly, star formation in the outer part of the disk is believed to be an inevitable consequence (Kolykhalov & Syunyaev 1980; Loose et al. 1982; Shlosman & Begelman 1987; Collin & Zahn 1999; Goodman 2003; Milosavljević & Loeb 2004; Collin & Zahn 2008). Wang et al. (2010) showed that turbulence driven by SN explosion plays a key role in the outward transportation of angular momentum; meanwhile, metals are produced consequently. In this context, a quantitative explanation for the relation between metallicity and luminosity/Eddington ratio becomes feasible. More than 40 yr ago, the notion of starbursts in galactic centers was advocated by Krügel et al. (1983), Terlevich & Melnick (1985), and Terlevich et al. (1992, 1995) to explain AGN phenomenon, but this idea is not favored by many aspects (e.g., Rees 1984). It turns out that this intriguing and feasible idea of starbursts in the central regions of galactic nuclei hardly replaces the standard model of AGNs (powered by accretion onto SMBHs). On the other hand, the activity of star formation in the vicinity of SMBHs is clearly necessary to produce metal elements in AGN BLRs and to build up the way

---

[8] Ionization parameter defined as a ratio of the radiation pressure to gas sets up a constraint on temperature and density, and thus iron spectral fluxes should tend to be an indicator of abundance, rather than turbulence in light of observations.

to supply accretion onto SMBHs for the global scenario of AGNs.

Many thanks are given to RM campaigns (Kaspi et al. 2000; Peterson et al. 2004; Du et al. 2014; Shen et al. 2015; U et al 2022; Shen et al. 2023), showing that the typical radii of AGN BLRs are of the order of $10^3 \sim 10^5 R_g$ (where $R_g$ is the gravitational radius of the central SMBH; see Figure 6 in Du et al. 2016), which is in good agreement with the self-gravitating regions (Goodman 2003; Sirko & Goodman 2003). The idea of star formation is extended to connect with the BLR metallicity by Wang et al. (2010, 2011, 2012) for properties of metallicity with SMBH masses, accretion rates, and luminosity (see also Qi et al. 2022; Fan & Wu 2023). This is also supported by evidence from the correlation between accretion rates and star formation rates (Chen et al. 2009; Zhuang & Ho 2020). In addition, Artymowicz et al. (1993) suggested the interaction of an AGN accretion disk and stars in the nuclear star cluster as another source of metallicity in quasars, and additional details of the evolution of captured stars were discussed (e.g., Cantiello et al. 2021). Stars accreting from the dense environment of AGN disks tend to be top-heavy (Goodman & Tan 2004) and are rapidly evolving (Artymowicz et al. 1993; Cheng & Wang 1999; Collin & Zahn 1999; Cheng & Loeb 2022). However, these accreting stars as a population are poorly understood in AGN disks. We suggest a new terminology for these stars, denoting them "accretion-modified stars" (AMSs)[9], which describes stars accreting from the AGN disks and whose evolutionary tracks are greatly modified by accretion due to the extremely dense environment of AGN disks. The accreting stars could be main-sequence stars, white dwarf stars, neutron stars, or black holes (BHs; Wang et al. 2021, 2021). It is a timing job to make a global budget for metal production and total energy releases of star formation in AGNs.

Recently, there has been mounting tentative evidence independent of the above classical observations for AMSs in AGN accretion disks. LIGO-detected gravitational waves of a merger of a black hole binary composed of $85^{+21}_{-14}\,M_\odot$ and $66^{+17}_{-18}M_\odot$, named GW190521 (Abbott et al. 2020), which are far beyond the direct remanent of massive stars (Woosley et al. 2002). In 2019, an electromagnetic counterpart (EMC) candidate for GW190521 was reported by Graham et al. (2020), and recently more candidates were discussed by Graham et al. (2022). Compact objects (white dwarfs, neutron stars, and stellar black holes) can be produced by massive stars in AGN disks, and gravitational waves emerge from mergers of these compact objects (Cheng & Wang 1999; Tagawa et al. 2020), in particular, stellar-mass BHs can easily grow up to be $\sim 100\,M_\odot$ through accreting gas of AGN disks (Wang et al. 2021). This potential connection favors AGN disk origins of these high-mass black holes (McKernan et al. 2012). The event has inspired many activities as a renewed field. Once one EMC candidate is confirmed by joint observations of LIGO and EMC observations, this will provide a new clue to understand undergoing physics of AGN disks connecting metal productions, emissions of accretion disks, and gravitational waves.

This is the third paper of the series based on Wang et al. (2010). The objective of this series is to explore whether self-gravity, star formation, and the evolution of stellar populations in a gaseous disk could jointly govern the formation of AGN structures, which are known as torus ($\gtrsim 10^5 R_g$), BLRs ($\sim 10^{3-5} R_g$), and the central part of accretion disks emitting optical-ultraviolet and X-rays ($\lesssim 10^2 R_g$). The first two papers focus on the effects of star formation (inside the BLRs) on metallicity gradients of the BLRs (Wang et al. 2011, Paper I) and episodic appearances of the BLRs (Wang et al. 2012, Paper II), respectively. In this paper, we build up iron production through stellar evolution in accretion disks of AGNs and quasars. We emphasize the roles of the fast-growing stars in iron production through accretion, which rapidly drives the Salpeter IMF to evolve into a top-heavy MF. Only massive stars produce significant iron through pair-instability (Woosley et al. 2002). The very dense environment of accretion disks greatly enhances top-heavy MFs. Meanwhile, the AMS populations produce significant emissions contributing to the observed SEDs of AGNs. Section 2 provides descriptions of MF evolution through the continuity equation, and results are presented in Section 3. In Section 4, we calculate iron productions, and in Section 5 we calculate SEDs. Section 6 provides some discussions on the present models, and Section 7 draws conclusions.

## 2. Populations of Accretion-modified Stars

Figure 1 shows the global working scenario of AGNs with AMS within 1 pc CNRs. Using the Toomre parameter of $Q = 1$ and following approaches of Goodman (2003), we have the self-gravitating radius $R_{SG}/R_g = 3.1 \times 10^3\,\alpha_{0.1}^{2/9} \ell_{Edd}^{4/9} M_8^{-2/9}$, where $\alpha_{0.1} = \alpha/0.1$ is the viscosity parameter, $M_8 = M_\bullet/10^8 M_\odot$ is the central SMBH mass, $R_g = GM_\bullet/c^2 = 1.5 \times 10^{13} M_8$ cm is the gravitational radius of the SMBH, $G$ is the gravitational constant, $c$ is the speed of light, $\ell_{Edd} = L_{Bol}/L_{Edd}$ is the Eddington ratio, $L_{Bol}$ is the bolometric luminosity, and $L_{Edd}$ is the Eddington limit luminosity. As previously mentioned, RM campaigns of broad H$\beta$ line and near-infrared (NIR) emissions reliably demonstrate the distance scales of the BLR and dusty torus, respectively. The BLR size (Kaspi et al. 2000; Bentz et al. 2013) revised by strong Fe II lines is given by $R_{H\beta} \approx 36.3 (L_{5100}/10^{44}\,\mathrm{erg\,s^{-1}})^{0.55} 10^{-0.35 \mathcal{R}_{Fe}}$ ltd (Du & Wang 2019), and the NIR region is $R_{NIR} \approx 237$ (or 0.2 pc)$(L_V/10^{44}\mathrm{erg\,s^{-1}})^{0.4}$ ltd (Koshida et al. 2014; Minezaki et al. 2019), where $L_{5100}$ and $L_V$ are the 5100 Å and V-band luminosity, respectively, and $\mathcal{R}_{Fe}$ is the flux ratio of optical iron lines to H$\beta$. To illustrate the consequence of the evolving population of AMS in this paper, we focus on the physics around the radii of $10^5 R_g$ with enough gas for star formation (see Equation (10)).[10] This region is likely situated within the classical torus where the tidal force exerted by the central SMBH considerably restricts the accretion of AMS, resulting in distinct physical properties compared to those discussed earlier. In Section 6, we further explore the radial distribution of stellar populations and its observational effects.

Star formation in AGN accretion disks invokes microphysics and macrophysics, which are different from normal molecular clouds (see an extensive review of McKee & Ostriker 2007, hereafter), and more complicated than the latter. The formation generally deals with highly nonlinear physical processes: turbulence, magnetic fields, self-gravity, and cooling of regions, while the angular momentum of AGN disks additionally works on accretion onto stars. There are mainly two modes of forming stars: (1) direct collapse of clouds, and (2) continuous accretion of dense cores to form high-mass stars (Zinnecker & Yorke 2007). Vital questions remain open: how are magnetic fluxes and angular momentum of the gas lost to make the growth of protostars? In the current context of AGN accretion disks (e.g., Collin & Zahn 1999, 2008; Derdzinski & Mayer 2023), the self-gravity of highly dense disks makes collapses more efficient; on the other hand, the strong turbulence of accreting gas with differential rotation around the central SMBHs prevents collapse, especially if the magnetorotational instability still works efficiently enough in this star-forming region. In addition, radiation fields from the dissipation of the gravitational energy of accreting gas bound by the SMBHs lead to lower star formation efficiencies. Furthermore, IMFs depend on many factors of the disk environments (e.g., gas density and metallicity). We do not seek to solve these questions. Rather, we introduce simple parameterizations to investigate the evolution of the accreting stellar populations formed in AGN disks and the resultant consequences.

---

[9] In this paper, all AMSs are referred to the case in which the accreting objects are normal stars unless they are pointed out specifically. The central accreting objects could be neutron stars (Chen et al. 2023; Li et al. 2023), black holes (Wang et al. 2021; Fan & Wu 2023; Wang et al. 2023), or white dwarfs in published papers.

[10] Some discussions of the $10^4 R_g$ region are made by Wang et al. (2012), where the physical conditions are different from the present regions.

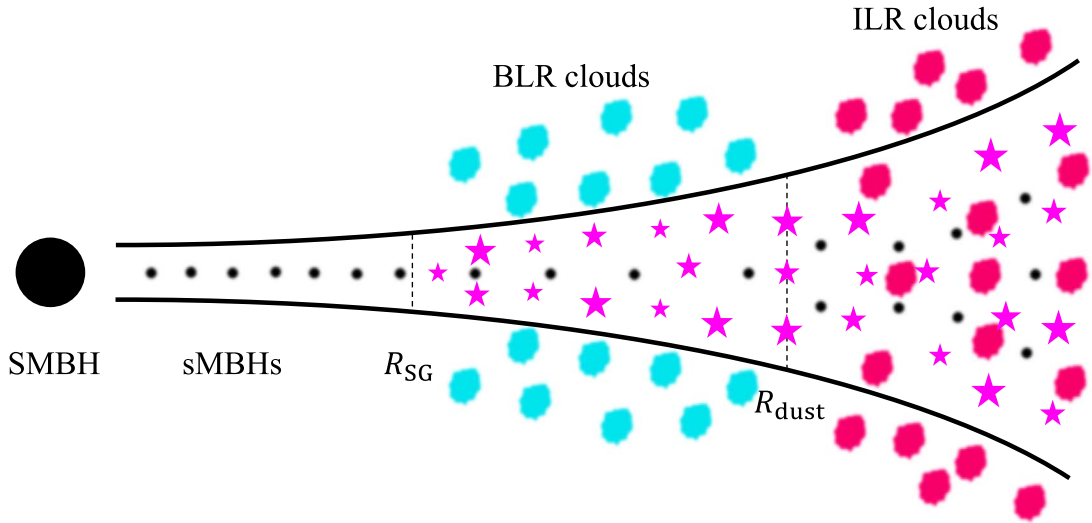

**Figure 1.** The global scenario of parsec-scale CNRs. $R_{SG}$ is the self-gravitating radius, and $R_{dust}$ is the sublimating radius of dust particles. The black dots represent stellar-mass black holes (sMBHs), ★ symbols does accreting stars, and their sizes indicate magnitude of mass. The shallow blue and red clouds are ionized gaseous clouds in the normal BLRs and intermediate-line regions (ILRs; i.e., beyond $R_{dust}$). The observed emission line spectra comprise two components with varying widths that originate from the BLR and ILR. The CNR is a complex environment consisting of accreting SMBHs, stars, and SNe. Both sMBHs and stars are accreting gas from the AGN disk. All of them are called AMSs in this series of papers. The inner part of accretion disks is responsible for powerful radiation, and the outer part acts as an Alchemy furnace of metals.

### 2.1. Governing Equations

We use $\Psi(m_*, t)$ to denote the MF of stars, where $m_*$ is stellar masses and $t$ is the time. It is defined by $\Psi(m_*, t) = \Delta N_*(m_*, t)/\Delta m_*$, where $\Delta N_*(m_*, t)$ is the numbers of stars at time $t$ between $m_*$ and $m_*+\Delta m_*$. Stellar populations can be described by the continuity equation in "mass space" (e.g., Zinnecker 1982), which allows us to include the ongoing physics, to investigate the evolution of AMS populations. The MFs are governed by three factors: (1) the evolution of AMS (accelerate evolution and SN explosion) due to mass gains from the self-gravitating part of accretion disks; (2) stellar winds driven by radiation pressures (Castor et al. 1975); and (3) the input of stars formed in the accretion disks or captured from nuclear star clusters. The governing equation of MFs is then written as

$$\frac{\partial}{\partial t}\Psi(m_*, t) + \frac{\partial}{\partial m_*}[\dot{m}_*\Psi(m_*, t)] = -\dot{\mathcal{R}}_{SN}(m_*, t) + \dot{\mathcal{S}}_{AD}(m_*, t),$$
(1)

where $\dot{m}_*$ is the net increasing rates of stars through accretion and stellar winds, $\dot{\mathcal{R}}_{SN}$ is supernovae rates, and $\dot{\mathcal{S}}_{AD}(m_*, t)$ is star formation rates in the self-gravitating part of the disk including capture rates of stars from nuclear star cluster (e.g., Artymowicz et al. 1993). In principle, $\dot{\mathcal{R}}_{SN}$ is related to star formation history coupled with the MFs, and thus is an integral term of $\Psi$. We assume $\dot{\mathcal{S}}_{AD}$ to be nonzero in a range of $[m_1, m_2]$, which are evaluated by the following arguments. Equation (1) can be numerically solved with the fully implicit difference scheme proposed by Chang & Cooper (1970), which is a popular numerical grid for the type of the continue equation.

AMSs are accreting gas from AGN disks mainly through the equatorial plane. Meanwhile, they are losing mass through winds in the polar regions. We thus have net increases with a rate of

$$\dot{m}_* = \dot{m}_{ac} - \dot{m}_w,$$
(2)

where $\dot{m}_{ac}$ is accretion rates of the AMS, and $\dot{m}_w$ is mass rates of the stellar winds. Figure 2 shows the structure of AMSs. The stellar winds may fall back to the core stars because the AGN disks are so dense that the winds are choked. See Section 2.3 for a brief discussion on this issue. The fallback rates may depend on AGN disk density and stellar mass. Detailed calculations are beyond the scope of this paper, and we absorb the fallback into accretion terms for a simple treatment. Here, we emphasize the roles of $\dot{m}_*$ leading to fast shifts of the population to the high-mass end and a pile-up of stellar mass distributions.

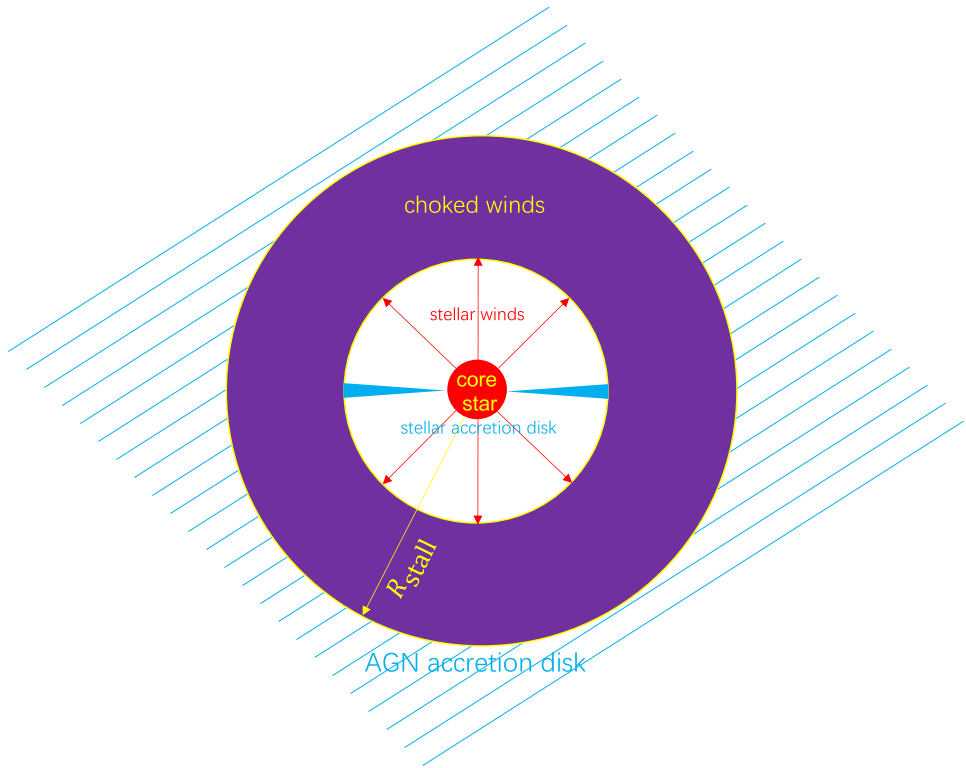

**Figure 2.** Characteristic structure of AMSs formed in AGN accretion disks. The AGN disk is so dense that the stellar winds of the host stars are choked, forming a massive envelope. The envelope stalls somewhere and returns to the host stars. The shocked (choked) envelope could be hot enough to trigger hydrogen burning, leading to an increase in the helium cores of the stars due to the falling back of helium. We note that due to the high number density of AMSs, their envelopes can overlap and become clumpy.

The theoretical calculation of SN rates involves the detailed processes of stellar evolution and star formation. This deals with the delays driven by stellar evolution during the main sequence (as well as during the giant phase), which has been extensively studied about SN Ia rates (Greggio & Renzini 1983; Ruiz-Lapuente & Canal 1998; Kobayashi et al. 2000; Greggio 2005). In the current cases, we take the simplest approach as follows:

$$\dot{\mathcal{R}}_{\rm SN}(m_*, t) \approx \begin{cases} \dfrac{\Psi(m_*, t)}{\tau_{\rm evo}(m_*)}, & (t \geqslant \tau_{\rm evo}) \\ 0, & (t < \tau_{\rm evo}) \end{cases} \quad (3)$$

where $\tau_{\rm evo}$ is the lifetime of stellar main sequences. The approximation means that stars with masses of $m_*$ disappear from the stellar population with $\tau_{\rm evo}$. This formulation for the SN rates allows us to obtain the characteristics of the continuity equation for the first-order approximation, avoiding the difficulties of solving the integral-differential equation. This is similar to the treatment of nonthermal electron spectral evolution in light of the continuity equation (e.g., Equation (1) in Chiaberge & Ghisellini 1999). Moreover, we neglect the evolutionary phases beyond the main-sequence, and approximate that the SN explosion instantaneously happens once departing from the main consequences.

There are plenty of theoretical papers on stellar evolution (e.g., Lamers & Levesque 2017). We plot the main-sequence lifetime–mass, luminosity–mass, and temperature–mass relations in Appendix A. The models are consistent with each other at a level of 10% accuracy. The main-sequence lifetime and luminosity are almost independent of metallicity since they are mainly determined by the hydrogen masses of stars. Therefore, AMS lifetime could be only weakly dependent on the AGN disk chemical composition. However, effective temperatures of the main-sequence stars are sensitive to their metallicity, thereby influencing SEDs of AMS populations. We use the lifetime of main-sequence stars

$$\log\left(\frac{\tau_{\rm evo}}{\rm Myr}\right) = 4.01 - 4.43x + 2.09x^2 - 0.48x^3 + 0.04x^4, \quad (4)$$

for a wide range of stellar masses ($\sim 1000\,M_\odot$; Tout et al. 1996; Schaerer 2002), where $x = \log(m_*/M_\odot)$. It should be noted that $\tau_{\rm evo}$ is insensitive to metallicity.

In this paper, we just discuss the evolution of AMS during a single AGN episode. Its lifetime ($t_{\rm AGN}$) is still a matter of debate, and is suggested to be in a wide range spanning from $10^6 \sim 10^8$ yr (e.g., Martini 2004; Kelly et al. 2010; Khrykin et al. 2021), even shorter in local Seyfert galaxies (Schawinski et al. 2015; but this may be only flickering timescales). As we show subsequently, $t_{\rm AGN}$ cannot be shorter than a few million years in light of iron production. Equation (1) describes the evolution of an AMS population for metal enrichment in one single episode of AGNs.

### 2.2. Accretion-modified Stars

In this paper, we focus on stars in the regions of $10^5\,R_{\rm g}$. There are several models of self-gravitating disks (e.g., Goodman 2003; Sirko & Goodman 2003). For simplicity, we employ the standard model of accretion disks (e.g., Shakura & Sunyaev 1973) to show the environments of AMS to first order to analytically discuss issues of the AMSs. The dimensionless accretion rate of the central SMBH is defined by $\mathscr{\dot{M}} = \dot{M}_\bullet/\dot{M}_{\rm Edd}$, where $\dot{M}_{\rm Edd} = L_{\rm Edd} c^{-2}$ is the Eddington rate, the Eddington luminosity $L_{\rm Edd} = 1.3 \times 10^{46} M_8$ erg s$^{-1}$, and $\dot{M}_\bullet$ is the accretion rate of the central SMBH. The half-thickness, density, mid-plane temperature, and radial velocity of the outer part of the SMBH disk are

$$\begin{cases} H = 5.7 \times 10^{15} \alpha_{0.1}^{-1/10} M_8^{9/10} \mathscr{\dot{M}}^{3/20} r_5^{9/8} \text{ cm}, \\ \rho_d = 0.92 \times 10^{-12} (\alpha_{0.1} M_8)^{-7/10} \mathscr{\dot{M}}^{11/20} r_5^{-15/8} \text{ g cm}^{-3}, \\ T_c = 0.82 \times 10^3 (\alpha_{0.1} M_8)^{-1/5} \mathscr{\dot{M}}^{3/10} r_5^{-3/4} \text{ K}, \\ v_r = 1.5 \times 10^2 \alpha_{0.1}^{4/5} M_8^{-1/5} \mathscr{\dot{M}}^{3/10} r_5^{-1/4} \text{ cm s}^{-1}, \end{cases} \quad (5)$$

respectively (e.g., Kato et al. 2008). Here the dimensionless quantities are the viscosity parameter $\alpha_{0.1} = \alpha/0.1$, $r_5 = R/10^5 R_g$. In the outer part of the disks, gas pressure dominates over radiation pressures, and the opacity is mainly contributed by bound–bound, bound–free, and free–free absorptions, while electron scattering can be neglected. The Kramers' approximation is given by $\kappa_R \approx 1.6 \times 10^2 \rho_{\bar{1}2} T_3^{-7/2}$ cm$^2$ g$^{-1}$, and optical depth $\tau_{\rm AD} \sim 10^{5-6}$, which is extremely optically thick for optical UV photons.

*2.2.1. Star Formation*

Collin & Zahn (1999, 2008), Davies & Lin (2020), and Derdzinski & Mayer (2023) discussed some details of star formation in AGN disks, which are very different from normal molecular clouds (e.g., McKee & Ostriker 2007). Three aspects are stressed here: (1) AGN disks are very dense and warm compared with the normal star formation regions; (2) there is tidal interaction of the central SMBHs with protostars; and (3) AGN disk gas has significant angular momentum. Star formation is still insufficiently understood thus far, so we keep its simplest version of initial star formation in light of the Jeans instability. Gravitationally bound clouds as protostars are formed with radii, timescales, and mass scales of

$$R_G = \frac{c_s}{(G\rho_d)^{1/2}} = 1.1 \times 10^{15} T_3^{1/2} \rho_{\bar{1}2}^{-1/2} \text{ cm}, \quad (6)$$

$$t_G = \frac{1}{(G\rho_d)^{1/2}} = 1.2 \times 10^2 \rho_{\bar{1}2}^{-1/2} \text{ yr}, \quad (7)$$

and

$$m_G = \frac{c_s^3}{(G^3 \rho_d)^{1/2}} = 0.69 T_3^{3/2} \rho_{\bar{1}2}^{-1/2} M_\odot, \quad (8)$$

(see Equations (13) in McKee & Ostriker 2007), where $c_s = 2.9 \times 10^5 T_3^{1/2}$ cm s$^{-1}$ is the sound speed, $T_3 = T/10^3$ is the temperature of the gas, and $\rho_{\bar{1}2} = \rho_d/10^{-12}$ g cm$^{-3}$ is the density of gas. $R_G$ is comparable to the height of the disks at $10^5 R_g$, which is consistent with the Hill radius of $R_H \approx (m_G/M_\bullet)^{1/3} R \approx 3 \times 10^{15} M_8^{2/3} r_5$ cm due to tidal interaction with the central SMBHs. Moreover, the tidal timescale is about $t_{\rm tid} \sim 500 M_8 r_5^{3/2}$ yr, which is comparable with $t_G$. Therefore, tidal interaction efficiently constrains protostar formation. Since the AGN disk density is much higher than normal regions of star formation, the Jean masses (Equation (8)) are about $1 M_\odot$. Consequently, the gravity instability tends to form low-mass protostars rather than massive objects (e.g., McKee & Ostriker 2007); however, the dense environments may offer opportunities for low-mass protostars to rapidly grow subsequently. Outflows from the protostars could be choked by the dense environments (see Section 2.3). The protostars are corotating with surrounding gas around the central SMBHs, and thus we neglect possible collisions or mergers after their formation (e.g., Tan 2000). On the other hand, the intensive radiation field of the AGN disk disfavors the formation of very low-mass stars because radiation pressure prevents the ongoing collapse as shown by both analytical studies (Krumholz & McKee 2008) and detailed numerical simulations (e.g., Tanvir et al. 2022). Moreover, the high-metallicity environments tend to be the Salpeter (see Figure 4 in Marks et al. 2012). The high-density context of AGN disks favors the formation of low-mass stars ($\sim 1-8\,M_\odot$; see Derdzinski & Mayer 2023),[11] which is taken as the input star formation (i.e., the term $\dot{S}_{\rm AD}$ in Equation (1)).

Equations (5) yields the surface density of the AGN disks

$$\Sigma_{\rm gas} = 5.0 \times 10^7 \alpha_{0.1}^{-4/5} M_8^{1/5} \mathscr{\dot{M}}^{7/10} r_5^{-3/4} \, M_\odot \text{ pc}^{-2}, \quad (9)$$

their mass

$$M_{\rm disk} = 5.8 \times 10^7 \alpha_{0.1}^{-4/5} M_8^{11/5} \mathscr{\dot{M}}^{7/10} r_5^{5/4} \, M_\odot, \quad (10)$$

and radial motion timescale

$$t_{\rm AGN} = \frac{R}{v_r} = 3.3 \times 10^8 \alpha_{0.1}^{-4/5} M_8^{6/5} \mathscr{\dot{M}}^{-3/10} r_5^{5/4} \text{ yr}, \quad (11)$$

which is usually regarded as the typical lifetime of AGNs. We find $t_{\rm AGN}$ is consistent with the timescale of iron production. The disk size ($10^5 R_g$) corresponds to 0.5 pc. We note that this surface density could be regarded as averaged values of clumps driven by the self-gravity of this part. The density is much higher than the known ones in starburst galaxies (see Figure 15 in Kennicutt & Evans 2012).

As far as we know, there are no direct formulations to estimate star formation rates from observations in such dense regions (though roughly estimated by Collin & Zahn 1999). The Kennicutt-Schmidt (KS) law is established for regions with a surface density less than $\lesssim 10^5\,M_\odot$ pc$^{-2}$ in various kinds of galaxies (Kennicutt 1998). In the context of AGN disks, the surface density is extremely high ($\sim 10^8\,M_\odot$ pc$^{-2}$), so that the validity of the KS law comes into question. Star formation rates could be regulated by feedback when $\Sigma_{\rm gas}$ is high enough (Ostriker et al. 2010). If we use the KS law, we have a surface rate of star formation $\dot{\Sigma}_* \approx 40 (\Sigma_{\rm gas}/10^8\,M_\odot$ pc$^{-2})^{1.4}\,M_\odot$ yr$^{-1}$ pc$^{-2}$, (see Equation (7) in Kennicutt 1998). All of the gas (Equation (10)) will be converted into stars on a timescale of $\Sigma_{\rm gas}/\dot{\Sigma}_* \sim 7$ Myr, which is much shorter than the episodic lifetime of AGNs (comparing with $t_{\rm AGN}$ in Equation (11)). A more serious issue is that stars contribute more emissions so that they are even brighter than that of accretion onto SMBHs. There is growing evidence that star formation efficiency is in a wide range from 0.15% in G0.253 +0.016, and Sgr B1-off in CNRs of the Milky Way (Lu et al. 2019) or from numerical simulations (Grudić et al. 2022) to 20%–50% (Walker et al. 2018).

---

[11] As usual, we refer to 8–100 $M_\odot$ as massive stars, and $\gtrsim 100\,M_\odot$ as very massive in light of Figure 12 in Woosley et al. (2002).

In such a context of accretion disks, the efficiency of star formation could be depressed by magnetic fields, turbulence, and temperature (see Equation (54) in Collin & Zahn 1999). We keep the power-law index of the KS law (which is determined by the dependence of gravitational collapse) but keep the zero-point open ($\epsilon^*_{\rm AGN}$). Star formation rates can be expressed by

$$\dot{\Sigma}_* = \epsilon^*_{\rm AGN} \dot{\Sigma}_{\rm KS}$$
$$= 15.1\, \epsilon^*_{\rm AGN} \alpha_{0.1}^{-1.12} M_8^{0.28} \dot{\mathcal{M}}^{0.98} r_5^{-1.05}\, M_\odot\, {\rm yr}^{-1}\, {\rm pc}^{-2}, \quad (12)$$

with the Salpeter IMF within mass ranges of $m_1 \leqslant m_* \leqslant m_2$. In this paper, we assume that star formation has the Salpeter function, but stars grow through accretion inside AGN accretion disks to be top-heavy. Star formation rates in this region are given by

$$\dot{\mathcal{R}}_* = \int 2\pi R \dot{\Sigma}_* dR \approx 23.0\, \epsilon^*_{\rm AGN} \alpha_{0.1}^{-1.12} M_8^{2.28} \dot{\mathcal{M}}^{0.98} r_5^{0.95}\, M_\odot\, {\rm yr}^{-1}. \quad (13)$$

Here $\epsilon^*_{\rm AGN}$ is highly unknown since it depends on collisions, mergers, and fragmentation of protostars. We constrain $\epsilon^*_{\rm AGN}$ by the global budget of AGN output powers, where it cannot exceed critical values; otherwise, AGN accretion will lack gas. We would point out that the gas ejected by SNe will be recycled inside the AGN disk (e.g., Cheng & Loeb 2022). Including this recycled gas, $\epsilon^*_{\rm AGN}$ can be relaxed somehow, but we absorbed this effect and star formation efficiency into $\epsilon^*_{\rm AGN}$ in this paper.

### 2.2.2. Growth of AMS

Stars formed inside AGN disks are assumed to couple with the disk gas corotating around the central SMBH. The validity of this assumption is supported by extensive discussions on star interaction with the gaseous disk by Artymowicz et al. (1993). Neglecting the relative motion of stars with respect to the disk, the stars are accreting but must be with rates different from the Bondi rate. Since the sound speed velocity of the SMBH-disk gas is $c_s \approx 2.9\, T_3^{1/2}\, {\rm km\, s^{-1}}$, much slower than the Keplerian, we take the simplest form of $\dot{m}_* = 4\pi (Gm_*)^2 \rho_{\rm d}/c_s^3 \sim 0.14\, M_\odot {\rm yr}^{-1}$, which is too high to be feasible in practice.[12] The AMS growth in AGN disks is strongly influenced by several processes, as we mentioned previously, which are more complicated than the case of massive protostars (Tan & McKee 2004; McKee & Tan 2008). The accretion of stars inside AGN disks has been studied by including feedback to the AGN disk (Dittmann et al. 2021); however, the accretion rates of stars are still described by a phenomenological prescription (Jermyn et al. 2021; Dittmann et al. 2022; Toyouchi et al. 2022). In Section 6.3.1, we briefly discuss the accretion rates. In this paper, we follow the phenomenological prescription of Toyouchi et al. (2022) to investigate the evolution of AMS. We take a simple form of AMS accretion rates

$$\dot{m}_{\rm ac} = \dot{m}_{\rm ac}^0 \left(\frac{m_*}{M_\odot}\right)^2 \left(1 + \frac{m_*}{m_c}\right)^{-\beta}, \quad (14)$$

---

[12] Accreting Population III stars have been studied by several authors (e.g., Omukai & Palla 2003; Ohkubo et al. 2009); these stars are supplied by gas in H I regions. Their accretion rates can be of the order of $\sim 10^{-3} M_\odot\, {\rm yr}^{-1}$ with metallicity of $Z \sim 10^{-5}$–$10^{-6}\, Z_\odot$. The current AMSs are in very different environments from the Population III stars.

where $\dot{m}_{\rm ac}^0$ is a characteristic value depending on AGN disk environments, and $m_c$ is the critical mass of stars. This formulation couples the Bondi-like dependence with other effects on massive stars by $\beta$ (e.g., Toyouchi et al. 2022). The $\dot{m}_{\rm ac}^0$ is dependent on AGN disk density; for example, AMSs can grow up to $10^3 M_\odot$ within $10^6$–$10^7$ yr for densities of $10^{-11}$–$10^{-15}\, {\rm g\, cm}^{-3}$ of AGN disks (Jermyn et al. 2021). Parameters $\beta$ and $m_c$ control the accretion of massive stars avoiding too much growth. We take $\dot{m}_{\rm ac}^0 = 10^{-7} M_\odot\, {\rm yr}^{-1}$ as a typical value in this paper. We adjust three parameters ($\beta, \dot{m}_{\rm ac}^0, m_c$) to test the dependence of AMS accretion rates on environments. As found subsequently, $\beta$ plays a key role in the determination of the massive part of the MFs. We stress that $\beta$ as a comprehensive parameter covers the effects of the accretion dropout process. Equation (14) covers Bondi-like accretion onto low-mass stars, modifying MFs (as discussed by Zinnecker 1982), and Bondi-like modified accretion rates (see discussions in clusters by Adams & Fatuzzo 1996; Basu & Jones 2004; Dib et al. 2010), even with accretion dropout (mass lost from protostar disks; Hoffmann et al. 2018; Essex et al. 2020), which have been applied to explain the power-law tails of MFs in star clusters.

### 2.2.3. Stellar Winds

Massive stars are blowing strong winds through radiation pressure, and losing a large fraction of their masses (e.g., reviews of Castor et al. 1975; Abbott & Conti 1987). If there is no accretion, however, massive stars are not able to enrich the metallicity of their environments because the carbon-oxygen cores are too small to generate a significant fraction of metals (see Maeder 1992; e.g., Figure 16 in Woosley et al. 2002; Yusof et al. 2022), except for PISNe of Population III stars (however, they hardly exist in the nuclear regions). As a result, metal-rich BLRs in AGNs are not explained according to the general theory of massive star evolution.

The present scenario of AMS deals with accretion and wind loss. The stars prefer accretion in the equatorial plane while they are losing mass through winds in the polar direction. The maximum mass of an AMS could be determined by the balance between accretion and stellar winds or stellar evolution. There are very comprehensive studies on stellar winds as a function of metallicity and luminosity (e.g., Nugis & Lamers 2000; Vink et al. 2001), and rotation of stars (Maeder & Meynet 2000; for additional details on stellar winds, see Gräfener & Hamann 2008). In order to show the dependence of AMS evolution on the stellar winds, we take the simple form of stellar winds (Langer 1989)

$$\dot{m}_{\rm w} = \dot{m}_{\rm w}^0 \left(\frac{m_*}{M_\odot}\right)^{\gamma_m} \left(\frac{Z}{Z_\odot}\right)^{\gamma_z}, \quad (15)$$

where $\gamma_m \approx 2.5$ and $\gamma_z = 0.5$–$1$ are indexes and $\dot{m}_{\rm w}^0 \approx 10^{-7} M_\odot\, {\rm yr}^{-1}$. For low-metallicity stars, winds can be neglected. We then have the cutoff mass of AMSs due to stellar winds when $\dot{m}_{\rm ac} = \dot{m}_{\rm w}$ (Artymowicz et al. 1993; Collin & Zahn 1999)

$$\frac{m_{\rm cut}}{M_\odot} = \left(\frac{\dot{m}_{\rm ac}^0}{\dot{m}_{\rm w}^0}\right)^{1/(\beta+\gamma_m-2)} \left(\frac{Z}{Z_\odot}\right)^{-\gamma_z/(\beta+\gamma_m-2)} \left(\frac{m_c}{M_\odot}\right)^{\beta/(\beta+\gamma_m-2)}, \quad (16)$$

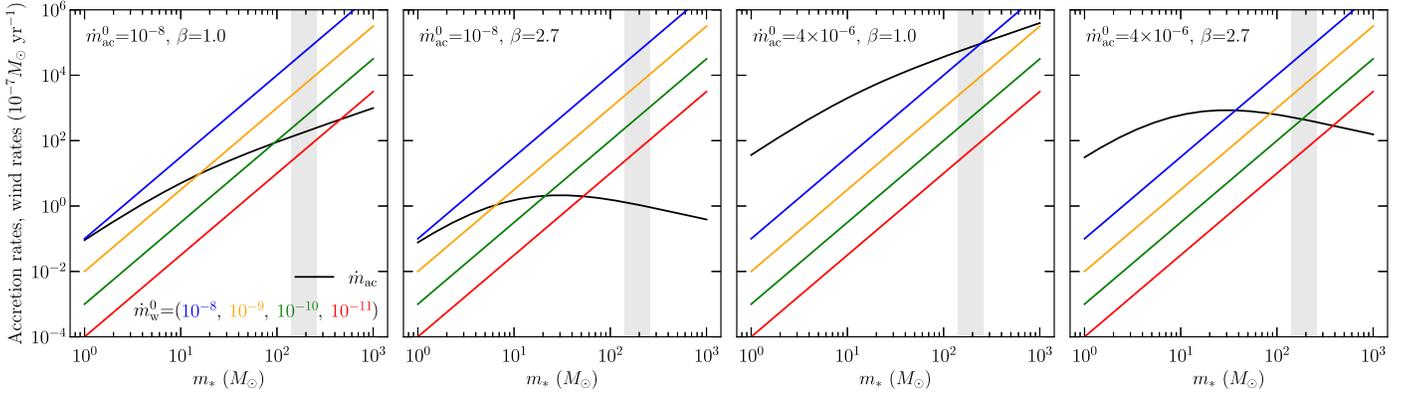

**Figure 3.** Growth rate curves with different characteristic parameters. We fix stellar wind rates but adjust the accretion of the AMS. We plot stellar winds for four $\dot{m}_w^0$ values (in units of $M_\odot$ yr$^{-1}$: very-high-winds in blue, high-winds in yellow, mid-winds in green, and low-winds in red, respectively). The shaded regions are the ranges of PISNe.

when the AMS system can reach a stationary state. Figure 3 shows the $m_{cut}$ for different values of parameters. It is expected to have a cutoff distribution of the continuity equation with a peak at $m_{cut}$.

However, if the stellar winds are not strong, the maximum masses of stars could depend on the balance between accretion and the rapid evolution of massive stars. For AMSs (see Section 2.3), from $\tau_{evo} = \tau_{acc}$, we have

$$m_{max}(\text{stationary}) = \left(\frac{\tau_0 \dot{m}_{ac}^0}{M_\odot}\right)^{1/(\beta-1)} \left(\frac{m_c}{M_\odot}\right)^{\beta/(\beta-1)} M_\odot, \quad (17)$$

where we approximate $\tau_{evo} = \tau_0$ for very massive stars. This is very sensitive to $\beta$, which is only valid for massive stars ($\tau_0 \approx 2 \times 10^6$ yr). For $\beta = 2.0$ and $\dot{m}_{ac}^0 = 5 \times 10^{-7} M_\odot$ yr$^{-1}$, we have the maximum masses of stars $m_{max} = (20, 16, 14) M_\odot$, respectively. It is expected to see stars weighing a few 1000 $M_\odot$ in populous galaxies from numerical simulations (e.g., Elmegreen 2000). The observed stars have initial masses of up to $m_* \approx 300 M_\odot$ (Crowther et al. 2010; Schneider et al. 2014), which are expected to appear in AGN disks.

Finally, we would like to point out that the number of AMSs is determined by the injection of $\dot{S}_{AD}$ (actually by $\epsilon_{AGN}^*$), but the shapes of MFs mainly depend on AMS accretion rates. In order to clarify all parameters used in this global model of AGNs with AMS, we summarize in Table 1, which covers parameters of the central SMBH, stellar, and AMS physics.

### 2.3. Choked Stellar Winds of Massive Stars

Considering that the AGN disk is very dense, we explore the situation of choked stellar winds of massive stars in this subsection, suggesting a new scenario for the metal production of massive stars with solar metallicity. The present idea of choked stellar winds of massive stars goes back to Shull (1980a), who suggested the fast winds will be slowed down by radiation loss and stall somewhere depending on their surroundings and radiation. Stellar winds drive strong shocks through interaction with their surrounding medium. The typical speed of wind is $V_w \sim 10^3$ km s$^{-1}$ whereas the sound speed of an AGN disk is $c_s \approx 2.9 T_3^{1/2}$ km s$^{-1}$. The Mach number is $\mathcal{M} \approx V_w/c_s \approx 300$. The temperature of the post-shock medium is about $T_h \approx \mathcal{M}^2 T_d \approx 10^8 T_3$ K, and the density is four times the pre-shock medium. The cooling time of the post-shock medium is $t_h = 1.7 \times 10^2 T_8^{1/2} n_{12}^{-1}$ s through free–free cooling, where $n_{12}$ is the number density in units of $10^{12}$ cm$^{-3}$. However, the passing through timescale of the shock is $t_{sh} \sim R_w/V_w \sim 7 \times 10^5 R_{100} V_3^{-1}$ s. This indicates that the fast wind will be rapidly thermalized through radiative cooling and stall somewhere. This is the choked winds.

The simplest model of stellar winds with efficient cooling has been studied for an ambient gas density of $\sim 10^5$ cm$^{-3}$ by Shull (1980b), who considered the bremsstrahlung emissions, but it is still much lower than that of AGN disks $\sim 10^{12}$ cm$^{-3}$. The stellar winds evolve into four-zone structures: (a) an inner region of stellar wind; (b) a hot region of shocked wind material and AGN disk medium; (c) a thin and dense shell containing most of the swept-up interstellar matter; and (d) ambient mediums of the AGN disks. We start from stellar wind kinematic energy given by

$$L_w = \frac{1}{2}\dot{m}_w V_w^2 = 1.28 \times 10^{37} \dot{m}_{w,\bar{5}} V_{2000}^2 \text{ erg s}^{-1}, \quad (18)$$

where $\dot{m}_{w,\bar{5}} = \dot{m}_w/10^{-5} M_\odot$ yr$^{-1}$ for $m_* = 10^2 M_\odot$ stars, and the wind velocity $V_{2000} = V_w/2000$ km s$^{-1}$. As shown by Shull (1980b), after the cooling timescale

$$t_{cool} \approx 3.6 L_{37}^{-1/8} \rho_{12}^{-9/8} \text{ hr}, \quad (19)$$

where $L_{37} = L_w/10^{37}$ erg s$^{-1}$, the winds rapidly evolve into the momentum-driven phase from the energy-driven phase. The self-similar solution of the radius of the shell and velocity at this phase (see Equations 12.18 and 12.19 in Lamers & Cassinelli 1999) is given by

$$\frac{R_s(t)}{R_\odot} = 2.0 \times 10^4 L_{37}^{1/4} T_3^{-1/8} \rho_{12}^{-1/4} t_1^{1/2}, \quad (20)$$

and

$$V_s = 21.8 L_{37}^{1/4} T_3^{-1/8} \rho_{12}^{-1/4} t_1^{-1/2} \text{ km s}^{-1}, \quad (21)$$

where we take $V_\infty = c_s$ and $t_1 = t/10$ yr. Considering that the escaping velocity of the AMS is given by

$$V_{esc} = 43.7 \left(\frac{m_*}{10^2 M_\odot}\right)^{1/2} \left[\frac{R_s(t)}{10^4 R_\odot}\right]^{-1/2} \text{ km s}^{-1}, \quad (22)$$

**Table 1**
Parameters of the Global Model of AGN with AMS

| Parameters | Meanings | Values General | Values Galactic Center |
|---|---|---|---|
| SMBH physics | | | |
| $M_\bullet(M_\odot)$ | masses of SMBHs | $10^8$ | $4 \times 10^6$ |
| $\dot{\mathcal{M}}$ | dimensionless accretion rates | 1.0 | 1.0 |
| $t_{\rm AGN}(10^7{\rm yr})$ | AGN episodic lifetime determined by radial motion of AGN disks | 3 | 1 |
| $R(10^5 R_{\rm g})$ | radii of AMS formation regions | 1 | 25 |
| Stellar physics | | | |
| $\alpha_s$ | index of the Salpeter IMFs | 2.35 | 2.35 |
| $m_1(M_\odot)$ | lower limit of stellar population of the Salpeter function | 1 | 1 |
| $m_2(M_\odot)$ | upper limit of the Salpeter MF | 10 | 10 |
| $\dot{m}_w^0(10^{-7}M_\odot{\rm yr}^{-1})$ | mass rates of the stellar winds | 1 | 0 |
| $\gamma_m$ | index of stellar winds on stars masses | 2−2.5 | … |
| $\gamma_Z$ | index of stellar winds on metallicity | 0.5−1 | … |
| AMS physics | | | |
| $\epsilon^*_{\rm AGN}$ | efficiency of star formation rates in AGN disks with respect to the KS | $10^{-2}$ | $8 \times 10^{-4}$ |
| $\beta$ | index of correction of AMS accretion rates due to environment | 1.5−2.7 | 3.0 |
| $\dot{m}_{\rm ac}^0(10^{-7}M_\odot{\rm yr}^{-1})$ | accretion rates of 1 $M_\odot$ AMS | 1 | 6 |
| $m_c(M_\odot)$ | a critical mass of stars with depressed accretion in the context of radiation environments | 10 | 10 |
| $q$ | fraction of AMS accreted masses to the AGN disk masses | [0.01, 0.1] | [0.01, 0.1] |

**Note.** Considering that stellar physics is well understood in principle, we separate it from the AMS physics to clarify the AMS uncertainties.

**Table 2**
Comparison of Accretion-modified Stars with Known Stellar Populations

| Population | X | Y | Z | $m_*/M_\odot$ | Age/Spectra | Wind Properties | Born |
|---|---|---|---|---|---|---|---|
| I | 0.6 ∼ 0.74 | 0.22 ∼ 0.39 | 0.01−0.04 | massive | young (O, B, A) | strong | the Galactic disk, spiral arms |
| II | 0.9 | 0.099 | 0.001 | low mass | old (cold) | weak | spheroid, globular cluster |
| III | 0.75 | 0.25 | 0 | massive or very massive | young | no wind | earlier universe |
| AMS | | arbitrary | | massive or very massive | arbitrary | weak | AGN accretion disks |

**Note.** X, Y, and Z are hydrogen, helium, and metal abundances, respectively.

we have the stalling timescale

$$t_{\rm stall} = 2.4\, L_{37}^{3/2} T_3^{-3/4} \rho_{12}^{-3/2} \left(\frac{m_*}{10^2\, M_\odot}\right)^{-2} {\rm yr}, \quad (23)$$

and the corresponding radius

$$\frac{R_{\rm stall}}{R_\odot} = 9.6 \times 10^3\, L_{37} T_3^{-1/2} \rho_{12}^{-1} \left(\frac{m_*}{10^2\, M_\odot}\right)^{-1}, \quad (24)$$

when $V_s = V_{\rm esc}$, implying that the winds are cumulating at the stalling radius.[13] We note that the stalling radius is just comparable to the vertical thickness of the AGN disks. For more powerful stellar winds, the AGN disks could be not dense enough to choke them, so the winds will be partially choked. The AMS envelopes could overlap with each other inside the disks. We discuss this issue in Section 6.

The above discussions briefly outline the characteristics of the stalled winds. The winds accumulate with time because the diffusion timescale ($R_{\rm stall}/c_s \sim 500$ yr) is much longer than $t_{\rm stall}$, and $R_{\rm stall}$ decreases then but more slowly than $t_{\rm stall}$. The stalled medium forms a very massive envelope around the core star (see Figure 2). Once the envelope accumulates sufficient mass, it may be heated to a temperature at which hydrogen and helium gas begin to burn, and some of the gas falls toward the core of the star under the influence of gravitational attraction, and increases the mass of the core. These stratified structures of AMS would efficiently increase the CO core then, greatly enhancing the iron production through PISNe. However, the head-on colliding interaction between the stellar winds and the fallback gas of the envelope forms strong shocks (e.g., Stevens et al. 1992). When the wind kinematic luminosity is comparable with that of the fallback gas, the shocks produce high temperatures ($10^7 \sim 10^8$ K), and as a result, the high enough density in the envelope likely triggers hydrogen burning with significant luminosity. In order to distinguish AMS from commonly known stars, we briefly compare AMS with stellar populations in Table 2.

Moreover, the massive envelope could be photoionized by the powerful radiation from the core star. UV and optical emission lines will be absorbed, whereas the infrared lines will escape from the regions. The IR emission lines have intermediate width ($\sim 10^3\, r_5^{-1/2}$ km s$^{-1}$). We would like to point out the case of NGC 2992 showing compelling evidence for this signature. As a Seyfert 2 galaxy (optical emission lines only show narrow components), NGC 2992 shows such an intermediate-width Brackett$\gamma$ line (FWHM $\sim 1800$ km s$^{-1}$ estimated from Figure 1 in Davies et al. 2007). Detailed studies of the AMSs will be carried out in a separate paper.

---
[13] Shull (1980b) suggested that the winds stall when $V_w = c_s$, where host star's gravity is not considered. The current situation is different from this case because of the gravity of the host stars.

## 3. Results: Mass Functions

### 3.1. Model Setup

In order to determine the MF, we connect the injection rates with star formation as follows:

$$\int_{m_1}^{m_2} m_* \dot{\mathcal{S}}_{AD}(m_*) dm_* = \dot{\mathcal{R}}_*, \quad (25)$$

which is a constant injection of stars formed inside the accretion disk. We take the IMF in the disk as the Salpeter law, namely, $\dot{\mathcal{S}}_{AD} = \dot{s}_0 (m_*/M_\odot)^{-\alpha_s}$ with a mass range of $1 \leqslant m_*/M_\odot \leqslant 10$, where $\alpha_s = 2.35$ and $\dot{s}_0$ is the rates of star formation. $m_1 = 1 M_\odot$ corresponds to the Jeans masses (Equation (8)) and $m_2 = 10 M_\odot$ to whose main-sequence lifetime is about the AGN episodic lifetime (Equation (11)). As argued by Kroupa & Jerabkova (2021), the IMF takes a probability distribution function if the turbulent-fragmentation process dominates, whereas it takes an optimal distribution function if a self-regulated growth process works. The IMF is poorly understood for the AGN disks. In principle, $\alpha_s$ is a function of metallicity and stellar mass (Bastian et al. 2010; Kroupa & Jerabkova 2021). We can also try other forms of IMF; fortunately, the fast evolution of stars smears out the initial condition. For example, the stationary MFs are not dependent on a log-normal or a simple power-law IMF. We assume that the IMF of the low-mass star formation is

$$\dot{s}_0 = \frac{(2 - \alpha_s) M_\odot^{-2} \dot{\mathcal{R}}_*}{(m_2/M_\odot)^{2-\alpha_s} - (m_1/M_\odot)^{2-\alpha_s}}. \quad (26)$$

Two parameters mainly determine the evolution of the MFs: $(\dot{m}_{ac}^0, \beta)$, where $\dot{m}_{ac}^0$ depends on $\dot{\mathcal{M}}$ of the AGN disks. In principle, $\dot{m}_*$ determines the shifting timescale of the MF (reaching the stationary states), whereas $\beta$ determines the shape of the massive part of the MFs. The accretion rates and star formation efficiency should be limited by

$$M_* = \int_0^{t_{AGN}} dt \int_{m_1}^{m_{max}(t)} \dot{m}_* \Psi(m_*, t) dm_* = q M_{disk}. \quad (27)$$

namely, $\dot{m}_{ac}^0$ and $\epsilon_{AGN}^*$ are then limited by $q$. $\epsilon_{AGN}^*$ in an AGN disk remains open in light of both theory and observations, but it can be constrained by the total emissions from the AMS populations as well as the total AMS in AGN disks.

Considering the three competing processes, (1) injection of stars formed from the AGN accretion disks ($\tau_{SF}$), (2) accretion from the disks ($\tau_{acc}$), and (3) stellar evolution ($\tau_{evo}$), we have the stationary solution when $\partial \Psi(m_*, t)/\partial t = 0$

$$\Psi(m_*) = \begin{cases} \frac{1}{\mathscr{F}(m_*)} \int \dot{\mathcal{S}}_{AD}(m_*) \dot{m}_*^{-1} \mathscr{F}(m_*) dm_*, & (m_1 \leqslant m_* \leqslant m_2), \\ \psi_0 m_*^{-1} \exp\left[-\int [\dot{m}_* \tau_{evo}(m_*)]^{-1} dm_*\right], & (m_* \geqslant m_2), \end{cases} \quad (28)$$

where $\mathscr{F}(m_*) = \exp\left\{\int dm_* [2m_*^{-1} + (\dot{m}_* \tau_{evo})^{-1}]\right\}$, and $\psi_0$ is determined by the connection with $\Psi$ at $m_2$. The initial condition is $\Psi(m_*, 0) = 0$. This formal solution can be analytically expressed if given the complicated functions of $\tau_{acc}$ and $\tau_{evo}$. The stationary functions have a cutoff of $\exp[-\dot{m}_{ac}^0 \tau_0 m_*^{3-\beta} m_c^\beta/(3-\beta)]$ by $\tau_{acc}$ and $\tau_{evo}$. As we found subsequently, the MFs in some cases cannot reach a stationary state within AGN episodic lifetime ($\lesssim 3 \times 10^7$ yr).

In this paper, $\dot{m}_{ac}^0$ is in units of $10^{-7} M_\odot$ yr$^{-1}$. We take $m_c = 10 M_\odot$, which has accretion rates of $10^{-5} M_\odot$ yr$^{-1}$. It is then expected to grow up to $\sim 10^2 M_\odot$ during AGN episodic lifetime ($\sim 10^7$ yr). Though the accretion rates of AMS are $\sim 10^{2\sim 3} L_{Edd}/c^2$ (see Appendix B), the radiated luminosity is still much fainter than the stellar nuclear luminosity. Disk accretion configurations hold in AMS. The rates are consistent with the input in Jermyn et al. (2021) and Cantiello et al. (2021). In this section, we carry out numerical results.

### 3.2. Growth Rates of AMS

Accretion, stellar winds, and hydrogen burning jointly determine the MFs. In order to understand the evolution of MFs, we plot the growth rates of AMS in Figure 3 for different values of characteristic parameters. Though stellar winds are relatively well understood, the accretion of stars remains the most uncertain in AGN disks. As previously pointed out, low-mass stars accrete at Bondi-like rates, while accretion of massive stars is suppressed (by adjusting $\beta$). There is a cross point for given $(\beta, \dot{m}_{ac}^0)$, which is the maximum mass of AMSs determined by Equation (16). When accretion rates are not high enough, the AMSs tend to have a pile-up distribution at $m_*^{max}$. When $\dot{m}_{ac}^0$ is high enough, the maximum mass of AMSs is determined by the balance between the stellar evolution and accretion (Equation (17)).

### 3.3. Evolution from the Salpeter to Top-heavy and Top-dominate MFs

Except for the Salpeter IMF (also with several different $\alpha_s$), we also take a log-normal MF as an input in our calculations. We found that MFs are rapidly evolving insensitive to the IMFs. In order to understand the complicated evolution of MFs, we first focus on the case without stellar winds, which corresponds to the strong fallback of the stalling envelope. Figure 4 shows a general property that the MFs are rapidly evolving from inputs to top-heavy and extremely top-heavy MFs, whatever the inputs are. Only $\dot{m}_{ac}^0$ and $\beta$ jointly control the evolution of the MFs. In Figure 4, panels are arranged by increasing $\beta$ and $\dot{m}_{ac}^0$ from left to right, and from top to bottom, respectively.

#### 3.3.1. Dependence on $\dot{m}_{ac}^0$ and $\beta$

We test MF dependence on $\dot{m}_{ac}^0$ for a given $\beta$. Low-mass stars have rates of $\dot{m}_* \propto m_*^2$ whereas high-mass stars have $\dot{m}_* \propto m_*^{2-\beta}$. In Figure 4, panels from top to bottom show MF evolutions of AMS with accretion rates from $\dot{m}_{ac}^0 = (0.1 \sim 40) \times 10^{-7} M_\odot$ yr$^{-1}$. MFs are evolving slowly if $\dot{m}_{ac}^0 \ll 10^{-7} M_\odot$ yr$^{-1}$, keeping the Salpeter index of the IMF, although the number of stars increases with injection. Low-mass stars grow faster while high-mass ones grow more slowly. Increasing $\dot{m}_{ac}^0$, we find that high-mass stars begin to grow faster than low-mass ones. Consequently, we get faster-evolving MFs at the high-mass end, and the MFs rapidly approach to the top-heavy form. When $\dot{m}_{ac}^0 \gtrsim 10^{-6}\,M_\odot\,\mathrm{yr}^{-1}$, the MFs become top-dominant forms ($m_*\Psi \propto m_*^0$ and maximum mass of AMSs are larger than $10^3 M_\odot$ beyond the scope of the figures). Moreover, the numbers of AMSs decrease with $\dot{m}_{ac}^0$ from the top to bottom panels of Figure 4. This property can be attributed to the rapid growth of AMS, which results in a high rate of supernova explosions. The high explosion rates lead to fast decreases of AMS numbers in the region. It would be not feasible for AMSs with high $\dot{m}_{ac}^0$ and small $\beta$ values; otherwise, SN explosion rates could be too high (too many sMBHs appear in AGN disks).

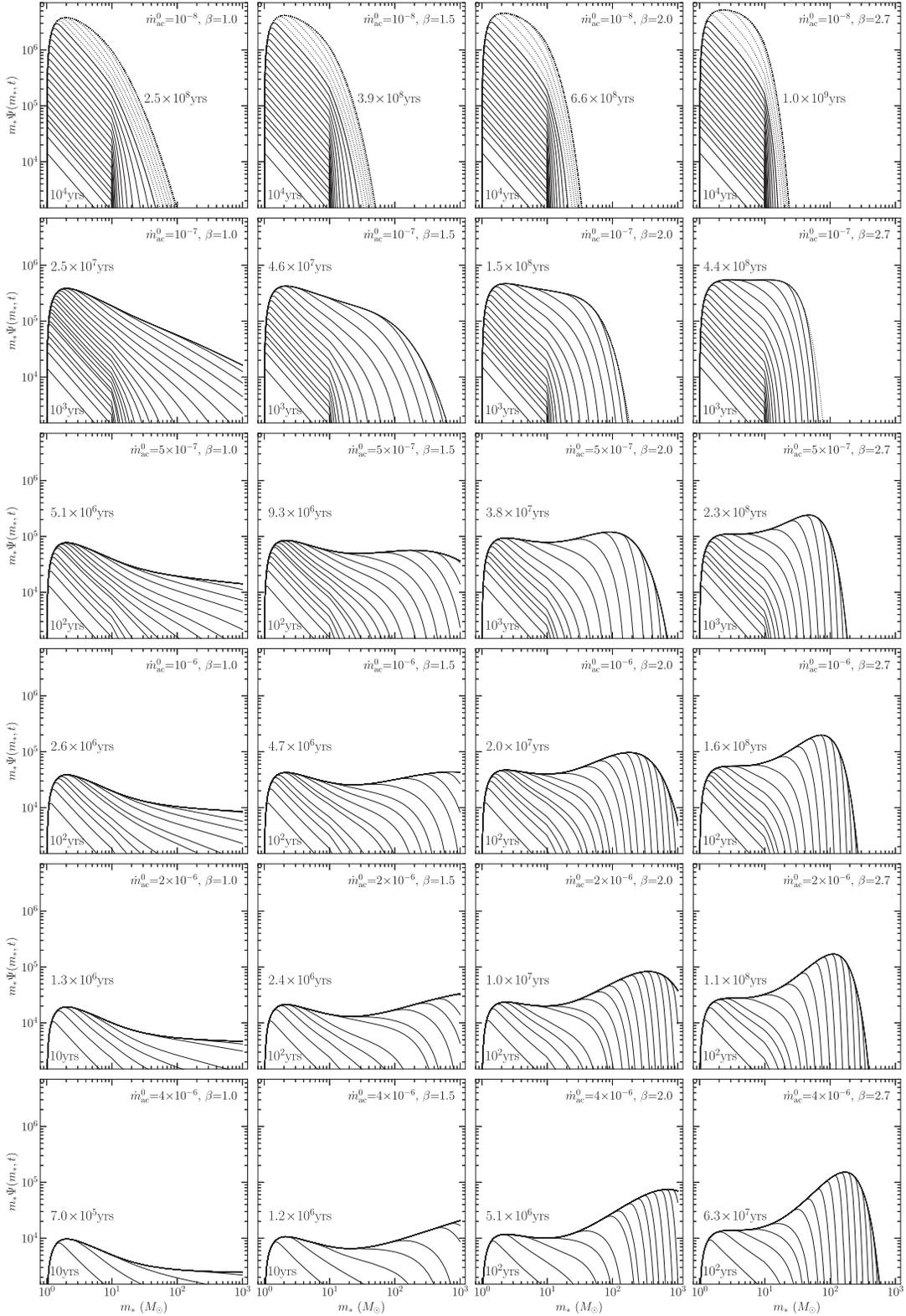

**Figure 4.** The evolving AMS MFs. The panels are arranged as follows: horizontal rows from left to right with increases with $\beta$ for a given $\dot{m}_0$, and the vertical with increases of $\dot{m}_0$ from upper to below. Stellar populations are evolving in the context of the dense environment of AGN accretion disks. Whatever the input stars are, the MFs rapidly evolve from the IMFs. AMS MF is significantly changed by their accretion, not only the slope but also the maximum mass of stars. The dotted lines mean MFs beyond $t_{AGN}$, namely, the MFs cannot reach a stationary state. The time in years marked in the panels is the time attaining the stationary state of MFs.

We fix $\dot{m}_{ac}^0$ to test dependence on $\beta$, namely, the suppression effects of environments on accretion onto stars. Detailed processes of the suppression of accretion onto massive stars are a complicated subject (see review in Zinnecker & Yorke 2007). We use the index $\beta$ as a presumed parameter to include all of the effects of the AGN disk environment and massive stellar winds on the accretion. Since high-mass stars have $\dot{m}_* \propto m_*^{2-\beta}$ for $m_* \gtrsim m_c$, the maximum mass of AMSs is limited by $\beta$; see Equation (17). For intermediate $\dot{m}_{ac}^0$ with very strong suppression, $m_*\Psi$ shows a very flattened form (see the case of $\dot{m}_{ac}^0 = 10^{-7} M_\odot \text{ yr}^{-1}$ and $\beta = 2.7$). With $\dot{m}_{ac}^0$ increasing, there appears to be a pile-up distribution around the maximum AMS, and MFs become the top-dominant form (see the cases of $\dot{m}_{ac}^0 \gtrsim 5 \times 10^{-7} M_\odot \text{ yr}^{-1}$ and $\beta \gtrsim 2.0$). The pile-up masses of AMS increase with $\dot{m}_{ac}^0$ and decrease with $\beta$. This pile-up distribution has obvious advantages: stars just correspond to the narrow window of PISN, which is the most efficient way to produce iron. When including the stellar winds, the pile-up phenomena will be sharper than the case without stellar winds.

Moreover, we find that the time-dependent solution rapidly reaches the stationary solution Equation (28) for most cases except for a few cases of stars with low accretion rates. For example, for massive stars $\tau_{\text{evo}} = \tau_0/m_*$, we have an analytical solution of the stationary MFs expressed by Equation (28), which is in the form of $m_*\Psi(m_*) \propto m_*^{0.3}$ for the massive part of the MF, which is consistent with the numerical results. The power index of MFs is determined jointly by the stellar evolution and accretion ($\beta$, $\tau_0$, $\dot{m}_{ac}^0$).

In this subsection, we find that stellar MFs become top-heavy and top-dominant due to accretion onto stars from AGN disks whatever the input of star's IMFs is. There is a pile-up of massive stars in most cases depending on accretion rates of AMSs and interaction with their surroundings.

### 3.3.2. Effects of Stellar Winds

Stellar winds are a powerful way to lose stellar mass for massive stars. We calculate the cases with wide ranges of physical parameters ($\dot{m}_{ac}^0$, $\beta$, and $\dot{m}_w^0$). Figures 5, 6, and 7 illustrate three different kinds of stellar winds ranging from weak to strong (note that the strength of these winds could depend on metallicity, thus requiring an adjustment of $\dot{m}_w^0$ to account for this effect). The most prominent feature is the pile-up with a cutoff at the high-mass part of the MFs. This cutoff feature is caused by the balance between accretion and winds, i.e., $\dot{m}_* = \dot{m}_{ac} - \dot{m}_w = 0$. When this happens, we stop the calculations for more massive stars (i.e., the cutoff mass). We simply set $\Psi = 0$ for $m_* > m_*^{\text{cut}}$. The cutoff masses decrease with an increase of $\dot{m}_w^0$. As shown by Figure 8, the cutoff masses cannot exceed a few tens of solar mass for high-winds so that metals are not able to be efficiently produced in such a case. It seems that such strong winds do not work in AGNs.

The cutoff pile-up of stellar mass distribution is simply due to the fact that AMS grows faster than hydrogen burning (main-sequence timescales). These pile-up properties provide a more accurate mechanism to produce an PISN in order to explain the unusual ratios of Mg II/Fe II fluxes. Strong winds also prevent the stars from a direct collapse into black holes. The fallback gas may mitigate the impact of stellar winds on the evolution of AMS, resulting in an increase in the maximum mass of stars. However, the overlapping of AMS envelopes introduces additional complexity (clumps form, resulting in uncertainties of fallback gas). All of the uncertainties could be explored by comparing them with the known stellar disks and G2 cloud in the Galactic nuclei.

### 3.4. Pile-up Peaks and $\dot{\mathcal{M}}$

As the prominent feature of MFs, we estimate the peak properties at the pile-up masses. Equating the masses from Equation (16) and (17), we have

$$\dot{m}_w^0 = \dot{m}_{ac}^{0 \frac{1-\gamma_m}{\beta-1}} \left(\frac{Z}{Z_\odot}\right)^{-\gamma_z} \left(\frac{m_c}{M_\odot}\right)^{\frac{\beta(1-\gamma_m)}{\beta-1}} \left(\frac{\tau_0}{M_\odot}\right)^{\frac{\beta+\gamma_m-2}{1-\beta}}, \quad (29)$$

providing a relation for the cutoff masses at the peak between the accretion rates and stellar winds. Moreover, $\dot{m}_* = 0$ and $\partial\Psi/\partial t = 0$ hold at the peak; thus, we have the pile-up MF at the peak from Equation (1)

$$\Psi_{\text{peak}}(m_{\text{cut}}) = \tau_{\text{evo}}(m_{\text{cut}}) \int_{m_1}^{m_2} \dot{\mathcal{S}}_{\text{AD}}(m_*) dm_* \propto \dot{\mathcal{M}}. \quad (30)$$

This implies that $\Psi_{\text{peak}}(m_{\text{cut}}, t)$ follows the accretion rates, i.e., metallicity increases with $\dot{\mathcal{M}}$. For iron production, it needs $m_{\text{cut}}/M_\odot \in [140, 260]$ required by PISN. This can naturally explain why narrow-line Seyfert 1 galaxies are so metal-rich, because of their high accretion rates (e.g., Boroson & Green 1992; Du & Wang 2019). The narrow windows for iron production driven by PISNe should also be satisfied (e.g., Equation (29)); otherwise, they could be iron-poor despite being metal-rich. When the stellar winds are strong enough, on the other hand, MFs seem to be infinite, as shown in Figures 7 and 8. See explanations in Appendix D.

### 3.5. Comparison with Observations

Bastian et al. (2010) summarized the IMF of star formation in various environments. In ultra-compact dwarf galaxies, $\alpha_s \sim 1.0$; namely $m_*\Psi \propto m_*^0$, as a top-heavy form, is found in massive part (Dabringhausen et al. 2009), where gas spatial distributions could be very compact to form a top-heavy MF. In the GC, $\alpha_s = 0.85$ is found based on 41 OB supergiant, giant, and main-sequence stars (Paumard et al. 2006). Interestingly, a top-heavy stellar disk with radii from 0.03–0.5 pc is found to have top-heavy MFs ($\Psi \propto m_*^{-0.45}$) different from the S-star cluster through observations of the SINFONI on the VLT (Bartko et al. 2010), which is generally consistent with the present calculations. Clearly, the GC stellar populations are significantly more top-heavy than those of the ultra-compact dwarf galaxies. X-ray observations of CNRs support the top-heavy MFs (Nayakshin & Sunyaev 2005). This consistency provides evidence that the GC had undergone such an AMS phase, which subsequently triggers the activity of the central SMBH. If metallicity around the GC can be reliably measured, this will provide evidence for the AMS history with top-heavy MFs. This is similar to the case of $\beta = 1.5$ and $\dot{m}_{ac}^0 \gtrsim 5 \times 10^{-7} M_\odot \text{ yr}^{-1}$ shown in Figure 4.

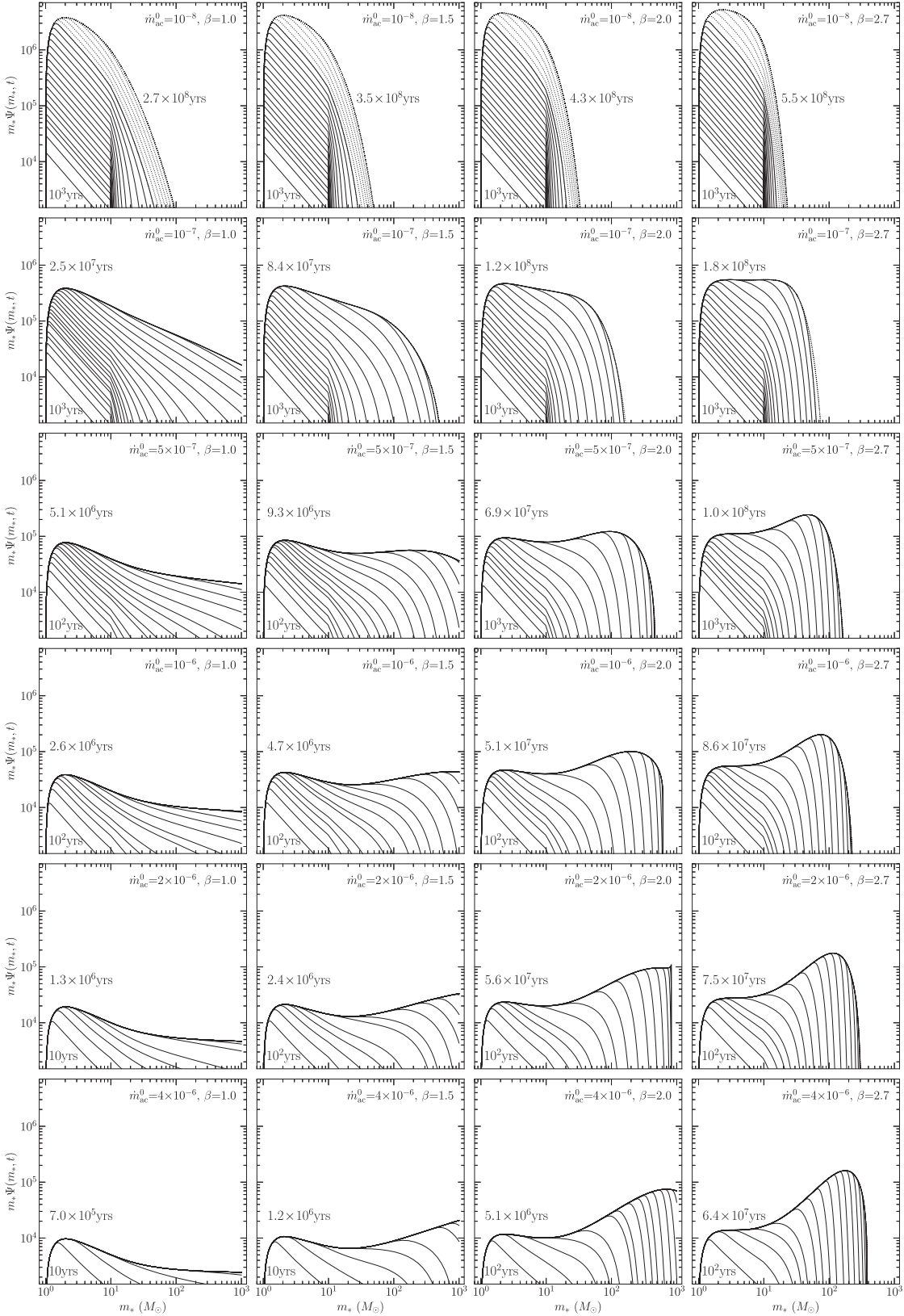

**Figure 5.** Same as Figure 4, but the AMS is undergoing low-winds of $\dot{m}_{\rm w}^0 = 10^{-11}\,M_\odot\,{\rm yr}^{-1}$. This is the case with weak winds, but the MFs begin to show the pile-up features with a cutoff, in particular, for the cases with high $\dot{m}_{\rm ac}^0$.

Given $\beta$, the high-mass MF is generally very sensitive to $\dot{m}_{\rm ac}^0$. When $\dot{m}_{\rm ac}^0 \sim 5 \times 10^{-7}\,M_\odot\,{\rm yr}^{-1}$, we have $m_*\Psi \propto m_*^{\sim 0}$ (i.e., a very flattened shape) as shown by Figure 4. Moreover, the present-day MFs ($m_*\Psi$) can increase with $m_*$ (i.e., $\beta \gtrsim 2.0$), which $m_*\Psi \propto m_*^{0.55}$ assembles to that of the GC (Bartko et al. 2010), even to more extreme cases of MF peaking at $\gtrsim 10^2\,M_\odot$. However, such top-heavy-MFs have not been reported as far as we know, except for this case in the GC. A simple application to GC stars is provided in Section 3.6, and a detailed version of the present model of the top-heavy stellar disk will be carried out separately. Multi-Unit Spectroscopic Explorer observations of M87 reveal a metal-rich core ($\lesssim 30''$) indicating an early and brief star formation history but a prominent negative IMF gradient (Sarzi et al. 2018). It seems that the M87 core has a distinct history from the GC. It is necessary to explore stellar MFs in accretion disks of a few nearby AGNs (e.g., Circinus galaxy, NGC 1068), potentially through spatially resolved observations of GRAVITY+/VLTI in order to construct the histories of stellar evolution and gas around the central SMBH (Sarzi et al. 2018).[14] Additionally, the stellar metallicity of these stars and the interstellar medium (ISM) in galactic nuclei will be crucial to test the stellar history.

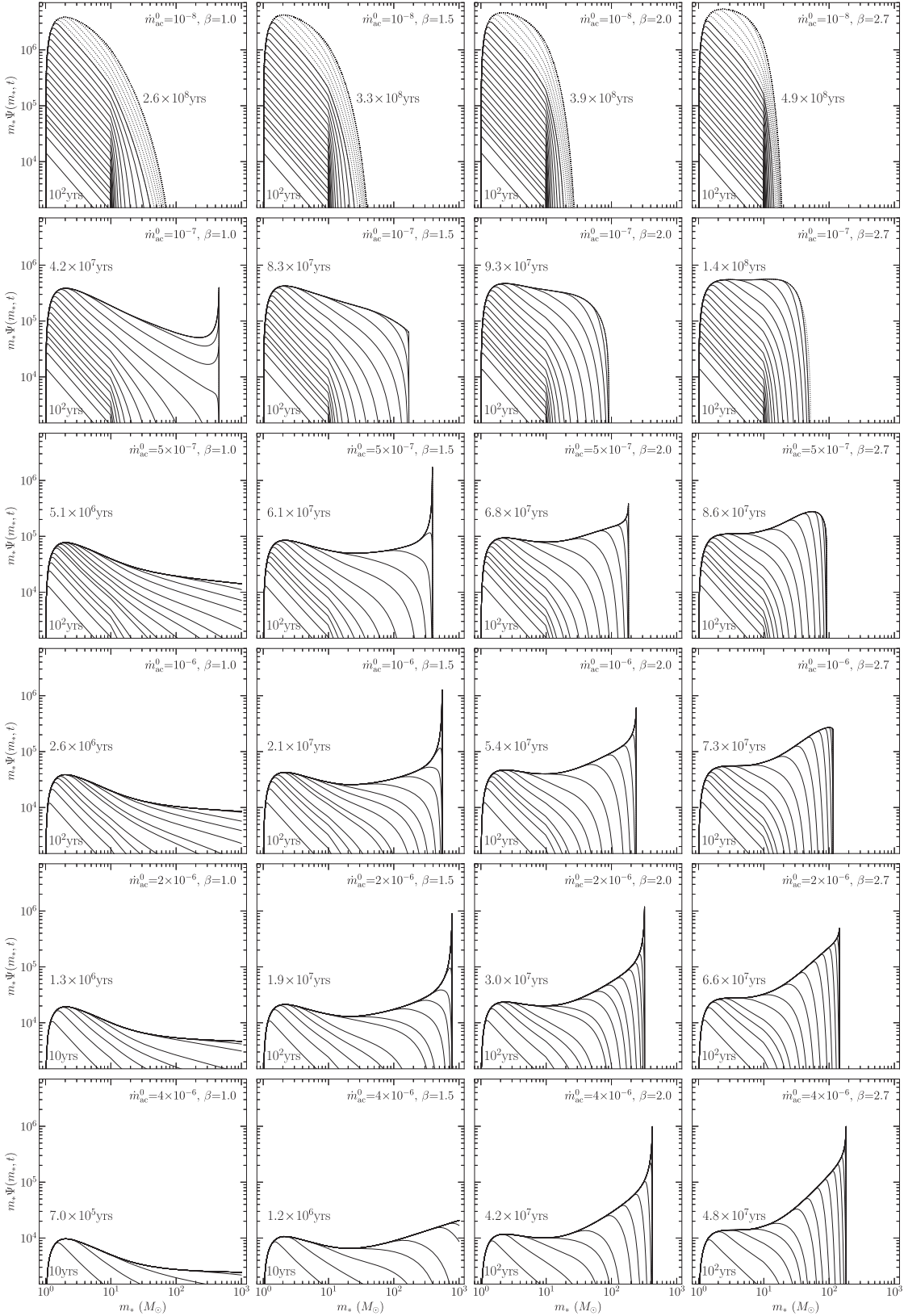

**Figure 6.** Same as Figure 4, but the evolving AMS has with mid-winds of $\dot{m}_w^0 = 10^{-10}\,M_\odot$ yr$^{-1}$. This still allows for the formation of very massive stars for iron.

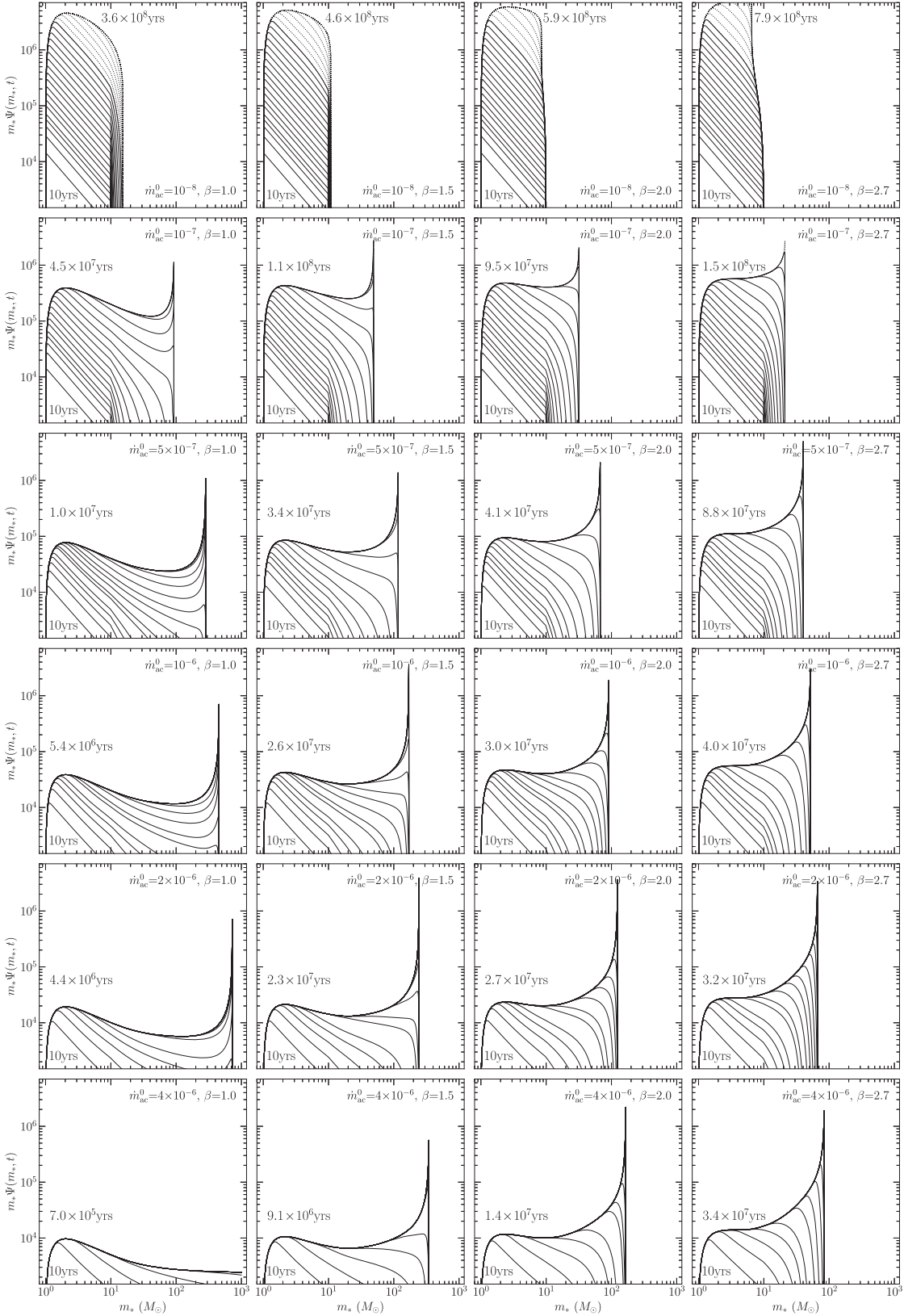

**Figure 7.** Same as Figure 4, but the evolving AMS has high-winds of $\dot{m}_w^0 = 10^{-9}\,M_\odot\,{\rm yr}^{-1}$, showing very strong cutoff features. The cutoff due to stellar winds is too strong to produce iron.

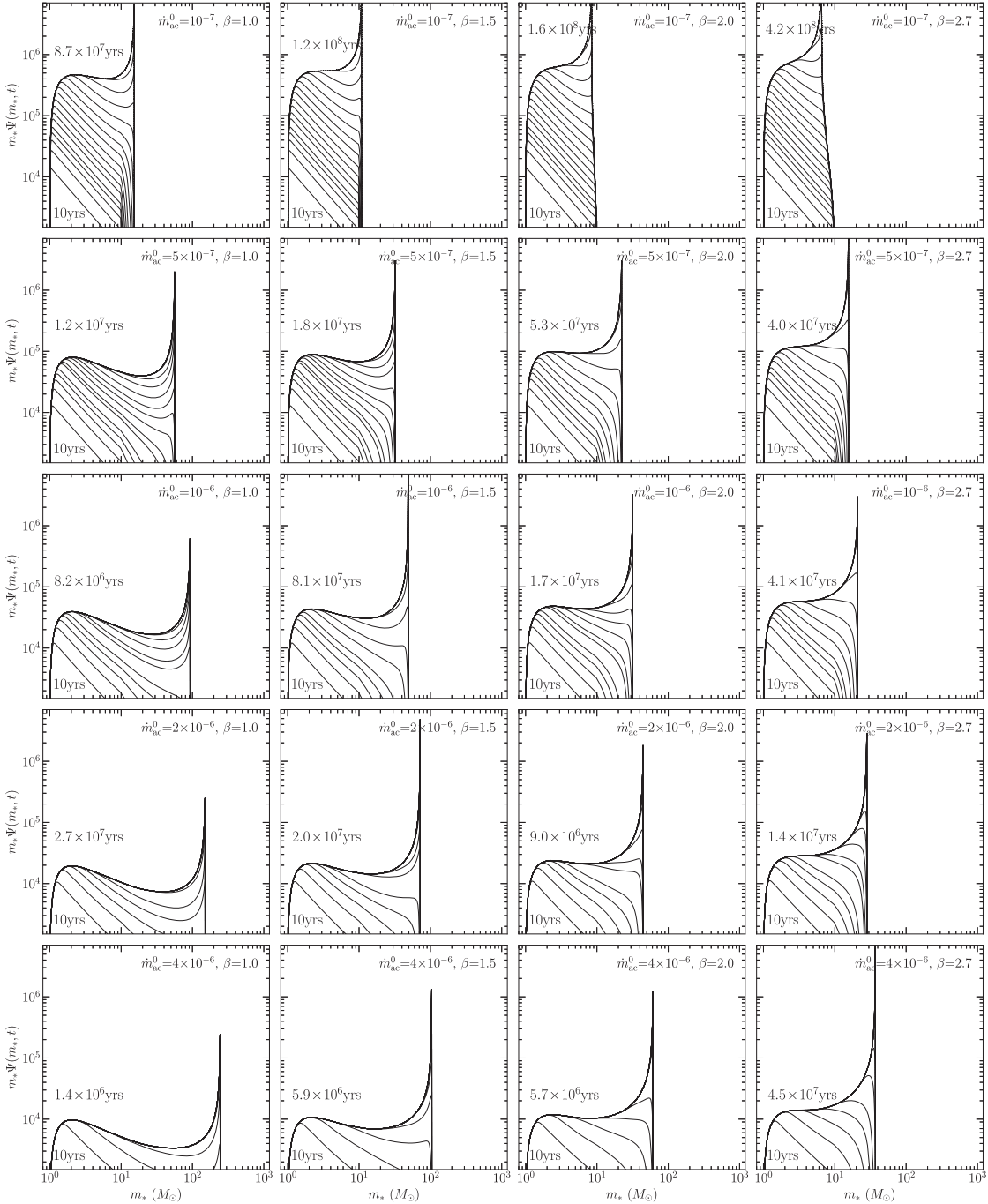

**Figure 8.** Same as Figure 4, but the evolving AMS has very-high-winds of $\dot{m}_w^0 = 10^{-8}\,M_\odot\,\mathrm{yr}^{-1}$. The cutoff due to stellar winds is too strong to produce abundant iron.

In order to further understand the AMS MFs in AGN disks, we simply compare them with the MFs of star clusters in the Milky Way. From deep observations of the Milky Way globular cluster using a principal-component-type analysis (e.g., Equation (15) in Marks et al. 2012), the MF can be expressed by $\Psi \propto m_*^{-\alpha_3}$ as (e.g., Kroupa & Jerabkova 2021)

$$\alpha_3 = 0.05 \log\left(\frac{Z}{Z_\odot}\right) - 0.41 \log\left(\frac{\rho_{\mathrm{gas}}}{10^6 M_\odot\,\mathrm{pc}^{-3}}\right) + 1.94 \quad (31)$$

where $Z$ is metallicity, and $\alpha_3$ is insensitive to the metallicity except for cases of extremely metal-rich regions. For the present case of AGN disks, $\rho_{\mathrm{gas}} = \rho_{\mathrm{d}} \approx 5 \times 10^8\,M_\odot\,\mathrm{pc}^{-3}\rho_{\bar{12}}$ and $Z \approx 10\,Z_\odot$, we have $\alpha_3 \approx 0.88 \ll 2.35$ from Equation (31); namely, $m_* \Psi \propto m_*^{0.12}$, which is much flatter than the Salpeter, but still steeper than the top-heavy stellar disk in the GC. This implies that the GC stellar disks were not formed through the stellar evolution that happened in globular clusters. Moreover, the present MF is much more top-heavy than the globular cluster (e.g., von Fellenberg et al. 2022). A simple application to this issue is given in Section 3.6. This comparison supports that there is an AMS evolution to form a top-heavy star formation in the GC. It is then reasonably expected for the AMS mechanism in AGN disks.

---

[14] We note the different structures of the M87 core from the GC. One may speculate that this could be extended as a corollary that cores of elliptical galaxies (SMBHs only have low accretion rates) generally have systematically different stellar structures from spirals. The former had undergone major mergers whereas the latter had not. Stellar disks around the central SMBHs could be strongly affected by the SMBH companion and totally changed by the major mergers. More highly spatially resolved IFU observations are needed.

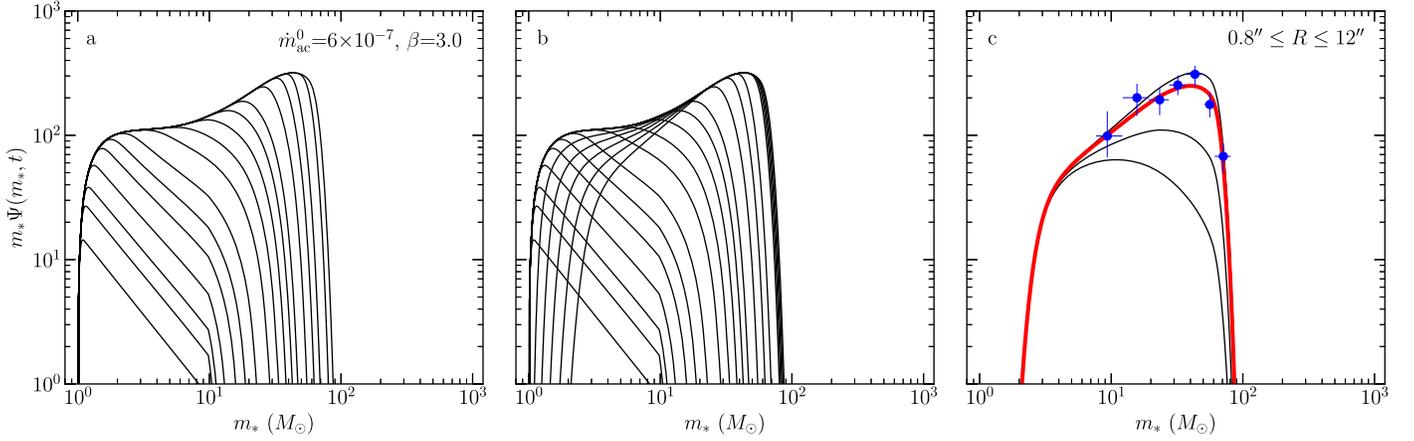

**Figure 9.** An application to the young stellar disk in the GC. Panel (a) shows the MF evolution of AMSs when the central SMBH was active. Panel (b) shows the star injection stopped at $t_* = 8.5 \times 10^6$ yr. Panel (c) shows the evolution of AMSs after the SMBH quenches accretion, where the red line represents an interval of $10^6$ yr after $\dot{\mathcal{M}} = 0$.

In our calculations, the maximum masses of AMS can reach $\sim 10^3 \, M_\odot$ depending on accretion and stellar winds (by two parameters $\dot{m}_{ac}^0$ and $\beta$). Numerical simulations predict the maximum masses stars of a few $1000 \, M_\odot$ in populous galaxies (e.g., Elmegreen 2000); however, the observed stars have initial masses of up to $m_* \approx 300 \, M_\odot$ in clusters (Crowther et al. 2010; Schneider et al. 2014). The maximum AMS are radiating extreme UV bands and should be tested by observations; however, radiations are converted into infrared bands since the environments of the AMS are dusty. The top-heavy and top-dominate MFs are predicted for galactic nuclei of nearby galaxies.

### 3.6. A Simple Application to GC Stars

In this subsection, we apply the present model to the GC stellar disk within $0.8'' - 12''$ in the nuclear regions (with a distance of 8 kpc to the GC). Inputs of the model include the central SMBH mass of $M_\bullet \approx 4 \times 10^6 \, M_\odot$ (e.g., Genzel et al. 2010), $\dot{\mathcal{M}} = 1$, and the characteristic radius is $R = 2.5 \times 10^6 \, R_g$ (i.e., the $0.8'' - 12''$ region) when the GC SMBH was active. Stars are injected with the Salpeter IMF for simplicity. In addition, we assume that the AGN episodic lifetime is $t_0 = 10^7$ yr. In this application, we assume a star injection period ($t_*$) in the model (i.e., the AGN disks are depleted to form the input stars, $\dot{\mathcal{S}}_{AD} = 0$) by either accretion onto the central SMBH or star formation). Table 1 lists all parameters. Once reaching the episodic lifetime, we set $\dot{\mathcal{M}} = 0$, such that it is as low as the present-day accretion. Consequently, the AMSs have $\dot{m}_* = 0$, and the AMS populations are undergoing pure stellar evolution in the analytical form of

$$\Psi(m_*, t) = \Psi(m_*, t_0) \exp\left[-\frac{t - t_0}{\tau_{\rm evo}(m_*)}\right], \quad (32)$$

for $\dot{\mathcal{S}}_{AD}(m_*, t) = 0$ ($t > t_*$) from Equation (1), where $(t - t_0)$ is the time since the central SMBH quenched accretion. Massive stars are rapidly evolving and eventually disappearing from the population.

We find that MFs reach a stationary state [$\Psi(m_*, t_0)$] after $3.2 \times 10^8$ yr from Equation (1), which is 1 order of magnitude longer than AGN episodic lifetime. Panel (a) of Figure 9 illustrates the evolution of AMS during the active phase of the central SMBH with continuous stellar injection (without the break of the injection), resulting in the formation of a top-dominant MF. Panel (b) shows the AMS evolution with an injection break at $t_* = 8.5 \times 10^6$ yr, and the low-mass end of the AMS exhibits a significant mass growth within $1.5 \times 10^6$ yr. Panel (c) displays the rapid evolution after $\dot{\mathcal{M}} = 0$, with observational data plotted (Bartko et al. 2010). The red line represents the MF of $t \approx 10^6$ yr evolution after the cessation of central SMBH activity, which matches the observed present-day MF very well. Furthermore, during this interval, more than 80 black holes are formed by SN explosion, roughly consistent with the population deduced from the spatial distribution of X-ray sources in the central parsec of the GC detected by Chandra (Hailey et al. 2018).

Moreover, as a consequence of SN explosion during stellar evolution, it impacts the GC medium of the Milky Way. Recent measurements of $\alpha$/Fe elements show the hot medium in the GC is supersolar (1−2 times solar abundance) observed by Chandra (Hua et al. 2023), which is consistent with those of the model calculated in Appendix C. Therefore, this simple application demonstrates that the present model can explain the present-day MFs in the GC and its evolutionary sequence (existing black holes in GC). This encourages us to make a detailed model by including the migration of stars for the young stellar disk of the GC. This will be carried out in a forthcoming paper.

In summary, stellar MFs are jointly governed by accretion from AGN disks and stellar evolution. Pile-up distributions are generally produced in AGN disks. This leads to PISNe dominated in AGN disks. Interestingly, we can recover the MF of the top-heavy stellar disk with radii of 0.03–0.5 pc in the GC. Details of explaining counter-rotating stellar disks remain an open question, as a subject of future study. In addition, MFs reach stationary states with timescales depending on accretion rates from AGN disks.

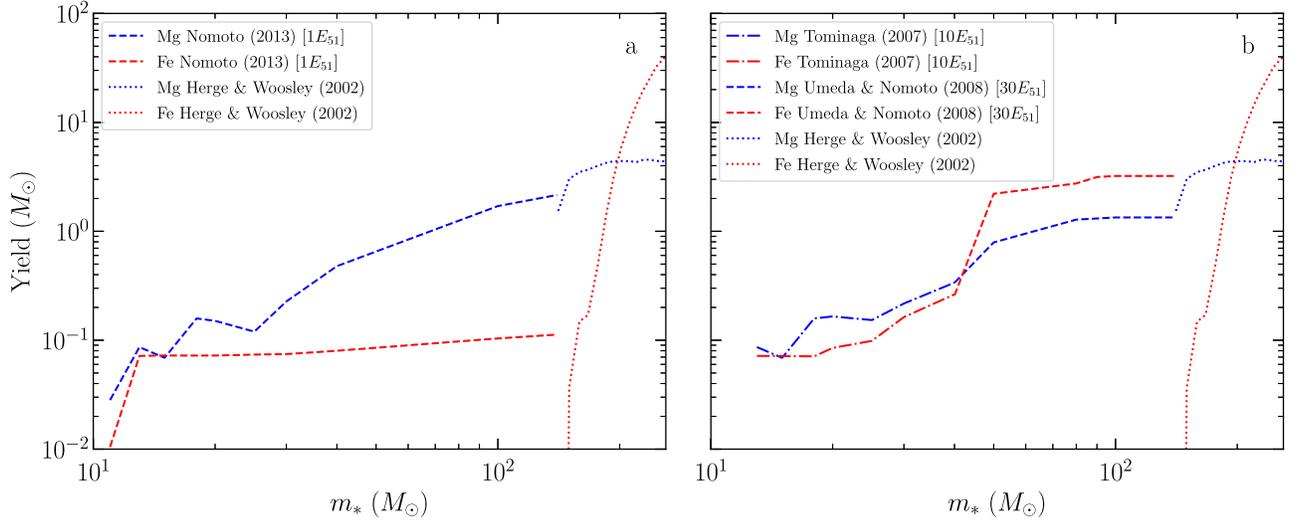

**Figure 10.** Yields from SN explosion. Iron yields are sensitive to explosion energy ($E_{SN}$). We use the numerical results for two extreme cases with $E_{SN} = (1, 30) \times 10^{51}$ erg for iron productions, respectively.

## 4. Iron Production and Abundances

Though the ratio of Mg II/Fe II as an exact proxy of metallicity is still controversial, it represents enrichment of metals in AGNs. Mg II and Fe II have similar ionization potentials (7.6 and 7.9 eV, respectively) and are expected to originate in the same partially ionized zone of the excited gas, which can be justified by the reverberation lags of Fe II (Hu et al. 2015) and Mg II (Yu et al. 2021, 2022; though the $R_{Mg}-L$ relation has large scatters). Therefore, the density, temperature, and turbulence of the Mg II and Fe II regions should hold as approximately the same. The ratios of these two lines provide a valuable measure to estimate the $\alpha$-element versus iron abundance ratio, at least about a relative scale (e.g., Wills et al. 1985; Hamann & Ferland 1999). Fe II(UV) is additionally sensitive to turbulence; fortunately, the optical Fe II is a better indicator of metallicity (Baldwin et al. 2004; Sarkar et al. 2021), and moreover optical Fe II lines indeed do reverberate with varying continuum (Hu et al. 2015). Nevertheless, it has been suggested that Fe II(tot)/Mg II could be a good indicator of abundance ratio, where Fe II(tot) is the total fluxes of UV and optical bands (e.g., Dietrich et al. 2002). Though the exact abundances of iron to magnesium remain a matter of debate, it is clear that classical SNe Ia are delayed too long to produce iron in the early universe. High ratios of Fe II/Mg II lines indicate a metal-rich environment in quasars. In particular, the recent observations of the $z \sim 7.54$ quasar suggest that the iron is produced by PISNe (Yoshii et al. 2022).

Yields of SNe are well understood for intermediate massive stars. Iron production strongly depends on progenitors of SN (see reviews of Woosley et al. 2002; Nomoto et al. 2013). It is still largely uncertain for massive stars ($\gtrsim 10\,M_\odot$). As shown in Figure 10, we adopt the iron and magnesium yield models taking two cases with $E_{SN} = 1\,E_{51}$ and $E_{SN} = 30\,E_{51}$ since yields depend on SN explosion energy ($E_{SN}$). We adopt the values of Nomoto et al. (2013) at $m_* = 10-140\,M_\odot$ for the former case, and those of Tominaga et al. (2007) at $m_* = 10-40\,M_\odot$ and Umeda & Nomoto (2008) at $m_* = 40-140\,M_\odot$ for the latter. Moreover, we adopt a yield model for PISNe at $m_* = 140-260\,M_\odot$ (Heger & Woosley 2002). Stars with masses over $260\,M_\odot$ (or $300\,M_\odot$) directly collapse into black holes, we thus adopt iron and magnesium yield to zero for them. It is found that iron production is mainly from very massive stars, whereas magnesium is about linear with progenitor mass. This difference provides a method to distinguish MFs of AMS in AGN accretion disks. Iron-rich BLRs indicate the history of very massive stars.

With the MFs, we have iron and magnesium production as

$$\Delta M_{\mathrm{Fe,Mg}}(t) = \int_0^t dt \int_{m_1}^{m_{\max}(t)} \mathcal{R}_{\mathrm{SN}}(m_*, t - \tau_{\mathrm{evo}}) Y_{\mathrm{Fe,Mg}}(m_*) dm_*, \quad (33)$$

where $Y_{\mathrm{Fe,Mg}}$ are the Fe and Mg yields from the stars, respectively. Here we would like to point out that the current approximations neglect the delay of stellar evolution and SN explosion. This works for massive stars, since their stellar evolution is short enough compared with the typical timescale of an AGN lifetime. Calculations of metallicity stop at the AGN lifetime. We simply assume that the AGN disk gas is depleted until the AGN lifetime.

The metal abundances of Fe and Mg are given by

$$[\mathrm{Fe, Mg/H}](t) = \log_{10}\left[\frac{\Delta M_{\mathrm{Fe, Mg}}(t)}{\mathcal{A}_{\mathrm{Fe,Mg}} M_{\mathrm{disk}}}\right] - \log_{10}\left(\frac{N_{\mathrm{Fe,Mg}}}{N_{\mathrm{H}}}\right)_\odot. \quad (34)$$

Metal abundances of Fe and Mg are sensitive to $\dot{m}_*$, but the relative abundance is given by

$$[\mathrm{Fe/Mg}](t) = \log_{10}\left[\frac{\mathcal{A}_{\mathrm{Mg}} \Delta M_{\mathrm{Fe}}(t)}{\mathcal{A}_{\mathrm{Fe}} \Delta M_{\mathrm{Mg}}(t)}\right] - \log_{10}\left(\frac{N_{\mathrm{Fe}}}{N_{\mathrm{Mg}}}\right)_\odot, \quad (35)$$

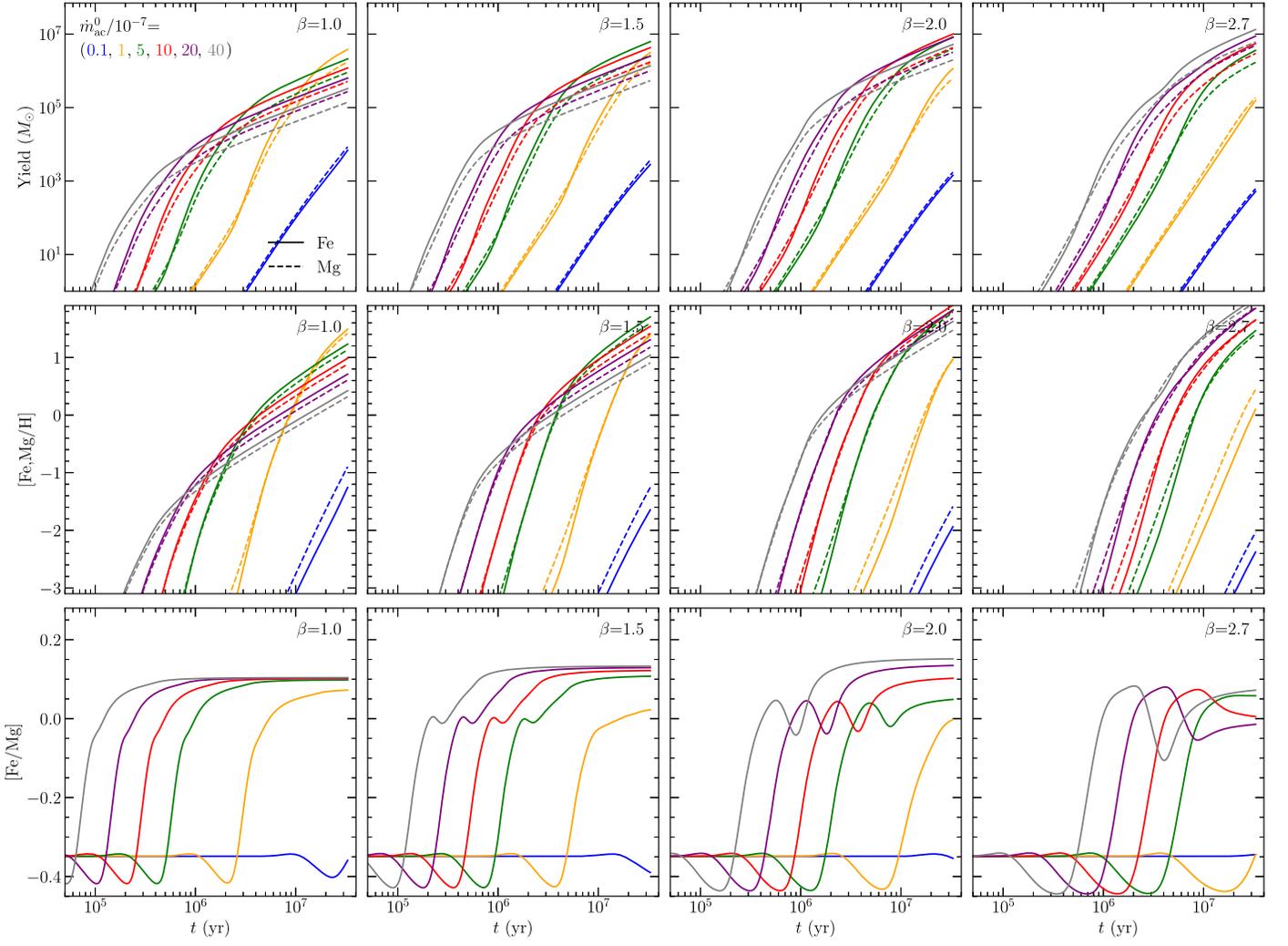

**Figure 11.** Upper panels: yields of iron and magnesium from AGN accretion disks for $E_{SN} = 30 \times E_{51}$ according to Figure 4. Metals increase with episodic time; however, some of them will be swallowed by SMBHs along with the accretion flows. Middle panels: iron and magnesium abundances produced in AGN accretion disks. Iron and magnesium abundances increase with episodic time. Lower panels: iron abundances relative to magnesium. [Fe/Mg] reaches a saturated state after $\sim 10^7$ yr. The saturated values of [Fe/Mg] can reach $\sim 0.2$ depending on the accretion of AMS in AGN disks.

where the mass number of magnesium and iron isotopes is $\mathcal{A}_{Mg} = 24$ and $\mathcal{A}_{Fe} = 56$, respectively, and $(N_{Fe,Mg}/N_H)_\odot$ is the solar abundance (Asplund et al. 2009). Yields of SN explosion are shown by Figure 10.

Figures 11–14 show the iron and magnesium productions with time for the model of $E_{SN} = 30 \times E_{51}$. Yields depend on accretion rates ($\dot{m}_{ac}^0$ and $\beta$), but nonmonotonically, as shown by the figures. Given $\beta$, high-$\dot{m}_{ac}^0$ leads to fast metal production at the beginning but is suppressed later, since most gas is converted into very massive stars ($\gtrsim 260 M_\odot$ stars, which directly collapse into black holes). However, such suppression can be relaxed by increasing $\beta$, avoiding the formation of very massive stars. Generally, iron production can reach masses of $10^5$–$10^6 M_\odot$ within a few megayears, but it enriches more slowly in the latter (see upper panels of Figure 11–14). The abundances of iron and magnesium show similar properties as shown in the middle panels of Figures 11–14. The abundance can be many times that of solar. Therefore, we expect to measure iron and magnesium abundances to determine AGN and quasar episodic ages.

Interestingly, the relative abundances of iron to magnesium tend to be very saturated ([Fe/Mg]$_{sat} \sim 0.2$) after a few megayears to 10 Myr, as shown by Figures 11 and 12. This is led by the relatively smooth shape of MFs compared to the cases with winds. Before the saturated [Fe/Mg], the ratio is sensitive to star formation processes. Observational line ratios of Fe II/Mg II can be conveniently measured from spectra, which are in the range of 0.5 ~ 4 with a mean value of ~2 (see Figure 5 in Onoue et al. 2020; Wang et al. 2022). As we mentioned in Section 1, the line flux ratios of Fe II/Mg II, which depend on ionizing source SED, turbulence, and density, are not directly equal to the relative abundance of [Fe/Mg]. Some detailed calculations of CLOUDY show the averaged [Fe/Mg] $\approx 0.18 \pm 0.31$ over a redshift range of from $z = 0$ to 3 given by Table 4 in Sameshima et al. (2020). Considering the large scatters of averaged [Fe/Mg] values, the present model is consistent with the observations. For AGNs and quasars with [Fe/Mg] $\lesssim 0.2$, we can easily explain them through the present model, as shown in Figure 11 and 12. For those with [Fe/Mg] $\gtrsim 0.2$, the overabundant iron can be obtained through conditions that the MFs have a sharp peak between $140-260 M_\odot$ as shown by Figures 6 and 7, but the case of more extreme winds in Figure 8 shifts to lower masses that are not able to produce iron. This narrow window of AMS for the extreme cases of [Fe/Mg] $\gg$ 0.2 in principle can be satisfied by accretion and choked winds; for example, $(\dot{m}_{ac}, \dot{m}_w, \beta) = (5 \times 10^{-7}, 10^{-9}, 1)$ in Figure 13, and $(\dot{m}_{ac}, \dot{m}_w, \beta) = (40 \times 10^{-7}, 10^{-8}, 1)$ in Figure 14, and other combinations of these parameters can also produce the overabundant iron to magnesium. However, it is hard to make an exact estimation owing to the uncertainties of $\dot{m}_{ac}$ and $\dot{m}_w$.

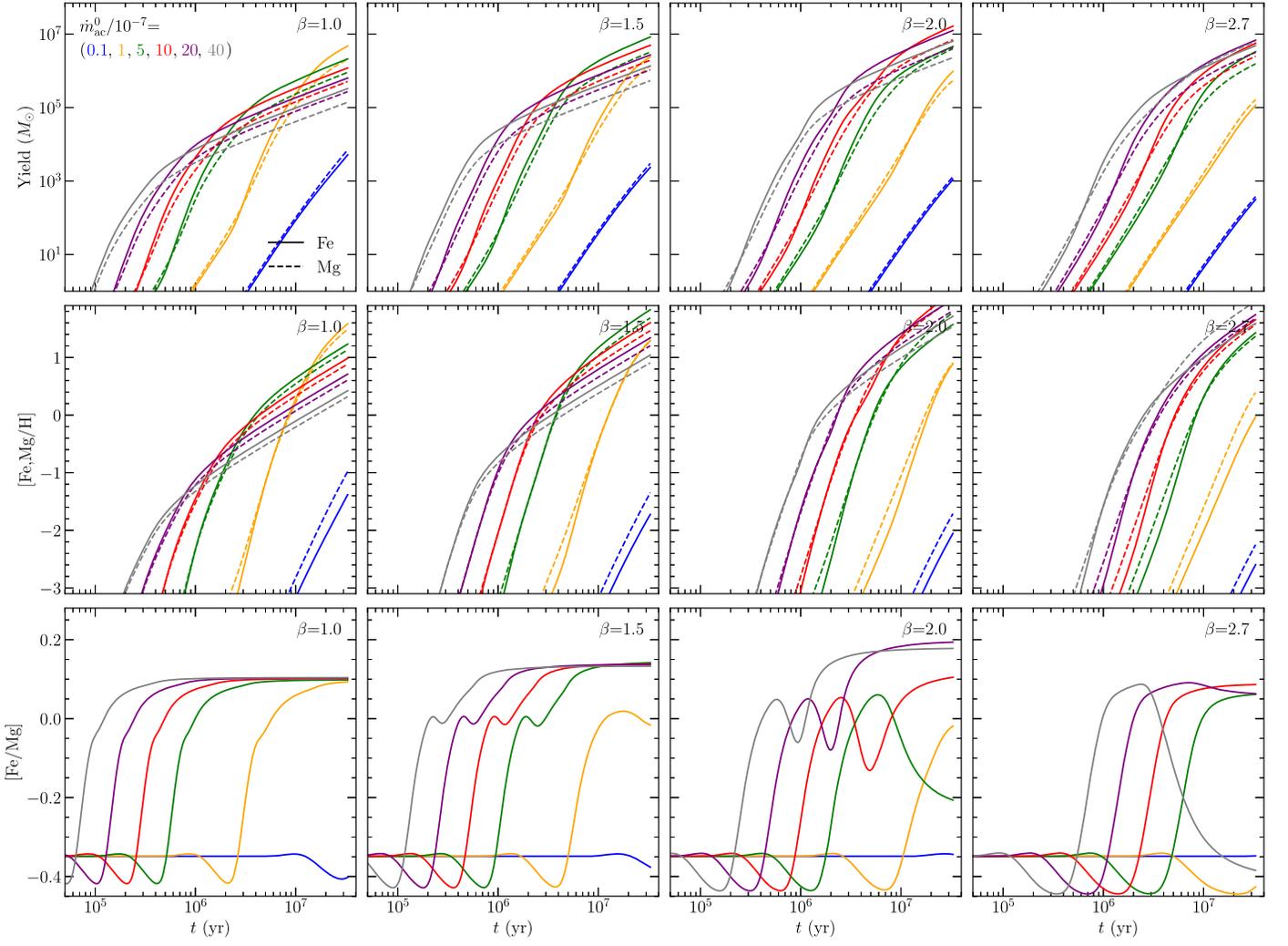

**Figure 12.** The same parameters of the AMS as Figure 11, but $E_{SN} = 30 \times E_{51}$ according to Figure 6 (mid-winds: $\dot{m}_w^0 = 10^{-10} \, M_\odot \, \text{yr}^{-1}$).

Stellar winds lead to some complications in metal production, seen in Figures 12–14 for the cases from mid-winds to very-high-winds. MFs are changed by the stellar winds, and thus the metal production. Yields of iron and magnesium are similar to the dependence on $\dot{m}_{ac}^0$ and $\beta$, as shown by Figure 11 (without stellar winds), but their values significantly decrease with winds. Moreover, the relative abundance of iron to magnesium shows more upon-down variations with evolutionary processes. This is caused by the yields ($Y_{Fe}(m_*)$ and $Y_{Mg}(m_*)$).

Figure 15 shows the case with $E_{SN} = 1 \times E_{51}$. Yield models generally produce much less metal than the case of $E_{SN} = 30 \times E_{51}$. Except for very-high-winds cases, this model hardly generates enough metals. High-$E_{SN}$ explosions eject more metals. Therefore, $E_{SN} = 30 \times E_{51}$ is preferred according to AGN metallicity, though $E_{SN}$ is uncertain in theoretical calculations (Nomoto et al. 2013).

Recently, Schawinski et al. (2015) estimated the episodic lifetime of local Seyfert galaxies from the Sloan Digital Sky Survey (SDSS) in light of echoes of the [O III] emission line. The lifetime is surprisingly as short as $10^5$ yr, see also some arguments from Martini (2004). This kind of flickering lifetime could be related to AGN disk instability, leading to chaotic accretion (e.g., King & Nixon 2015) and spindown evolution of the central SMBHs (Wang et al. 2009; Li et al. 2012; Volonteri et al. 2013), rather than the episodic lifetime of cosmic accretion growth. Counter-rotating stellar disks discovered by von Fellenberg et al. (2022), similar to the case of counter-rotating gas disks in the galactic center of NGC 1068 (Impellizzeri et al. 2019), support the chaotic accretion onto the central SMBH. The metal abundance of AGNs and quasars can be used as a clock for ages of AGNs and quasars. The episodic lifetime of AGNs should be traced by metals.

## 5. Spectral Energy Distribution

### 5.1. Intrinsic SED

In principle, we should employ stellar population synthesis to get the SED (see an extensive review of the stellar synthesis of Conroy 2013) through encoding all of the properties of the stellar populations, such as star formation history, abundance, dust (mass, grain size, and geometry distributions), and interstellar radiation. In this paper, we outline a simplified scheme to calculate SEDs to show the characterized features to first-order approximations. With the help of simple stellar synthesis, we have the SED of the core stars

$$L_*(\lambda, t) = \int_{m_{\min}}^{m_{\max}} \ell_\lambda(m_*)\Psi(m_*, t)dm_*$$
$$= \frac{\pi}{\sigma_{\rm SB}} \frac{2hc^2}{\lambda^5} \int_{m_1}^{m_{\max}(t)} \frac{\ell_*}{T_{m_*}^4} \frac{\Psi(m_*, t)dm_*}{\exp(hc/\lambda k T_{m_*}) - 1}, \quad (36)$$

where $\sigma_{\rm SB}$ is the Stefan-Boltzmann constant, $h$ is the Planck constant, respectively, $\lambda$ is the wavelength, and $\ell_*$ is the bolometric luminosity of stars with $m_*$. We use the mass–luminosity relation

$$\log(\ell_*/L_\odot) = 0.06 + 4.58x - 0.78x^2 - 0.04x^3 + 0.02x^4, \quad (37)$$

and effective temperature

$$\log(T_{m_*}/{\rm K}) = 3.49 + 1.90x - 0.92x^2 + 0.21x^3 - 0.02x^4, \quad (38)$$

from Schaerer (2002), which are shown by Figure 20 in Appendix A, to calculate the SEDs of the AMS populations. The effective temperatures tend to be saturated when stellar masses are large enough. While the saturated effective temperature of stars decreases with increasing metallicity, their radiated luminosity remains largely unaffected, as the primary energy source for stars is hydrogen. Increases of metallicity enhance the efficiency of cooling, which reduces the stellar temperatures, leading to a decrease in stellar radii (Schaerer 2002). Figure 16 and 17 show intrinsic SEDs of the AMS populations without and with stellar winds, respectively.

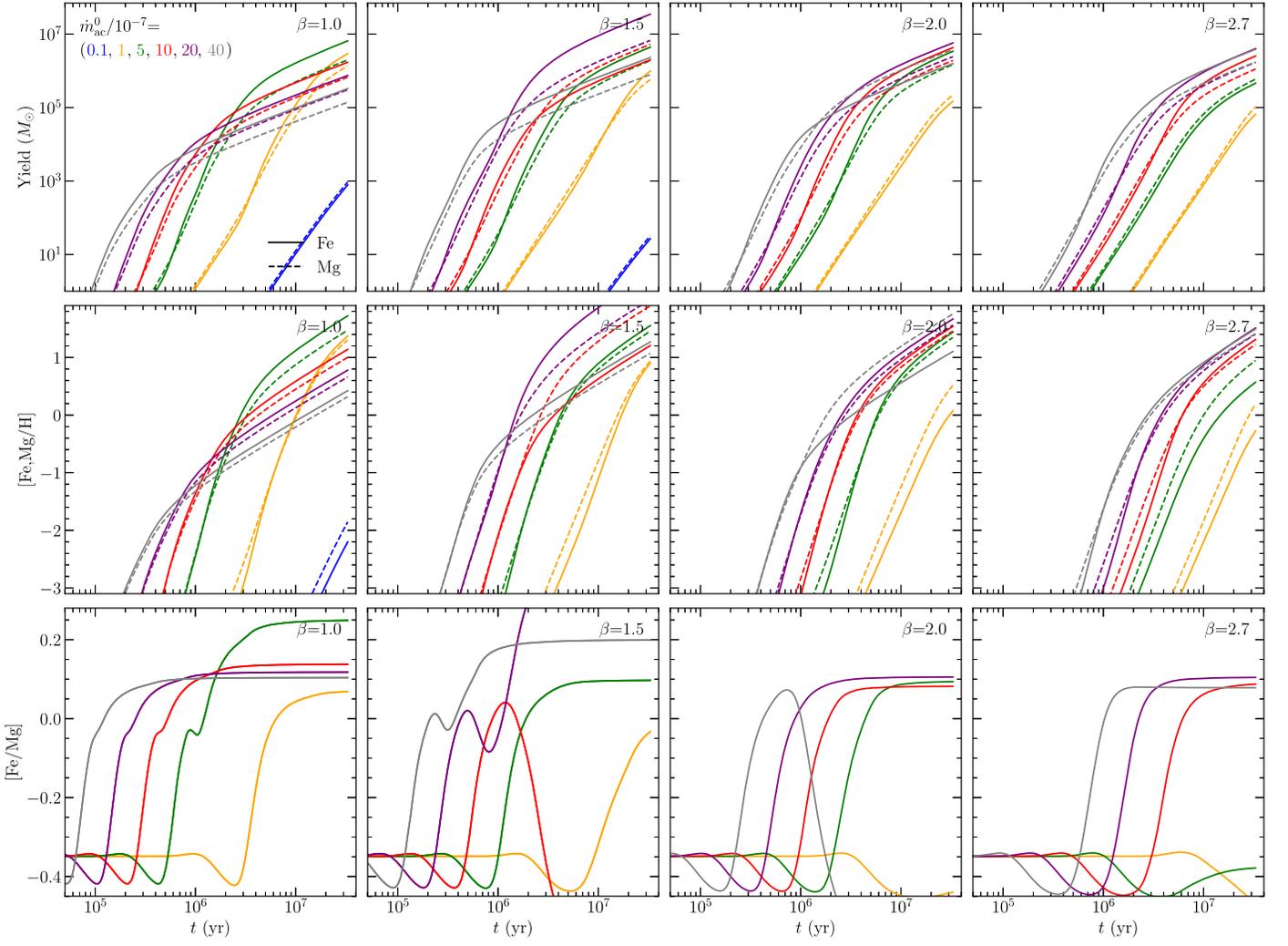

**Figure 13.** The same parameters of the AMS as Figure 11, but $E_{\rm SN} = 30 \times E_{51}$ according to Figure 7 (high-winds: $\dot{m}_w^0 = 10^{-9}\, M_\odot\, {\rm yr}^{-1}$). Comparing with cases with $E_{\rm SN} \sim 10^{51}\, {\rm erg\, s^{-1}}$, high-energy explosions eject more metals.

### 5.2. SED Emergent from Dusty Gas

Since star formation mainly happens around the mid-plane of the accretion disks, the emergent spectra from the dusty gas will differ from the intrinsic ones. In the dusty gas with temperature from $10^2 \sim 10^3$ K, the opacity is $\kappa_\lambda \sim 10.0$ as a rough constant (e.g., Bell & Lin 1994; Semenov et al. 2003), and the optical depth $\tau_\lambda \approx 1.8-6.0 N_{24}$, where $N_{24} = N_{\rm H}/10^{24}\, {\rm cm}^{-2}$ is the column density of the region. Emergent spectra from this region depend on the geometry and density distribution of its region owing to optical depth (and wavelength). The extreme UV photons from the massive stars will be absorbed and reprocessed into infrared ones. Detailed spectra depend on radiation transfer in the dusty gas region. This can be implemented in a similar way to Pier & Krolik (1992, 1993), or even in clumpy medium (e.g., Nenkova et al. 2008), but we leave this for a future paper.

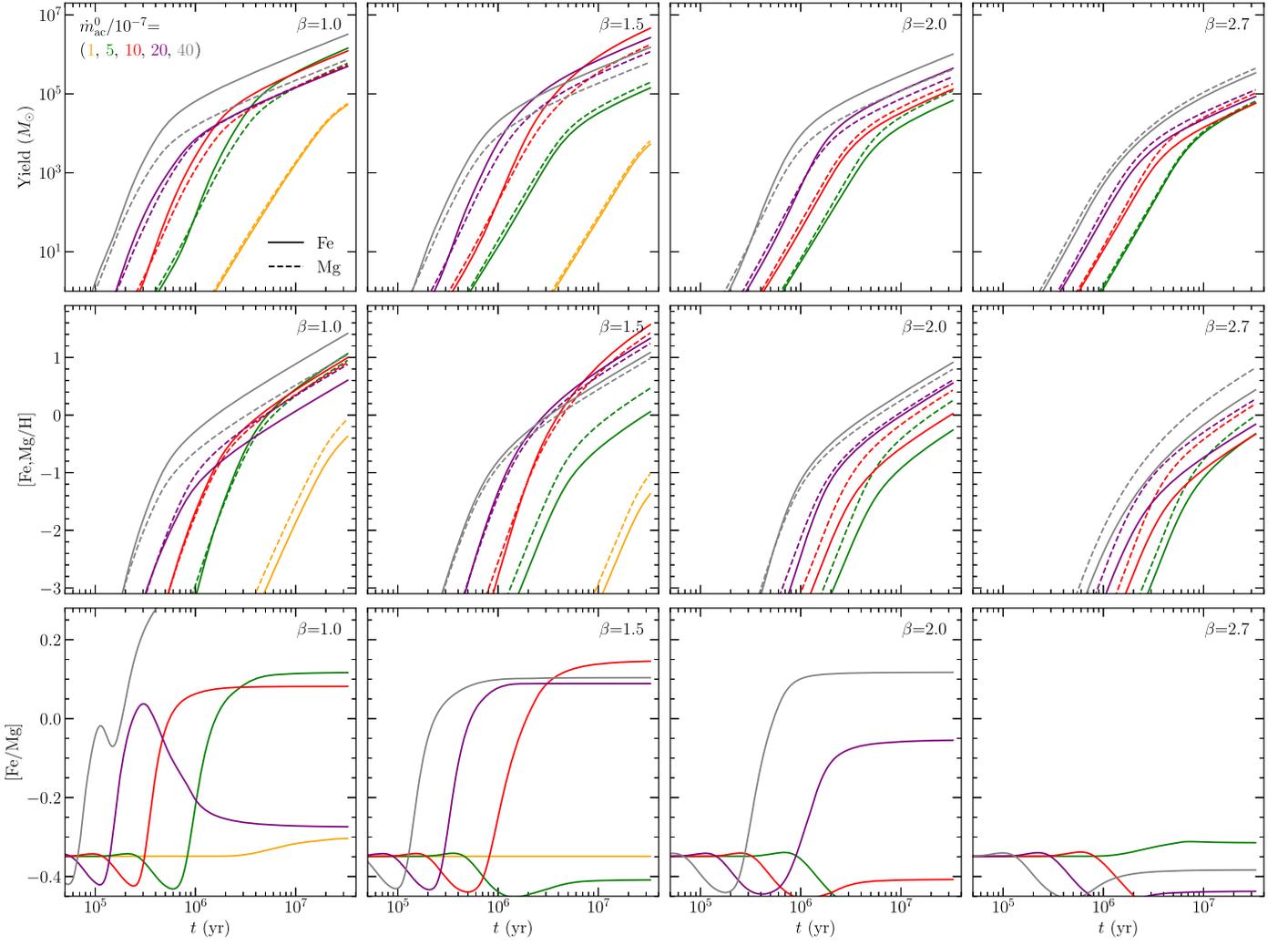

**Figure 14.** The same parameters of the AMS as Figure 11, but SN explosion energy is taken as $E_{SN} = 30 \times E_{51}$ according to Figure 7 (very-high-winds: $\dot{m}_w^0 = 10^{-8}\, M_\odot\, \text{yr}^{-1}$).

Radiation from the AMS population will be thermalized in the mid-plane, and we have an averaged temperature over the star formation region

$$T_{*,\text{eff}}^4(t) = \frac{1}{\pi R_{AMS}^2 \sigma_{SB}} \int L_{*\lambda}(t) d\lambda, \quad (39)$$

where $R_{AMS}$ is the radius of the AMS spatial distribution. The thermalized emissions are given by the Planck function, and we have its luminosity of

$$L_{*\lambda}^{AMS}(t) = \frac{2hc^2}{\lambda^5} \frac{\pi R_{AMS}^2/2}{\exp(hc/\lambda k T_{*,\text{eff}}) - 1}. \quad (40)$$

Here the factor of 1/2 refers to the luminosity from a single side of the region. This extra IR component is independent of emissions of the central AGN disks. The property helps to distinguish it from AGN disk emissions and reprocessed IR emissions. First, the present component of AMS radiation will not reverberate OUV (optical and ultraviolet) emissions from AGN accretion disks, leading to a long-term trend (much longer than that of the accretion disks powering the UV and optical emissions), namely, the difference between the OUV and infrared variations in the long term. This would observationally distinguish them from the reprocessed emissions of AGN accretion disks. Second, changing-look (CL) AGNs transiting between type 1 and 2 among some AGNs provide an opportunity to observationally test the present scenario (see details in Section 6.2). These tests would advance the understanding of star formation (and AMS evolution) in the CNRs of AGNs and quasars.

On the other hand, the reprocessed emissions of the central engine from the torus have been found by the NIR RM campaigns of small samples of AGNs (Suganuma et al. 2006; Koshida et al. 2014; Lyu et al. 2019; Minezaki et al. 2019) and a larger sample of quasars (Yang et al. 2020). A strong correlation of $\tau_{IR} \propto L_{\text{opt}}^{1/2}$ holds, where $L_{\text{opt}}$ is the optical luminosity (see a review of Lyu & Rieke 2022), but with quite large scatters in a large sample of quasars (Yang et al. 2020). In order to quantitatively obtain the temperature of the reprocessing torus, we have an energy balance equation determined by

$$\left(1 - \frac{\Delta\Omega}{4\pi}\right) \frac{L_{\text{AGN}}}{4\pi R^2 c} = \frac{1}{3} a T_{\text{torus}}^4, \quad (41)$$

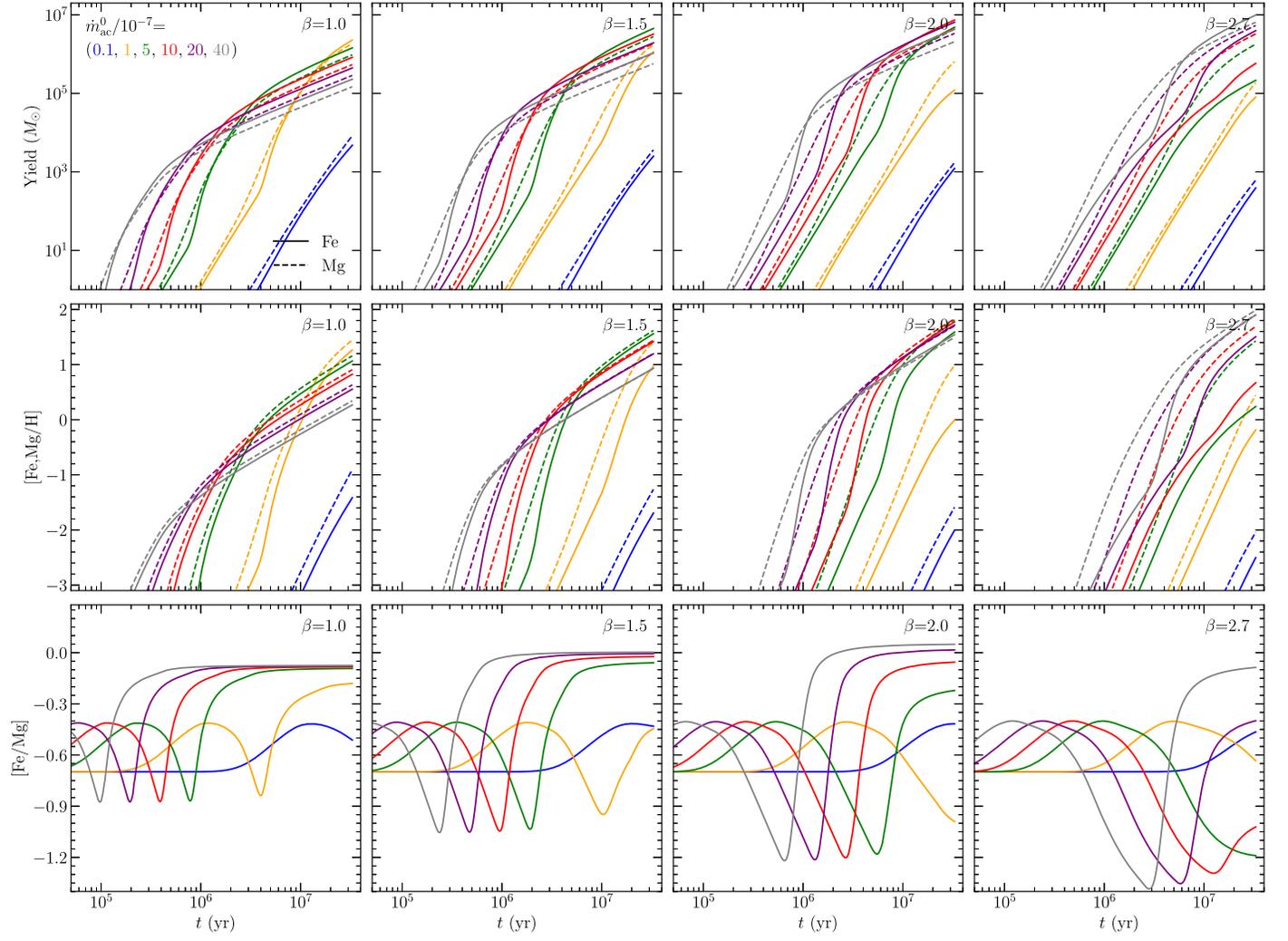

**Figure 15.** The same parameters of the AMS as Figure 11 but the SN explosion energy is $E_{SN} = 10^{51}$ erg. Comparing with the previous ones, we find that the metal production is lower than that in $E_{SN} = 30 \times E_{51}$ erg.

where $\Delta\Omega$ is the solid angle subtended by the torus, and $a$ is the blackbody constant. This equation holds under the assumption that the star formation regions are optically thick over the wavelength of the SED of the whole stellar population. This is the simplified version of Equation (2) in Pier & Krolik (1992). We then have the reprocessed SEDs of the torus

$$L_\lambda^{rep} = \int_{R_{in}}^{R_{out}} \frac{2hc^2}{\lambda^5} \frac{2\pi R dR}{\exp(hc/\lambda k T_{torus}) - 1}, \quad (42)$$

where $R_{in,out}$ are the inner and outer radii of the torus. The manipulations used above can be replaced by simple stellar synthesis for details (such as BC03 templates).

In order to characterize the contribution of the star formation region to SEDs, we use the Shakura-Sunyaev disk model and reprocessing model of the torus for emissions in infrared bands. The effective temperature of the accretion disks is given by $T_{AD,eff} = 0.586 \times 10^3 \mathscr{M}^{1/4} M_8^{-1/4} r_4^{-3/4}$ K (e.g., Kato et al. 2008). Figures 18 and 19 show three components: (1) AGN disks, (2) reprocessed emissions of the torus, and (3) AMS contribution (at stationary state). AMS contributions are comparable with the reprocessed emissions, but they vary with different timescales. According to the above discussions, we have the total emissions from the global scenario as

$$\lambda L_\lambda = \lambda L_{*\lambda}^{AMS} + \lambda L_\lambda^{rep} + \lambda L_\lambda^{AD}, \quad (43)$$

where $\lambda L_\lambda^{AD}$ is the emission from AGN disks in light of the standard accretion disk models (Shakura & Sunyaev 1973).

Figures 18 and 19 show various SEDs from the global model without and with stellar winds, respectively. The reprocessing of central AGN disk emission is quite well understood (e.g., Pier & Krolik 1992), which is re-radiated in a wide range from NIR, mid-infrared (MIR), to far-infrared bands. This is evidenced by the reprocessing physics through the NIR and MIR continuum reverberations (Suganuma et al. 2006; Koshida et al. 2014; Lyu et al. 2019; Minezaki et al. 2019; Lyu & Rieke 2022; Chen et al. 2023). AMS contributes emissions mostly in the NIR and MIR bands depending on AMS numbers and accretion rates. AMS contributions increase with $\dot{m}_{ac}^0$ but decrease when $\dot{m}_{ac}^0$ exceeds a critical value for a given $\beta$. Moreover, AMS contributes to SED with complicated behaviors. When accretion stays at relatively low rates ($\dot{m}_{ac}^0 \lesssim 5 \times 10^{-7} M_\odot$ yr$^{-1}$), the contribution monotonically decreases with an increase of $\beta$. However, when $\dot{m}_{ac}^0$ is relatively high, the contribution increases with $\beta$ but decreases after a maximum. Given its independence from AGN disk emissions, the new component is relatively nonresponsive to variations in the optical and UV emissions emanating from the AGN accretion disks but manifests as a stable component in NIR and MIR bands. Stellar winds significantly decrease the AMS contributions to infrared bands (we only provide the case of the mid-winds here).

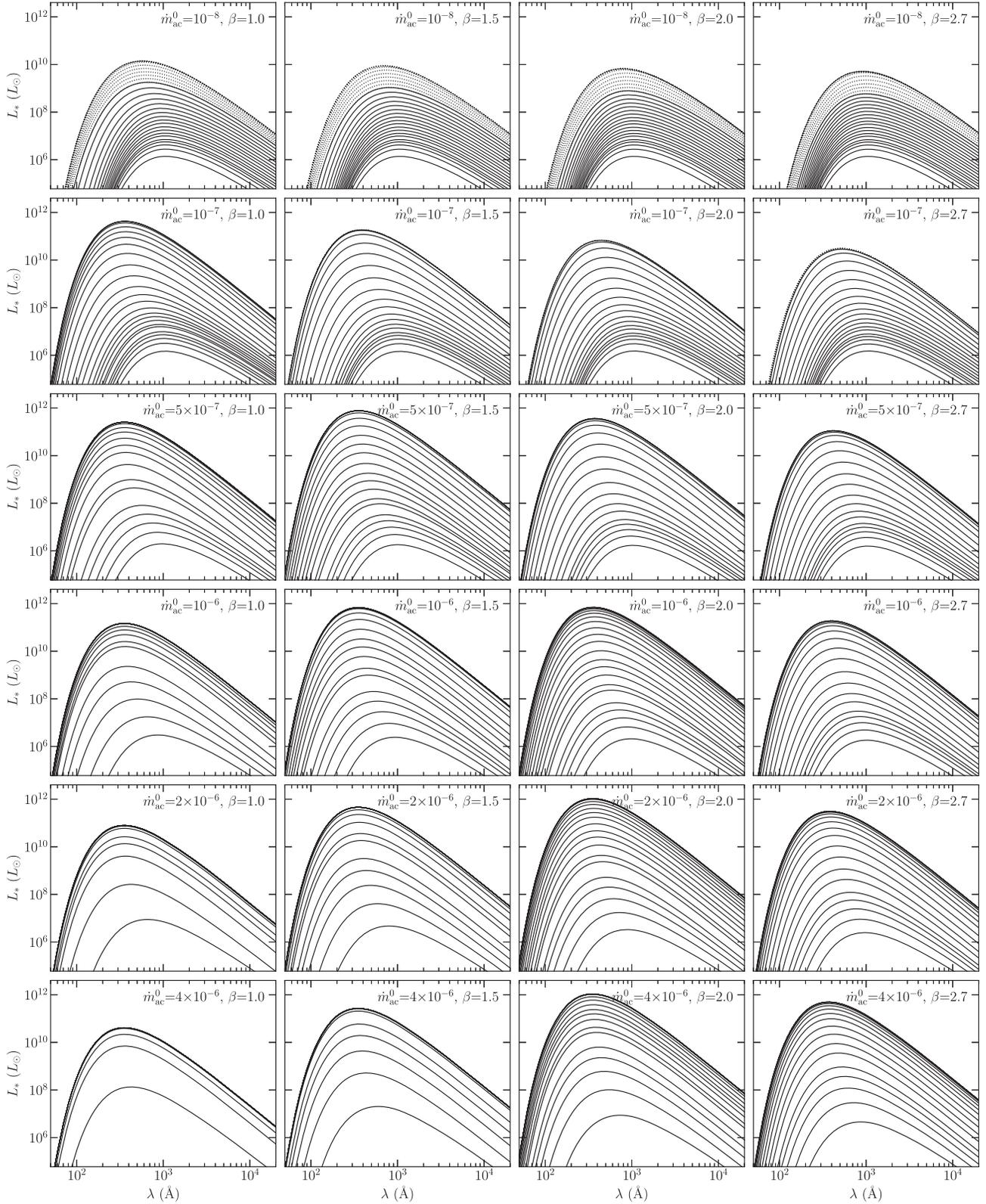

**Figure 16.** Intrinsic SEDs of the star formation corresponding to Figure 4. Peak fluxes at extreme ultraviolet bands are mainly contributed by the massive stars. The injected stars with the Salpeter IMF can be neglected for the SED.

We emphasize one point here. Emergent radiation from the AMS population, which produces extra energies inside the accretion disk, depends on radiation transfer. Moreover, the photoionization of clouds due to UV photons from AMSs will produce emission lines from optical to infrared bands, even ionized by the central engine (e.g., Mor & Netzer 2012).

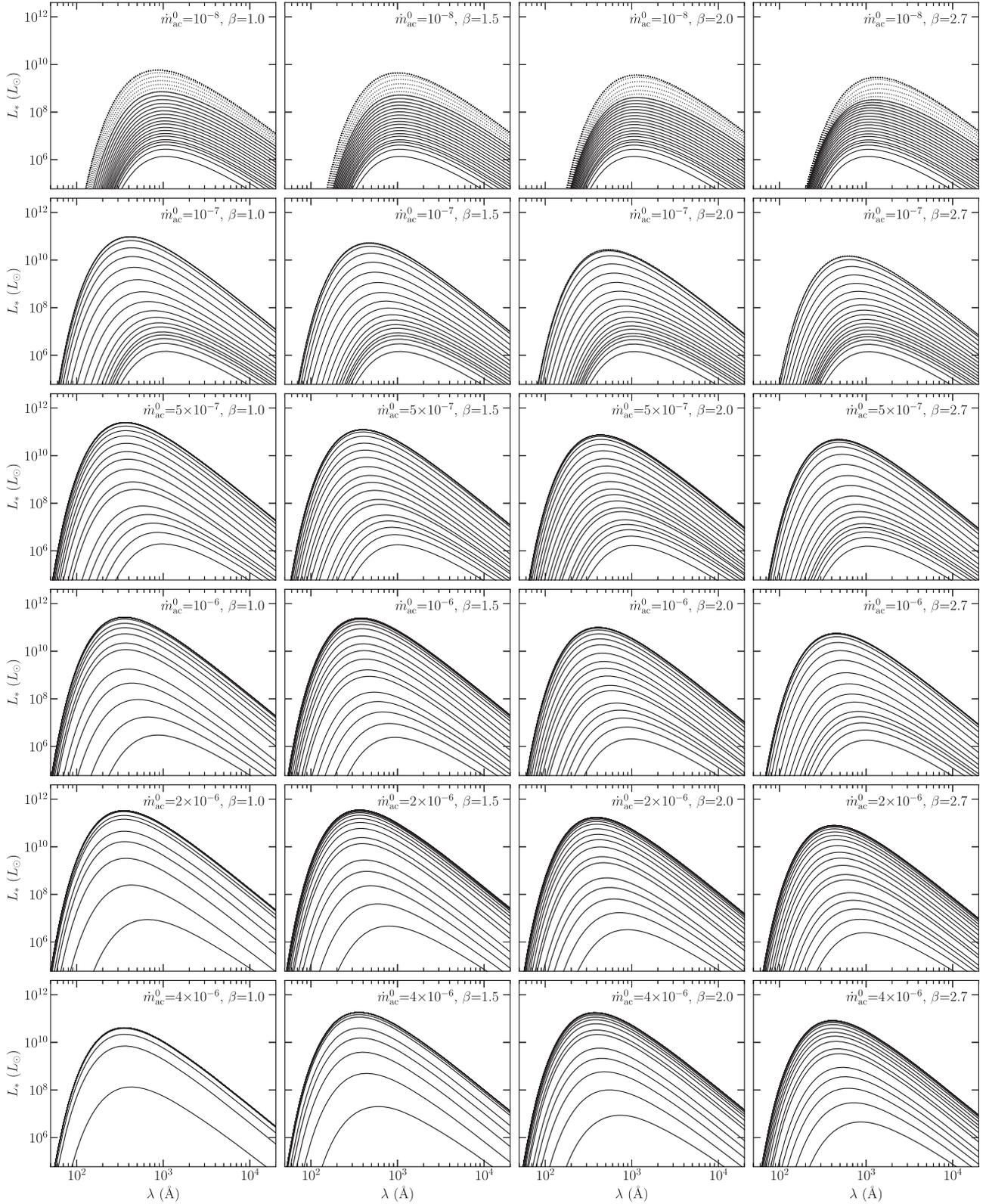

**Figure 17.** Intrinsic SEDs of the star formation corresponding to Figure 7 but $\dot{m}_w^0 = 10^{-9} M_\odot$ yr$^{-1}$. The SEDs are generally redder than that without stellar winds.

Calculations of emergent spectra deal with the spatial distributions of dusty clumps and chemical composition. For simplicity, we assume here that the AMS regions are completely optically thick over the whole wavelength of the SED of the stellar populations.

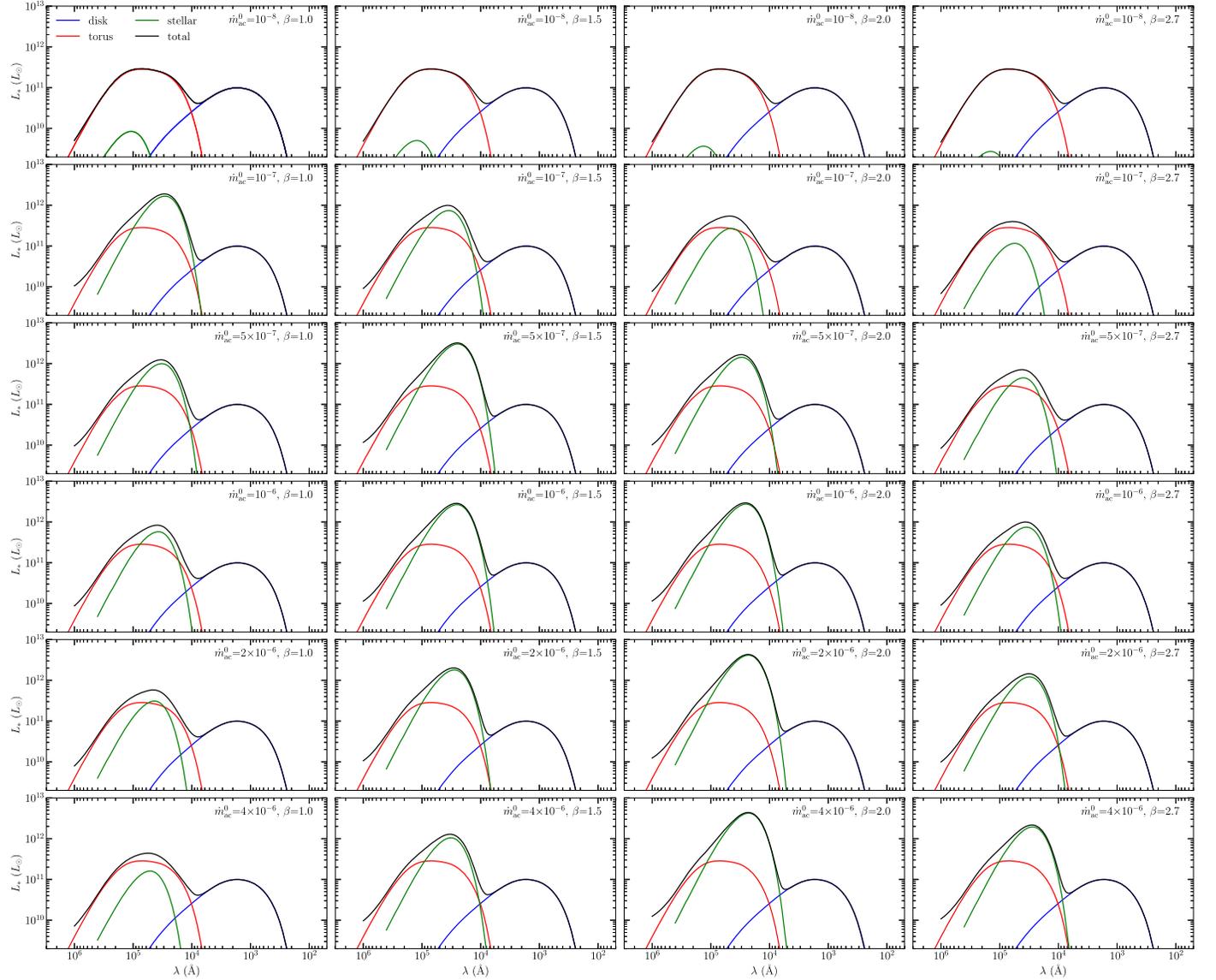

**Figure 18.** Global SED of AGNs with star formation with extinction, corresponding to Figure 4. We take $M_\bullet = 10^8\,M_\odot$ and $\dot{\mathcal{M}} = 1$ for AGN disks, and $\Delta\Omega = \pi/4$ for reprocessing emissions from the torus.

### 5.3. Supernovae Explosion: Support Dusty Torus

Accretion of AMS creates a large number of massive stars, as shown in Figures 4, 5, 6, and 7. We approximate the SN rates of $f_{\rm SN} \approx m_*\Psi/\tau_{\rm evo}$, which increase with accretion rates $\dot{m}_{\rm ac}^0$. There is increasing evidence for SNe in nearby AGNs (e.g., Villarroel et al. 2017, 2020). Moreover, NGC 3227 and another nine Seyfert galaxies have been spatially resolved by SINFONI observations within ∼10 pc, where starburst happened once, about a few tens of megayears ago (Davies et al. 2006, 2007). A dusty torus has been suggested to be formed by ejecta of SNe explosions of massive stars as obscuration of the central BLRs (Davies et al. 2006; Gohil & Ballantyne 2017). From our calculations, the averaged SN rates yield the kinematic energy luminosity, which can be obtained from

$$L_{\rm SN} = E_{\rm SN}\hat{\mathcal{R}}_{\rm SN} = 3.2\times 10^{42}\left(\frac{E_{\rm SN}}{10^{51}\,{\rm erg}}\right)\left(\frac{f_{\rm SN}}{0.1\,{\rm yr}^{-1}}\right){\rm erg\,s^{-1}}, \quad (44)$$

for general cases of AMS. This is another extra energy source inside the torus, which is in the form of kinematics, playing a role to support the geometry of the gaseous disks.

A geometrically thick torus plays a vital role in the unified model of AGNs (Antonucci 1993; Netzer 2015). It is generally believed that a dusty torus is composed of discrete clouds (or clumps; Nenkova et al. 2008) evidenced by observations of Tristram et al. (2007). However, collisions among clumps are unavoidable, and they then rapidly collapse (Krolik & Begelman 1988). Dissipation rates of clump collisions are estimated by Equation (18) in Krolik & Begelman (1988) as

$$\dot{E}_{\rm dis} = 2\times 10^{40}\,R_{\rm pc}\left(\frac{v_{\rm orb}}{200\,{\rm km\,s^{-1}}}\right)^3\epsilon_{\rm dis}N_{\rm cl,24}\,{\rm erg\,s^{-1}}, \quad (45)$$

where $v_{\rm orb}$ is orbiting velocity of clumps, $R_{\rm pc}$ is the radius of the torus in units of parsecs, $\epsilon_{\rm dis}$ is the dissipation efficiency on order of the unity, and $N_{\rm cl,24} = N_{\rm cl}/10^{24}\,{\rm cm}^{-2}$ is the column density of the torus, for the case with dispersion velocities of $\Delta v_z \sim \Delta v \sim v_{orb}$. SN explosions in the star formation regions can supply enough energy power to balance the dissipations, provided $\sim 1\%$ of the SN explosion energy can be used. This fraction of energies could be channeled into nonthermal emissions due to strong shocks of the explosion (see the last paragraph of this section). Moreover, the SN powers are higher than the dissipation of clump collisions, implying that the SN power could drive outflows. This naturally explains why the dusty torus generally exists in AGNs. The dusty torus obscuring the BLR in AGNs is the "dusty corona" of the mid-plane of gaseous disks. This is generally in agreement with the models suggested by Wada & Norman (2002), Wada (2012), and Hönig (2019), which are supported by observations of NGC 3227 (Davies et al. 2006). We leave this to a future paper about forming dust particles and dusty torus.

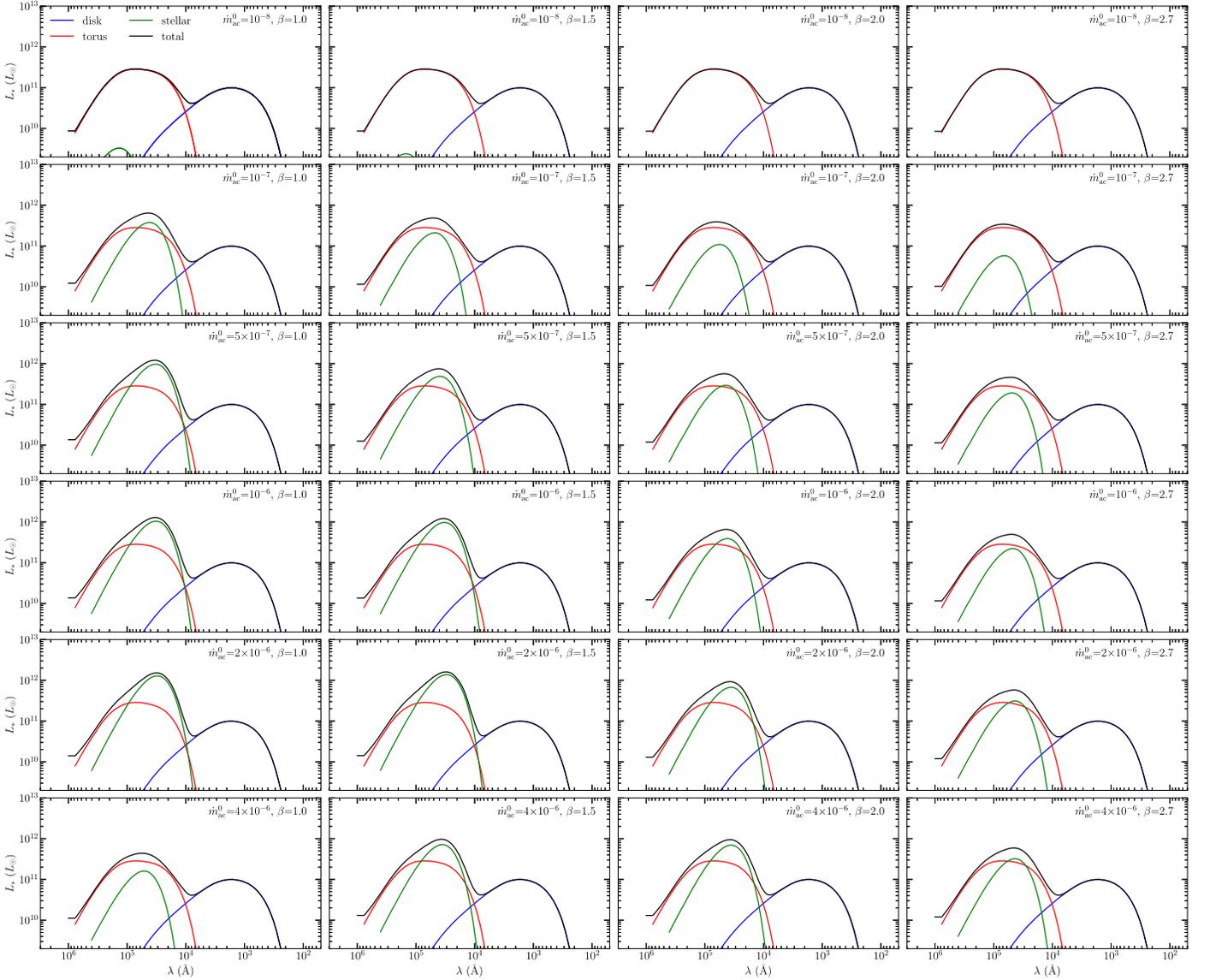

**Figure 19.** Global SED of AGNs with star formation with extinction, which corresponds to Figure 7 ($\dot{m}_w^0 = 10^{-9} M_\odot$ yr$^{-1}$). AGN disks have the same parameters as Figure 18.

Dust masses in the torus of PG quasars have been estimated to be in a wide range of $10^6 \sim 10^8 M_\odot$ by Haas et al. (2003), making use of Infrared Space Observatory observations. More careful estimations of dust masses of PG quasars are between $10^{6.2}$ and $10^{8.7} M_\odot$ with a mean value of $M_{dust} \sim 10^{7.6} M_\odot$ in light of the Two Micron All Sky Survey, Wide-field Infrared Survey Explorer (WISE), and Herschel, combined with Spitzer MIR (5–40 $\mu$m) spectra (Shangguan & Ho 2018). These masses may include a large fraction of dust beyond the present regions ($\gg$0.5 pc). Dust particles can be formed from the SN explosion of AMS in AGN disks, showing that dust mass could correlate with the metallicity of AGNs. Dust mass budget shows that the dusty torus, at least the compact part, can be supported by the present model.

In addition, the explosion energy of massive stars is uncertain, $E_{SN} = (1 - 50) \times 10^{51}$ erg (Woosley et al. 2002) or from observations (see the case of SN2016aps in Nicholl et al. 2020; Suzuki et al. 2021). High-energy SNe may drive fast outflows and sweep the ISM to produce $\gamma$-rays. This may lead to the $\gamma$-ray bubble discovered by Fermi observations in the Milky Way (Su et al. 2010). Other possible origins of the $\gamma$-ray bubble might be originally contributed by a starburst ring (Nguyen & Thompson 2022) or by a relativistic jet (Guo & Mathews 2012). The present model to the GC should be extended to a few hundred parsec to explain the Fermi bubble. Because there is an excess of massive stars in the CNRs of the Milky Way, we prefer that the γ-ray bubble could be driven by SNe of AMSs.

### 5.4. Variations Rising from SNs

When the AMS population reaches stationary states, the stellar contribution keeps constant powers in this torus region. It should be noted that there are frequent SN explosions with a frequency of

$$f_{\rm SN} \approx \frac{N_*}{\tau_{\rm evo}} = 0.1\, N_5 \tau_{\rm 3Myr}^{-1}\, {\rm yr}^{-1}, \quad (46)$$

where $N_5 = N_*/3 \times 10^5$ is the typical numbers of AMSs, and $\tau_{\rm 3Myr} = \tau_{\rm evo}(m_*)/3$ Myr is the main-sequence lifetime of $100\,M_\odot$ for typical values of the AMS populations shown in Figure 4. The SN rates estimated by Davies et al. (2007) of a few solar masses per year are consistent with the predictions of the present model. However, a thorough comparison with these observations could be performed as a subject of future research.

Recently, several superluminous SNs have been found, such as PS1-10adi (Kankare et al. 2017). SNe explosions have been studied for very dense and dusty environments (Shull 1980a; Wheeler et al. 1980). Most emissions are channeled into infrared bands in the dusty gas. However, detailed calculations of light curves of SNe explosions in a very dense medium, which should include geometric effects and scatters of electrons and dust particles, are beyond the scope of this paper. We emphasize that this power is an independent component of SED and does not reverberate with optical variations of AGN disks. We point out the example of I Zw1, which shows the different long trendings between the WISE light curve and the optical (the optical stays constant, but 3.4 μm and 4.6 μm are slowly increasing; see Figures 5 and A1 in Lyu et al. 2019 and Chen et al. 2023, respectively). This could be an indicator of an SN explosion in its torus. A systematic search for NIR reverberations with long-term trending will help us understand the top-heavy MF of AMS populations in AGNs (Y.-J. Chen et al. 2023, in preparation).

Lastly, we stress the following points. Metal enrichment in a single AGN episode due to massive stars also leads to some NIR and MIR emissions. The present model predicts a correlation between metallicity and infrared luminosity. It is worth testing such a correlation for a sample of AGNs.

## 6. Discussions
### 6.1. Implications to GW Detections of LIGO and LISA

Considering the discovery of stellar disks composed of ~200 young massive stars in the GC (e.g., Bartko et al. 2010; von Fellenberg et al. 2022), there must remain some sMBHs after the SN explosion of massive stars in the AGN accretion disks. Indeed, there is growing evidence for such a population of black holes (with a number of ~$2 \times 10^4$; e.g., Hailey et al. 2018; Gottlieb et al. 2020). Exploring the geometric and dynamic relation of those black holes (evolved from massive stars) with the young stars will advance the understanding of physics in the central regions.

Recently, this has drawn attention to the fact that the LIGO-detected GW190521 originating from a black hole merger (Abbott et al. 2020) is plausibly associated with an AGN J124942.3+344929 at $z = 0.438$ (Graham et al. 2020). The binary black holes are massive beyond the black hole formation hypothesis of stellar evolution (Woosley et al. 2002). AGN disks can serve as suitable wombs for the production of massive mergers (McKernan et al. 2012; Graham et al. 2020). This renews broad interest in the fates of compact objects formed from massive stars in AGN disks discussed by Cheng & Wang (1999) and extended by McKernan et al. (2012), Tagawa et al. (2020), and Toyouchi et al. (2022). Compact-object-AMSs as remnants of massive stars are accreting and producing new kinds of multiwavelength emissions (McKernan et al. 2014), bursts of collapsing neutron stars (Perna et al. 2021), and white dwarfs (Zhu et al. 2021), Bondi explosion of black hole hyper-accretion (Wang et al. 2021), and hierarchical merger of compact objects (Cheng & Wang 1999; Yang et al. 2019; Wang et al. 2021). In addition to GW190521, six new events potentially associated with AGNs have been claimed (Graham et al. 2022). The discovery of top-heavy stellar disks between 0.03 and 0.5 pc in the GC (Bartko et al. 2010) implies a significant population of sMBHs there (Nayakshin & Sunyaev 2005). Mergers of these black holes are highly desired for GW detection from LIGO observations. Moreover, sMBHs around the central SMBH as an extreme mass ratio spiral binary could be a source of microhertz GWs, which can be detected by LISA. With this implication, the potential associations of LIGO events with AGNs will be confirmed in the future, providing independent evidence of AMS roles in AGN disks (e.g., Vajpeyi et al. 2022).

### 6.2. GRAVITY+/VLTI Observations

Disentangling the nuclear spectra is a crucial step in unraveling the AGN-starburst (AMS) connections, particularly on the parsec scale, though the Baldwin-Phillips-Terlevich diagram can be used to diagnose. Fortunately, GRAVITY+/VLTI with a spatial resolution of 10 μas can resolve the AMS-ionizing regions. Seyfert 2s (identified by their optical spectra) only show narrow NIR emission lines (Riffel et al. 2006; Müller-Sánchez et al. 2018), but some type II AGNs show broad emission lines in NIR bands, such as intermediate-width Paα in NGC 7674 (Riffel et al. 2006), Brγ line in Mrk1210 (Müller-Sánchez et al. 2018) and NGC 2992 (Davies et al. 2007). GRAVITY+ observations of type II AGNs can show the signatures from different phases of the AMS-ionizing regions, whereas type I AGNs will show two components. In particular, NIR spectroscopic observations of CL AGNs at type II state may reveal the intermediate-width lines from AMS-ionizing clouds. Commonly studied CL AGNs are Mrk 590, Mrk 1018, NGC 1097, NGC 1566, NGC 2617, NGC 2992, NGC 3065, NGC 3516, NGC 7582, and NGC 7603 (e.g., Guolo et al. 2021), which are worth observing NIR spectra. NGC 5273 is the only object so far taken with NIR spectroscopic observations at the type II state (Neustadt et al. 2022); however, broad Paschen series lines disappeared with the Balmers.

Hu et al. (2008) found that there is a large fraction of AGNs with a redshifted intermediate component of Hβ emission line with a mean FWHM of about $10^3$ km s$^{-1}$, whose shifts correlate to the optical Fe II lines from the SDSS DR5 quasar sample. They explained this component is from the inner edge of the torus as the ILR. Interestingly, this width implies the intermediate region scale of approximately ~$10^5 R_{\rm g}$, a scale that is specifically discussed in the present paper. Considering that the torus is clumpy, we would think this component could rise from the AMS-ionizing clouds and escape from the gaps among the clouds. Actually, emission lines from the clumpy torus have been calculated by Mor & Netzer (2012), showing significant contributions. See their Figures 9 and 10 for H$\alpha$, H$\beta$, C IV, Fe II, and Mg II lines. Since the components of emission lines are attributed by AMS photoionization, they will not reverberate the optical-UV continuum variations staying relatively stable. RM campaigns of AGNs with the redshifted component selected from Hu et al. (2008) samples can conveniently test this scenario. Additionally, the two distinguished components will have different extinctions, which can be done by testing Balmer decrements. Fortunately, we find the extinction differences (H$\alpha$/H$\beta \approx$ 2.8 and 4.5 for the two components) in SDSS samples, supporting the presence of the intermediate-broad-line regions (Zhai et al. 2023, in preparation).

### 6.3. AMS Microphysics

#### 6.3.1. Accretion of Stars

Some discussions have been made by Jermyn et al. (2021) and Dittmann et al. (2021), and sophisticated accretion of AMS is needed for future studies. In this subsection, we highlight several aspects of AMS physics that require further investigation in future work. Accretion onto stars is a crucial factor influencing AMS evolution, yet it represents the primary source of uncertainty in current models. Accretion onto young stars is very complicated (e.g., see the review in Hartmann et al. 2016); nevertheless, the situation is more complicated in the AGN disk environment. It could be controlled by several factors: (1) AGN disk density and temperature, and their radial and vertical distributions; (2) radiation fields of the environment including the stars themselves; (3) angular momentum of AGN disks; and (4) fallback gas from the choked winds. Factors (2) and (3) decrease the accretion rates whereas (4) increases the rates of AMSs. On the other hand, the adjustment of equilibrium and radiation of AMSs is to be also discussed similar to that of Population III stars (see details in Omukai & Palla 2003).

Moreover, AMSs may lead to engulfment to open a gap[15] in the AGN disk when its sound speed is smaller than the orbital period around the central SMBH, or by the tidal torque of the interaction between a planet and the disk (see the pioneering studies of Lin & Papaloizou 1979; Goldreich & Tremaine 1980; Lin & Papaloizou 1993). It has been demonstrated that the gap properties are very sensitive to the mass ratios (denoted as $q$, as AMS over the central SMBH), disk relative height (as $H/R$), and viscosity of the disks ($\alpha$), and even the orbits of AMSs (see review in Kley & Nelson 2012). Most of the studies focus on the geometrically thin disks (i.e., $H/R \ll 1$) in the published literature. For example, Fung et al. (2014) and Duffell & MacFadyen (2013) derived analytical formulations of the gap density and engulfment conditions, and they found that the gas density is proportional to $q^{-2.2}\alpha^{1.4}(H/R)^{6.6}$ for the mass ratios of $q \sim 10^{-4}$. However, the current cases of $H/R \sim 1$ (because of the support of the SN explosions) and $q \sim 10^{-7}$ are very different from the cases of planets, which have not been studied. It is urgent to investigate the present cases of AMSs in AGN disks in order to obtain the accretion rates of AMSs in the gap.

---
[15] Here the "gap" could be a tunnel in the AGN disk, since it (i.e., the torus part) is geometrically thick. Within the tunnel, the AMS accretion is still suppressed. Such a configuration is to be demonstrated through numerical simulations.

We note that the formation of massive stars through competitive accretion (e.g., Bonnell et al. 2001) in environments of star clusters is very different from the present case. Bonnell's mode is for the formation of the initial population of young stars. In a normal mode of star formation, this is also the only episode of star formation. That is, after the cluster forms, the stars in it just evolve normally on the HR diagram, just one episode. The whole idea of AMS is that after initial formation, which, by the way, can very well proceed by competitive accretion, the newly formed stellar population can *continue* to evolve by accretion from the environment. Our scenario, therefore, is distinct from that of Bonnell et al. (2001).

#### 6.3.2. AGN Disks and Spatial Distributions

For simplicity, we still use the Shakura-Sunyaev model to estimate the AGN disk in the $10^5 R_g$ regions. However, the AMS radiation could dominate the released gravitational energy bound by the central SMBH. Taking into account all sources of energy, Thompson et al. (2005) developed a model that introduced the Toomre parameter $Q = 1$ for self-gravitating disk structures and radiation, but the model did not include the stellar MFs. We need to improve the disk model by including AMS radiation, SN explosion, and the Toomre criterion for a self-consistent model in the future. This allows us to build a new model of the torus. Such a model with more sophisticated considerations is expected to apply to model M87 (Sarzi et al. 2018), M31 nuclear regions.

In this paper, we focus on the regions of $\sim 10^5 R_g$, located inside the dusty torus. For $\sim 10^{3-4} R_g$ regions, the radial motion timescales are $\lesssim 10^6$ yr. This timescale is too short for an AMS to finish its main-sequence evolution. In this case, the MFs cannot reach stationary states, so the maximum masses of AMS are significantly less than those in $\sim 10^5 R_g$. The intrinsic SED could be directly observed and contributes to SED in OUV bands, and would be interesting to compare with observations. As we show below, the accreting processes of AMSs are unlikely to be stationery because of the interaction of stellar winds with the Bondi accretion flows.

Equation (1) describes the entire regions shown in Figure 1. Wang et al. (2010, 2011, 2012) studied the radial structures of the star formation and SN explosion for transportation of the angular momentum and metal production by assuming a top-heavy IMF. It is imperative to integrate the current model with that proposed by Wang et al. (2010, 2011, 2012) for the global scenario of AGNs with AMS. This integration may change SED from the AMS due to their distribution dependence on the radius of the AGN disks.

#### 6.3.3. Recycled Stellar Winds

The present model includes the fallback of the recycled gas (the choked winds) through the accretion rates. However, we note that the fallback rates are subject to highly uncertain in the AMSs. In addition, the strong interaction between the stellar winds and infalling gas forms a high March number shock and probably triggers hydrogen burning. This additional energy may change the evolution of AMS. Moreover, star formation efficiency and the KS law could be changed by metallicity (e.g., Shi et al. 2014). The yields of SN explosion also depend on the metallicity of the progenitors. The various effects of the recycled gas are expected to significantly impact the evolution of AGN-starburst systems, resulting in an increased production of metals and the generation of infrared excesses that go beyond the reprocessed emissions from the accretion disks.

## 7. Conclusions

We investigate the population evolution of AMS in AGN accretion disks for metal production in a single active episode. In the dense environment of AGN accretion disks, stellar winds are choked by the disk medium, forming a new kind of stellar population with massive envelopes. Accretion, stellar winds, and fallback of the choked winds to the core star determine the evolution and metal production. We show that the AMS mass distribution functions are generally quickly evolving through accretion regardless of the input MFs, leading to top-heavy MFs. From the numerical calculations, we find the following:

1. If the strength of stellar winds is not comparable to that of accretion (including fallback of the choked winds), stars exhibit a pile-up distribution around the maximum masses determined by the balance between supernova explosion after stellar evolution and accretion.
2. Stellar MFs show cutoff pile-up distributions when the stellar winds are strong enough. Stars pile up around the cutoff masses determined by the balance between wind loss and accretion gain. This balance provides a new mechanism to generate a narrow window of massive stars for the efficient production of iron elements through PISNe in AGNs. It should be noted that the pile-up masses are to be theoretically fixed through detailed considerations of AMS accretion rates and wind losses.
3. Except for pile-up distributions, the MFs generally show top-heavy distributions as $m_*\Psi \propto m_*^{\sim 0}$ from low-mass to several $10^2\,M_\odot$ whatever the IMFs of the input. The top-heavy MFs conveniently explain the high metallicity of quasars from low- to high-$z$. We apply this simple model to the GC, showing an excellent agreement with the observed stellar disk composed of young stars in the GC.
4. Meanwhile, the gas disk may experience vertical expansion due to SN explosions. The random motion of dusty clumps can be supported by the supply of SN explosions, naturally forming a dusty torus for the AGN unification scheme. We expect to link the metallicity and the torus.
5. SN explosion of these massive stars produces significant radiation absorbed by the dusty gas and converted into infrared bands. We show that this generates an NIR/MIR bump. Monitoring AGNs in infrared and optical bands can generally detect delayed responses of the infrared continuum. AMS-contributed IR emissions as an independent component lead to a long-term trend in IR variations. We predict that there is a correlation between iron and excess infrared emission beyond the reprocessing of the emissions from the AGN disk.
6. We expect that super-Eddington AGNs are iron-rich objects, which show strong Fe II lines in the optical band. This expectation generally corresponds to the case of AMS with high accretion rates in the present scenario. For high-$z$ quasars, AMS in accretion disks can generate iron, and they would be super-Eddington accretors with enhanced optical Fe II lines and infrared emissions observed by JWST.

The present model can be observationally tested by two approaches. Intermediate-broad lines are allowed from the clumpy torus, which originates from the photoionization of AMS, and are expected not to reverberate with the optical and UV continuum variations. GRAVITY+/VLTI is able to spatially resolve the ILR. Moreover, redshifts of the intermediate H$\beta$ line would correlate with the geometrical thickness of the torus, and thus with the iron abundance. These effects would be conveniently detected. Remnants of massive stars may leave numerous compact objects, such as sMBHs, or themselves form binaries or normal stars in galactic centers. Prospective examinations of the current model could be performed on the GC and some type II AGNs, in particular, revealing 100 and $10^{-3}$ Hz gravitational waves through LIGO, LISA, or Taiji/Tianqin observations.

### Acknowledgments

J.M.W. thanks X.-F. Zhang, S.-L. Bi, and Z. Han for useful discussions on stellar physics. The authors also thank members of the IHEP AGN Group for helpful discussions. We acknowledge the support from the National Key R&D Program of China (2020YFC2201400, 2021YFA1600404), NSFC (NSFC-11991050, -11991054, -11833008), the International Partnership Program of the Chinese Academy of Sciences (113111KYSB20200014) and the science research grants from the China Manned Space Project with NO. CMS-CSST-2021-A05. Y.R.L. acknowledges financial support from the NSFC through grant Nos. 11922304 and 12273041 and from the Youth Innovation Promotion Association CAS. P.D. acknowledges financial support from NSFC grants NSFC-12022301, and 11991051. L.C.H. was supported by the NSFC (11721303, 11991052, 12011540375, and 12233001), the National Key R&D Program of China (2022YFF0503401), and the China Manned Space Project (CMS-CSST-2021-A04, CMS-CSST-2021-A06).

## Appendix A
## Stellar Evolution

Massive stars just keep a small fraction of their initial masses when they reach the last evolutionary stage due to strong stellar winds. Theoretical calculations of the evolution of Population I and II stars are limited to $\sim 120\,M_\odot$ due to the effect of stellar winds (e.g., Maeder 1992; Tout et al. 1996; Ekström et al. 2012), and calculations predict a direct collapse into black holes without significant ejections (Woosley et al. 2002). Schaerer (2002) investigated the evolution of Population III stars ($\sim 10^3\,M_\odot$), and only Belkus et al. (2007) generally studied the evolution of very massive stars (up to $10^3\,M_\odot$). Unlike stars in ordinary environments, however, the fates of stars in the extremely high-density environments of AGN disks may be fundamentally modified, as discussed in the present paper.

We employ the theoretical results of main-sequence lifetimes from Ekström et al. (2012) and Georgy et al. (2013), but it only applies to $\lesssim 120\,M_\odot$. However, its extension to very massive stars is consistent with the results of Schaerer (2002). We employ the effective temperatures and mass–luminosity relation of stars from Schaerer (2002) and the mass–luminosity relation from the same reference. The mass–luminosity relation approximately follows $L_* \propto m_*^4$. For the present goal of this paper, these fitting results of stellar physics are accurate enough and are shown in Figure 20.

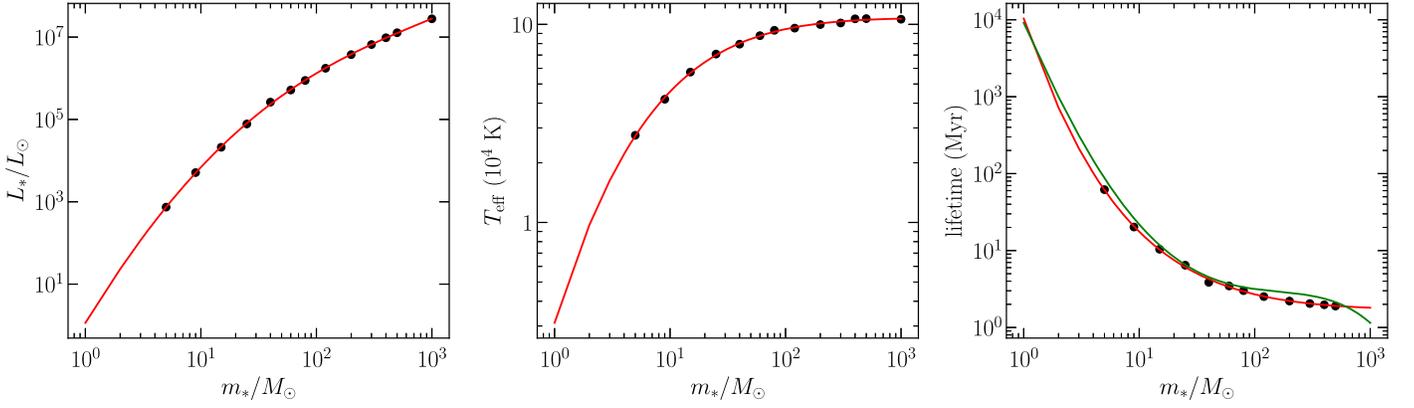

**Figure 20.** Mass–luminosity relation, main-sequence lifetime, and effective temperatures of stars from Schaerer (2002). Formulations of the fitting results are given in the main text. The green line (Ekström et al. 2012; Georgy et al. 2013; invalid for $\gtrsim 120 M_\odot$) is in good agreement with that of Schaerer (2002).

## Appendix B
## Radiation from Accretion onto Stars

The power of accretion onto a star is given by

$$\frac{L_*}{L_\odot} = 3.5 \times 10^{-2} \dot{\mathcal{M}} \left(\frac{m_*}{M_\odot}\right)\left(\frac{R_*}{R_\odot}\right)^{-1}, \quad \text{(B1)}$$

where we take the stellar surface as the inner radius of the accretion onto stars. This luminosity is much lower than the Eddington and significantly fainter than radiation power from a main-sequence star (see Equation (37)). Therefore, the accretion power of AMS is negligible even for $\dot{\mathcal{M}} \sim 10^2$, corresponding to $10^{-7} M_\odot \, \text{yr}^{-1}$ for a solar-mass star.

## Appendix C
## Medium Metallicity in GC

In order to compare with GC environments, we calculate the metals produced by massive stars after the AGN disk was depleted. In such a case, we calculate Fe abundance as follows:

$$[\text{Fe, Mg/H}] = \log_{10}\left[\frac{\sum_i \Delta M_{\text{Fe, Mg}}^i}{\mathcal{A}_{\text{Fe,Mg}} \sum_i (M_*^i - M_{\text{rem}}^i)}\right] - \log_{10}\left(\frac{N_{\text{Fe,Mg}}}{N_{\text{H}}}\right)_\odot. \quad \text{(C1)}$$

where $\Delta M_{\text{Fe,Mg}}^i$ is the iron or magnesium masses produced by the $i$th star with a mass of $M_*^i$, and $M_{\text{rem}}^i$ is its the final remnant mass. This is about the fraction of iron/magnesium mass to its initial star. It should be noted that this is different from Equation (34). Considering that $M_0$ is not well understood, we think that the [Fe/H] can be explained by the present model. Figure 21 shows the model for GC abundance. In the future, the abundance of other elements should be calculated for a detailed application to the GC young stellar disks.

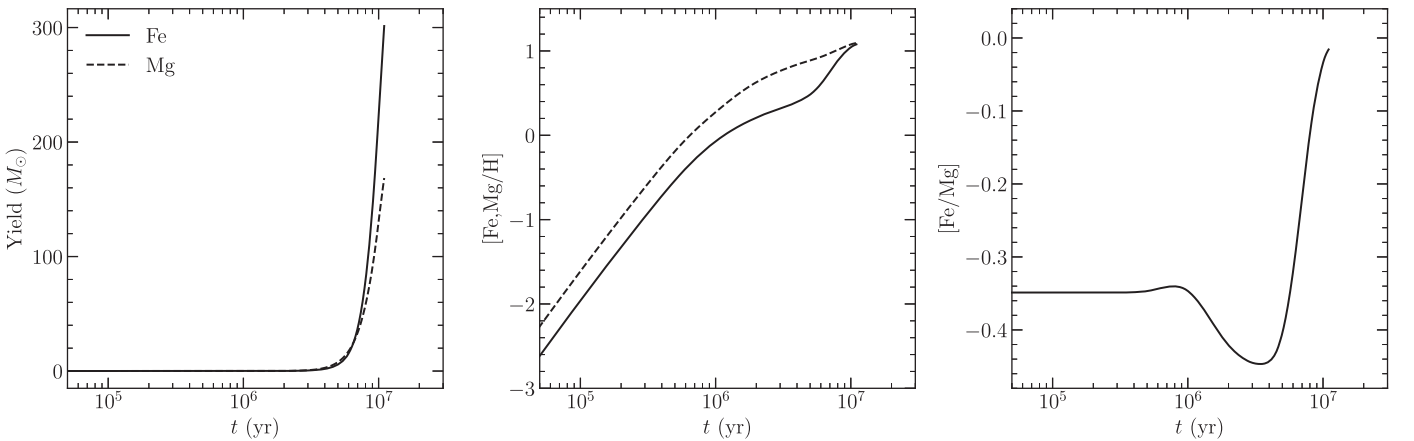

**Figure 21.** Yields of Fe and Mg elements produced by the present model are shown by the left panel. We take $E_{\text{SN}} = 30 \times 10^{51}$ ergs in the model. The Fe and Mg abundances are about 10 times that of the solar at $\sim 10^7$ yr, but Fe is just about the solar abundance before $\sim 5 \times 10^6$ yr. [$\alpha$/Fe] is just the solar value. This result is generally in agreement with the Chandra observations of Hua et al. (2023).

## Appendix D
## Properties of MFs at Cutoff Pile-up Mass

MFs at high-$\dot{m}_w$ seem to be infinite with time. This can be analytically investigated briefly. When growth timescale $t_{\rm grow} = m_*/\dot{m}_{\rm ac}$ is much shorter than the evolutionary $\tau_{\rm evo}$ (i.e., SN rates are very low), Equation (1) is reduced to

$$\frac{\partial \Psi}{\partial t} \approx \dot{\mathcal{S}}_{\rm AD}(m_*, t), \quad \text{(D1)}$$

where $\dot{\mathcal{R}}_{\rm SN}(m_*, t) \approx 0$, and $\dot{m}_* = 0$. We thus have

$$\Psi(m_{\rm cut}) = \int_0^t \dot{\mathcal{S}}_{\rm AD}(m_*, t') dt'. \quad \text{(D2)}$$

It thus tends to infinity at the cutoff pile-up mass, particularly in the cases with strong stellar winds (see Figures 7 and 8) unless the input of stars is quenched. After quenching the input, the stellar populations undergo pure evolution, similar to the case of the GC.

## ORCID iDs

Jian-Min Wang https://orcid.org/0000-0001-9449-9268
Yan-Rong Li https://orcid.org/0000-0001-5841-9179
Yu-Yang Songsheng https://orcid.org/0000-0003-4042-7191
Luis C. Ho https://orcid.org/0000-0001-6947-5846
Yong-Jie Chen https://orcid.org/0000-0003-4280-7673
Pu Du https://orcid.org/0000-0002-5830-3544
Ye-Fei Yuan https://orcid.org/0000-0002-7330-4756

## References

Abbott, D. C., & Conti, P. S. 1987, ARA&A, 25, 113
Abbott, R., Abbott, T. D., Abraham, S., et al. 2020, PhRvL, 125, 101102
Adams, F. C., & Fatuzzo, M. 1996, ApJ, 464, 256
Antonucci, R. 1993, ARA&A, 31, 473
Artymowicz, P., Lin, D., & Wampler, E. J. 1993, ApJ, 409, 592
Asplund, M., Grevesse, N., Sauval, A. J., & Scott, P. 2009, ARA&A, 47, 481
Bajtlik, S., Duncan, R. C., & Ostriker, J. P. 1988, ApJ, 327, 570
Baldwin, J. A., Ferland, G. J., Korista, K. T., et al. 2004, ApJ, 615, 610
Barth, A. J., Martini, P., Nelson, C. H., et al. 2003, ApJ, 594, L95
Bartko, H., et al. 2010, ApJ, 708, 834
Bastian, N., Covey, K. R., & Meyer, M. R. 2010, ARA&A, 48, 339
Basu, S., & Jones, C. E. 2004, MNRAS, 347, L47
Belkus, H., van Bever, J., & Vanbeveren, D. 2007, ApJ, 659, 1576
Bell, K. R., & Lin, D. N. C. 1994, ApJ, 427, 987
Bentz, M., Denney, K. D., Frier, K., et al. 2013, ApJ, 767, 149
Bonnell, I. A., Bate, M. R., Clarke, C. J., et al. 2001, MNRAS, 323, 785
Bonnell, I. A., & Rice, W. K. M. 2008, Sci, 321, 1060
Boroson, T., & Green, P. 1992, ApJS, 80, 109
Cantiello, M., Jermyn, A. S., & Lin, D. N. C. 2021, ApJ, 910, 94
Castor, J., McCray, R., & Weaver, R. 1975, ApJL, 200, L107
Chang, J. S., & Cooper, G. 1970, JCoPh, 6, 1
Chen, Y.-J., Liu, J.-R., Zhai, S., et al. 2023, MNRAS, 522, 3439
Chen, Y.-M., Wang, J.-M., Yan, C.-S., et al. 2009, ApJL, 695, L130
Cheng, K. S., & Wang, J.-M. 1999, ApJ, 521, 502
Cheng, S. J., & Loeb, A. 2022, arXiv:2208.04337
Chiaberge, M., & Ghisellini, G. 1999, MNRAS, 306, 551
Collin, S., & Zahn, J.-P. 1999, A&A, 344, 433
Collin, S., & Zahn, J.-P. 2008, A&A, 477, 419
Conroy, C. 2013, ARA&A, 51, 393
Crowther, P. A., Schnurr, O., Hirschi, R., et al. 2010, MNRAS, 408, 731
Dabringhausen, J., Kroupa, P., & Baumgardt, H. 2009, MNRAS, 394, 1529
Davies, M. B., & Lin, D. N. C. 2020, MNRAS, 498, 3452
Davies, R. I., Müller Sánchez, F., Genzel, R., et al. 2007, ApJ, 671, 1388
Davies, R. I., Thomas, J., Genzel, R., et al. 2006, ApJ, 646, 754
De Rosa, G., Decarli, R., Walter, F., et al. 2011, ApJ, 739, 56
De Rosa, G., Venemans, B. P., Decarli, R., et al. 2014, ApJ, 790, 145
Derdzinski, A., & Mayer, L. 2023, MNRAS, 521, 4522
Dib, S., Shadmehri, M., Padoan, P., et al. 2010, MNRAS, 405, 401
Dietrich, M., Appenzeller, I., Vestergaard, M., et al. 2002, ApJ, 564, 581
Dietrich, M., Hamann, F., Appenzeller, I., et al. 2003, ApJ, 596, 817
Dittmann, A., Cantiello, M., & Jermyn, A. S. 2021, ApJ, 916, 48
Dittmann, A. J., Jermyn, A. S., & Cantiello, M. 2022, ApJ, 946, 56
Du, P., Lu, K.-X., Zhang, Z.-X., et al. 2016, ApJ, 825, 126
Du, P., & Wang, J.-M. 2019, ApJ, 886, 42
Du, P., Wang, J.-M., Hu, C., et al. 2014, MNRAS, 438, 2828
Duffell, P. C., & MacFadyen, A. I. 2013, ApJ, 769, 41
Eilers, A.-C., Simcoe, R. A., Yue, M., et al. 2023, ApJ, 950, 68
Ekström, S., Georgy, C., Eggenberger, P., et al. 2012, A&A, 537, 146
Elmegreen, B. G. 2000, ApJ, 539, 342
Essex, C., Basu, S., Prehl, J., et al. 2020, MNRAS, 494, 1579
Fan, X., & Wu, Q. 2023, ApJ, 944, 159
Freudling, W., Corbin, M. R., & Korista, K. T. 2003, ApJ, 587, L67
Fung, J., Shi, J.-M., & Chiang, E. 2014, ApJ, 782, 88
Garnica, K., Negrete, C. A., Marziani, P., et al. 2022, A&A, 667, 105
Genzel, R., Eisenhauer, F., & Gillessen, S. 2010, RvMP, 82, 3121
Georgy, C., Ekström, S., Eggenberger, P., et al. 2013, A&A, 558, 103
Gohil, R., & Ballantyne, D. R. 2017, MNRAS, 468, 4944
Goldreich, P., & Tremaine, S. 1980, ApJ, 241, 425
Goodman, J. 2003, MNRAS, 339, 937
Goodman, J., & Tan, J. C. 2004, ApJ, 608, 108
Gottlieb, A. M., Eikenberry, S. S., Ackley, K., et al. 2020, ApJ, 896, 32
Gräfener, G., & Hamann, W. R. 2008, A&A, 482, 945
Graham, M. J., Ford, K., McKernan, B., et al. 2020, PhRvL, 124, 251102
Graham, M. J., McKernan, B., Ford, K., et al. 2022, ApJ, 942, 99
Greggio, L. 2005, A&A, 441, 1055
Greggio, L., & Renzini, A. 1983, A&A, 118, 217
Grudić, M. Y., Guszejnov, D., Offner, S. S. R., et al. 2022, MNRAS, 512, 216
Guo, F., & Mathews, W. G. 2012, ApJ, 756, 181
Guolo, M., Ruschel-Dutra, D., Grupe, D., et al. 2021, MNRAS, 508, 144
Haas, M., Klaas, U., Müller, S. A. H., et al. 2003, A&A, 402, 87
Hailey, C. J., Mori, K., Bauer, F. E., et al. 2018, Natur, 556, 70
Hamann, F., & Ferland, G. 1992, ApJ, 391, L53
Hamann, F., & Ferland, G. 1993, ApJ, 418, 11
Hamann, F., & Ferland, G. 1999, ARA&A, 37, 487
Hartmann, L., Herczeg, G., & Calvet, N. 2016, ARA&A, 54, 135
Heger, A., & Woosley, S. E. 2002, ApJ, 567, 532
Hobbs, A., & Nayakshin, S. 2009, MNRAS, 394, 191
Hoffmann, K. H., Essex, C., Basu, S., et al. 2018, MNRAS, 478, 2113
Hönig, S. 2019, ApJ, 884, 171
Hu, C., Du, P., Lu, K.-X., et al. 2015, ApJ, 804, 138
Hu, C., Wang, J.-M., Ho, L. C., et al. 2008, ApJ, 683, L115
Hua, Z., Li, Z., Zhang, M., et al. 2023, MNRAS, 522, 635
Impellizzeri, C. M. V., et al. 2019, ApJ, 884, L28
Iwamuro, F., Motohara, K., Maihara, T., et al. 2002, ApJ, 565, 63
Jermyn, A. S., Dittmann, A. J., Cantiello, M., et al. 2021, ApJ, 914, 105
Jiang, L., Fan, X., Vestergaard, M., et al. 2007, AJ, 134, 1150
Kankare, E., Kotak, R., Mattila, S., et al. 2017, NatAs, 1, 865
Kaspi, S., Smith, P., Netzer, H., et al. 2000, ApJ, 533, 631
Kato, S., Fukue, J., & Mineshige, S. 2008, Black-hole Accretion Disks: Towards a New Paradigm (Kyoto: Kyoto Univ. Press)
Kelly, B. C., & Shen, Y. 2013, ApJ, 764, 45
Kelly, B. C., Vestergaard, M., Fan, X., et al. 2010, ApJ, 719, 1315
Kennicutt, R. C., & Evans, N. J. 2012, ARA&A, 50, 531
Kennicutt, R. C., Jr 1998, ARA&A, 36, 189
Khrykin, I. S., Hennawi, J. F., Worseck, G., et al. 2021, MNRAS, 505, 649
King, A., & Nixon, C. 2015, MNRAS, 453, L46
Kley, W., & Nelson, N. P. 2012, ARA&A, 50, 211
Kobayashi, C., Tsujimoto, T., & Nomoto, K. 2000, ApJ, 539, 26
Kolykhalov, P. I., & Syunyaev, R. A. 1980, SvAL, 6, 357
Koshida, S., Minezaki, T., Yoshii, Y., et al. 2014, ApJ, 788, 159
Krolik, J. H., & Begelman, M. C. 1988, ApJ, 329, 702
Kroupa, P., & Jerabkova, T. 2021, arXiv:2112.10788
Krügel, E., Tutukov, A., & Loose, H. 1983, A&A, 124, 89
Krumholz, R., & McKee, C. F. 2008, Natur, 451, 1082
Kurk, J. D., Walter, F., Fan, X., et al. 2007, ApJ, 669, 32
Lamers, H., & Cassinelli, J. 1999, Introduction to Stellar Winds (Cambridge: Cambridge Univ. Press)
Lamers, H., & Levesque, E. M. 2017, Understanding Stellar Evolution (Bristol: Institute of Physics Publishing)
Langer, N. 1989, A&A, 220, 135
Levin, Y., & Beloborodov, A. M. 2003, ApJ, 590, L33
Li, F.-L., Liu, Y., Fan, X., et al. 2023, ApJ, 950, 161
Li, Y.-R., Wang, J.-M., & Ho, L. C. 2012, ApJ, 749, 187

Lin, D. N. C., & Papaloizou, J. C. B. 1979, MNRAS, 188, 191
Lin, D. N. C., & Papaloizou, J. C. B. 1993 (Protostars and Planets III) ed. E. H. Levy & J. I. Lunine (Tucson, AZ: Univ. Arizona Press), 749
Loose, H. H., Kruegel, E., & Tutukov, A. 1982, A&A, 105, 342
Lu, X., Zhang, Q., Kauffmann, J., et al. 2019, ApJ, 872, 171
Lyu, J., & Rieke, G. 2022, Univ, 8, 304
Lyu, J., Rieke, G., & Smith, P. S. 2019, ApJ, 886, 33
Maeder, A. 1992, A&A, 264, 105
Maeder, A., & Meynet, G. 2000, A&A, 361, 159
Maiolino, R., & Mannucci, F. 2019, A&ARv, 27, 3
Marconi, A., Risaliti, G., Gilli, R., et al. 2004, MNRAS, 351, 169
Marks, M., Kroupa, P., Dabringhausen, J., & Pawlowski, M. S. 2012, MNRAS, 422, 2246
Martini, P. 2004, in ASP Conf. Ser. 311, AGN Physics with the Sloan Digital Sky Survey, ed. G. T. Richards & P. B. Hall (San Francisco, CA: ASP), 389
Marziani, P., & Sulentic, J. W. 2014, MNRAS, 442, 1211
Matteucci, F., & Greggio, L. 1986, A&A, 154, 279
Mazzucchelli, C., Bañados, E., Venemans, B. P., et al. 2017, ApJ, 849, 91
McKee, C. F., & Ostriker, E. C. 2007, ARA&A, 45, 565
McKee, C. F., & Tan, J. C. 2008, ApJ, 681, 771
McKernan, B., Ford, K. E. S., Kocsis, B., et al. 2014, MNRAS, 441, 900
McKernan, B., Ford, K. E. S., Lyra, W., & Perets, H. B. 2012, MNRAS, 425, 460
Milosavljević, M., & Loeb, A. 2004, ApJL, 604, L45
Minezaki, T., Yoshii, Y., Kobayashi, Y., et al. 2019, ApJ, 886, 150
Mor, R., & Netzer, H. 2012, MNRAS, 420, 526
Müller-Sánchez, F., Hicks, E. K. S., Malkan, M., et al. 2018, ApJ, 858, 48
Nagao, T., Marconi, A., & Maiolino, R. 2006, A&A, 447, 157
Nayakshin, S., & Sunyaev, R. 2005, MNRAS, 364, L23
Negrete, C. A., Dultzin, D., & Marziani, P. 2018, A&A, 620, 118
Nenkova, M., Sirocky, M. M., Ivezić, Ž., & Elitzue, M. 2008, ApJ, 685, 147
Netzer, H. 2015, ARA&A, 53, 365
Neustadt, J. M. M., Hinkle, J. T., Kochanek, C. S., et al. 2022, MNRAS, 521, 3810
Nguyen, D. D., & Thompson, T. A. 2022, ApJ, 935, L24
Nicholl, M., Blanchard, P. K., Berger, E., et al. 2020, NatAs, 4, 893
Nomoto, K., Kobayashi, C., & Tominaga, N. 2013, ARA&A, 51, 457
Nugis, T., & Lamers, H. 2000, A&A, 360, 227
Ohkubo, T., Nomoto, K., Umeda, H., et al. 2009, ApJ, 706, 1184
Omukai, K., & Palla, F. 2003, ApJ, 589, 677
Onoue, M., Bañados, E., Mazzucchelli, C., et al. 2020, ApJ, 898, 105
Osterbrock, D., & Ferland, G. J. 2006, Astrophysics of Gaseous Nebulae and Active Galactic Nuclei (2nd ed.; Sausalito, CA: Univ. Science Books)
Ostriker, E. C., McKee, C. F., & Leroy, A. K. 2010, ApJ, 721, 975
Owen, J. E., & Lin, D. N. C. 2023, MNRAS, 519, 397
Paczyński, B. 1978, AcA, 28, 91
Panamarev, T., & Kocsis, B. 2022, MNRAS, 517, 6205
Paumard, T., et al. 2006, ApJ, 643, 1011
Perna, R., Tagawa, H., Haiman, Z., et al. 2021, ApJ, 915, 10
Peterson, B. M. 1993, PASP, 105, 247
Peterson, B. M. 2014, SSRv, 183, 253
Peterson, B. M., Ferrarese, L., & Gilbert, K. M. 2004, ApJ, 613, 682
Pier, E. A., & Krolik, J. 1992, ApJ, 401, 99
Pier, E. A., & Krolik, J. 1993, ApJ, 418, 673
Qi, Y.-Q., Liu, T., Cai, Z.-Y., & Sun, M. 2022, ApJ, 934, 1
Rees, M. J. 1984, ARA&A, 22, 471
Riffel, R., Rodríguez-Ardila, A., & Pastoriza, M. G. 2006, A&A, 457, 61
Ruiz-Lapuente, P., & Canal, R. 1998, ApJ, 497, L57
Sameshima, H., Yoshii, Y., & Kawara, K. 2017, ApJ, 834, 203
Sameshima, H., Yoshii, Y., Matsunaga, N., et al. 2020, ApJ, 904, 162
Sarkar, A., Ferland, G. J., Chatzikos, M., et al. 2021, ApJ, 907, 12
Sarzi, M., Spiniello, C., La Barbera, F., et al. 2018, MNRAS, 478, 4084
Schaerer, D. 2002, A&A, 382, 28
Schawinski, K., Koss, M., Berney, S., et al. 2015, MNRAS, 451, 2517
Schindler, J.-T., Farina, E. P., Bañados, E., et al. 2020, ApJ, 905, 51
Schneider, F. R. N., Izzard, R. G., de Mink, S. E., et al. 2014, ApJ, 780, 117
Semenov, D., Henning, T., Helling, C., et al. 2003, A&A, 410, 611
Shakura, N. I., & Sunyaev, R. A. 1973, A&A, 24, 337
Shangguan, J., & Ho, L. C. 2018, ApJ, 854, 158
Shemmer, O., & Netzer, H. 2002, ApJ, 567, L19
Shen, Y., Brandt, W. N., Dawson, K. S., et al. 2015, ApJS, 216, 4
Shen, Y., Grier, C. J., Horne, K., et al. 2023, arXiv:2305.01014
Shi, Y., Armus, L., Helou, G., et al. 2014, Natur, 514, 335
Shin, J., Nagao, T., Woo, J.-H., et al. 2019, ApJ, 874, 22
Shin, J., Woo, J.-H., Nagao, T., & Kim, S. C. 2013, ApJ, 763, 58
Shlosman, I., & Begelman, M. C. 1987, Natur, 329, 810
Shull, J. M. 1980a, ApJ, 237, 769
Shull, J. M. 1980b, ApJ, 238, 860
Sirko, E., & Goodman, J. 2003, MNRAS, 341, 501
Small, T. A., & Blandford, R. D. 1992, MNRAS, 259, 725
Śniegowska, M., Marziani, P., Czerny, B., et al. 2021, ApJ, 910, 115
Stevens, I. R., Blondin, J. M., & Pollock, A. M. T. 1992, ApJ, 386, 265
Su, M., Slatyer, T. R., & Finkbeiner, D. P. 2010, ApJ, 724, 1044
Suganuma, M., Yoshii, Y., Kobayashi, Y., et al. 2006, ApJ, 639, 46
Suzuki, A., Nicholl, M., Moriya, T. J., et al. 2021, ApJ, 908, 99
Tagawa, H., Haiman, Z., & Kocsis, B. 2020, ApJ, 898, 25
Tan, J. C. 2000, ApJ, 536, 173
Tan, J. C., & McKee, C. F. 2004, ApJ, 603, 383
Tanvir, T. S., Krumholz, M. R., & Federrath, C. 2022, MNRAS, 516, 5712
Terlevich, R., & Melnick, J. 1985, MNRAS, 213, 841
Terlevich, R., Tenorio-Tagle, G., Franco, J., et al. 1992, MNRAS, 255, 713
Terlevich, R., Tenorio-Tagle, G., Rozyczka, M., et al. 1995, MNRAS, 272, 198
Thompson, T. A., Quataert, E., & Murray, N. 2005, ApJ, 630, 167
Tominaga, N., Umeda, H., & Nomoto, K. 2007, ApJ, 660, 516
Tout, C., Pols, O. R., Eggleton, P. P., & Han, Z. 1996, MNRAS, 281, 257
Toyouchi, D., Inayoshi, K., Ishigaki, M. N., et al. 2022, MNRAS, 512, 2573
Tristram, K. R. W., et al. 2007, A&A, 474, 837
U, V., Barth, A., Vogler, H. A., et al. 2022, ApJ, 925, 52
Umeda, H., & Nomoto, K. 2008, ApJ, 673, 1014
Vajpeyi, A., Thrane, E., Smith, E., et al. 2022, ApJ, 931, 82
Verner, E., Bruhweiler, F., Verner, D., et al. 2004, ApJ, 611, 780
Verner, E., Bruhweiler, F., Verner, D., et al. 2013, ApJL, 592, L59
Villarroel, B., Imaz, I., Lusso, B., et al. 2020, MNRAS, 495, 4419
Villarroel, B., Nyholm, A., Karlsson, T., et al. 2017, ApJ, 837, 110
Vink, J. S., de Koter, A., & Lamers, H. 2001, A&A, 369, 574
Volonteri, M., Sikora, M., Lasota, J.-P., et al. 2013, ApJ, 775, 94
von Fellenberg, S. D., Gillessen, S., Stadler, J., et al. 2022, ApJ, 932, L6
Wada, K. 2012, ApJ, 758, 66
Wada, K., & Norman, C. A. 2002, ApJ, 566, L21
Walker, D. L., Longmore, S. N., Zhang, Q., et al. 2018, MNRAS, 474, 2373
Wang, J.-M., Chen, Y.-M., & Hu, C. 2006, ApJ, 637, L85
Wang, J.-M., Du, P., Baldwin, J. A., et al. 2012, ApJ, 746, 137, (Paper-II)
Wang, J.-M., Ge, J.-Q., Hu, C., et al. 2011, ApJ, 739, 3, (Paper-I)
Wang, J.-M., Hu, C., Li, Y.-R., et al. 2009, ApJ, 697, L141
Wang, J.-M., Liu, J.-R., Ho, L. C., & Du, P. 2021, ApJ, 911, L14
Wang, J.-M., Liu, J.-R., Ho, L. C., & Du, P. 2021, ApJ, 916, L17
Wang, J.-M., Yan, C.-S., Gao, H.-Q., et al. 2010, ApJ, 719, L148
Wang, M., Ma, Y., & Wu, Q.-W. 2023, MNRAS, 520, 4502
Wang, S., Jiang, L., Shen, Y., et al. 2022, ApJ, 925, 121
Warner, C., Hamann, F., & Dietrich, M. 2004, ApJ, 608, 136
Wheeler, J. C., Mazurek, T. J., & Sivaramakrishnan, A. 1980, ApJ, 237, 781
Wills, B. J., Netzer, H., & Wills, D. 1985, ApJ, 288, 94
Woosley, S. E., Heger, A., & Weaver, T. A. 2002, RvMP, 74, 1015
Yang, J., Wang, F., Fan, X., et al. 2021, ApJ, 923, 262
Yang, J., Wang, F., Fan, X., et al. 2023, ApJL, 951, L5
Yang, Q., Shen, Y., Liu, X., et al. 2020, ApJ, 900, 58
Yang, Y., Bartos, I., Gayathri, V., et al. 2019, PhRvL, 123, 181101
Yoshii, Y., Sameshima, H., Tsujimoto, T., et al. 2022, ApJ, 937, 61
Yu, Z., Martini, P., Penton, A., et al. 2021, MNRAS, 507, 3771
Yu, Z., Martini, P., Penton, A., et al. 2022, MNRAS, 522, 4132
Yusof, N., Hirschi, R., Eggenberger, P., et al. 2022, MNRAS, 511, 2814
Zhu, J.-P., Yang, Y.-P., Zhang, B., et al. 2021, ApJL, 914, L19
Zhuang, M., & Ho, L. C. 2020, ApJ, 896, 108
Zinnecker, H. 1982, NYASA, 395, 226
Zinnecker, H., & Yorke, H. W. 2007, ARA&A, 45, 481